\documentstyle[12pt,epsfig]{article}

\catcode`\@=11
\def\@citex[#1]#2{%
\if@filesw \immediate \write \@auxout {\string \citation {#2}}\fi
\@tempcntb\m@ne \let\@h@ld\relax \def\@citea{}%
\@cite{%
  \@for \@citeb:=#2\do {%
    \@ifundefined {b@\@citeb}%
      {\@h@ld\@citea\@tempcntb\m@ne{\bf ?}%
      \@warning {Citation `\@citeb ' on page \thepage \space undefined}}%
      {\@tempcnta\@tempcntb \advance\@tempcnta\@ne%
      \@tempcntb\number\csname b@\@citeb \endcsname \relax%
      \ifnum\@tempcnta=\@tempcntb 
        \ifx\@h@ld\relax%
          \edef \@h@ld{\@citea\csname b@\@citeb\endcsname}%
        \else%
          \edef\@h@ld{\ifmmode{-}\else--\fi\csname b@\@citeb\endcsname}%
        \fi%
      \else
        \@h@ld\@citea\csname b@\@citeb \endcsname%
        \let\@h@ld\relax%
      \fi}%
    \def\@citea{,\penalty\@highpenalty\,}%
  }\@h@ld
}{#1}}

\def\@citeb#1#2{{[#1]\if@tempswa , #2\fi}}
\def\@citeu#1#2{{$^{#1}$\if@tempswa , #2\fi }}
\def\@citep#1#2{{#1\if@tempswa , #2\fi}}

\def\bcites{         
        \catcode`\@=11
        \let\@cite=\@citeb
        \catcode`\@=12
}

\def\upcites{         
        \catcode`\@=11
        \let\@cite=\@citeu
        \catcode`\@=12
}

\def\plaincites{      
        \catcode`\@=11
        \let\@cite=\@citep
        \catcode`\@=12
}

\newcount\hour
\newcount\minute
\newtoks\amorpm
\hour=\time\divide\hour by 60
\minute=\time{\multiply\hour by 60 \global\advance\minute by-\hour}
\edef\standardtime{{\ifnum\hour<12 \global\amorpm={am}%
        \else\global\amorpm={pm}\advance\hour by-12 \fi
        \ifnum\hour=0 \hour=12 \fi
        \number\hour:\ifnum\minute<10 0\fi\number\minute\the\amorpm}}
\edef\militarytime{\number\hour:\ifnum\minute<10 0\fi\number\minute}

\def\draftlabel#1{{\@bsphack\if@filesw {\let\thepage\relax
   \xdef\@gtempa{\write\@auxout{\string
      \newlabel{#1}{{\@currentlabel}{\thepage}}}}}\@gtempa
   \if@nobreak \ifvmode\nobreak\fi\fi\fi\@esphack}
        \gdef\@eqnlabel{#1}}
\def\@eqnlabel{}
\def\@vacuum{}
\def\marginnote#1{}
\def\draftmarginnote#1{\marginpar{\raggedright\scriptsize\tt#1}}
\def\draft{
        \pagestyle{plain}
        \overfullrule=2pt
        \oddsidemargin -.5truein
        \def\@oddhead{\sl \phantom{\today\quad\militarytime} \hfil
        \smash{\Large\sl DRAFT} \hfil \today\quad\militarytime}
        \let\@evenhead\@oddhead
        \let\label=\draftlabel
        \let\marginnote=\draftmarginnote
        \def\ps@empty{\let\@mkboth\@gobbletwo
        \def\@oddfoot{\hfil \smash{\Large\sl DRAFT} \hfil}
        \let\@evenfoot\@oddhead}
        \def\@eqnnum{(\theequation)\rlap{\kern\marginparsep\tt\@eqnlabel}%
        \global\let\@eqnlabel\@vacuum}  }

\def\blackfonts{
        \font\blackboard=msbm10 scaled\magstep1
        \font\blackboards=msbm8
        \font\blackboardss=msbm6
}

\def\nblack{            
        \def\ZZ{{Z \n{10} Z}}
        \def\NN{{N \n{14} N}}
        \def\CC{{C \n{11} C}}
        \def\RR{{R \n{11} R}}
        \def\QQ{{Q \n{12} Q}}
}

\def\prep{         
        \catcode`\@=11
        \input art10.sty
        \catcode`\@=12
        
        \let\small\null
        \def\blackfonts{
                \font\blackboard=msbm10
                \font\blackboards=msbm7
                \font\blackboardss=msbm5
        }
        \let\sl\it
        \twocolumn
        \sloppy
        \voffset=-2.54truecm
        \hoffset=-2.54truecm
        \flushbottom
        \parindent 1em
        \leftmargini 2em
        \leftmarginv .5em
        \leftmarginvi .5em
        \marginparwidth 48pt
        \marginparsep 10pt
        \setlength{\columnsep}{2truecm}
        \setlength{\textwidth}{25.4truecm}
        \setlength{\textheight}{17truecm}
        \baselineskip=16pt
        \oddsidemargin .18truein
        \evensidemargin .17truein
}

\def\eqalign#1{\null\,\vcenter{\openup\jot\m@th
  \ialign{\strut\hfil$\displaystyle{##}$&$\displaystyle{{}##}$\hfil
      \crcr#1\crcr}}\,}
\def\eqalignno#1{\displ@y \tabskip\centering
  \halign to\displaywidth{\hfil$\@lign\displaystyle{##}$\tabskip\z@skip
    &$\@lign\displaystyle{{}##}$\hfil\tabskip\centering
    &\llap{$\@lign##$}\tabskip\z@skip\crcr
    #1\crcr}}

\def\section{\@startsection {section}{1}{\z@}{3.ex plus 1ex minus
 .2ex}{2.ex plus .2ex}{\large\bf}}
\def\subsection{\@startsection{subsection}{2}{\z@}{2.75ex plus 1ex minus
 .2ex}{1.5ex plus .2ex}{\bf}}        

\def\appendix{{\newpage\section*{Appendix}}\let\appendix\section%
        {\setcounter{section}{0}
        \gdef\thesection{\Alph{section}}}\section}

\def\abstract{\if@twocolumn
\section*{Abstract}
\else 
\begin{center}
{\bf Abstract\vspace{-.5em}\vspace{0pt}}
\end{center}
\quotation
\fi}

\catcode`\@=12

\newcommand{\beq}{\begin{equation}}
\newcommand{\eeq}{\end{equation}}
\newcommand{\beqa}{\begin{eqnarray}}
\newcommand{\eeqa}{\end{eqnarray}}

\newcommand{\Z}{{\bf Z}}

\newcommand{\R}{{\bf R}}
\newcommand{\C}{{\bf C}}
\newcommand{\e}{{\rm e}}

\newcommand{\z}{\zeta}

\newcommand{\dd}{{\rm d}}

\def\noj#1,#2,{{\bf #1} (19#2)\ }
\def\jou#1,#2,#3,{{\sl #1\/ }{\bf #2} (19#3)\ }
\def\ann#1,#2,{{\sl Ann.\ Physics\/ }{\bf #1} (19#2)\ }
\def\cmp#1,#2,{{\sl Comm.\ Math.\ Phys.\/ }{\bf #1} (19#2)\ }
\def\ma#1,#2,{{\sl Math.\ Ann.\/ }{\bf #1} (19#2)\ }
\def\ng#1,#2,{{\sl Nagoya.\ Math.\ J.\/ }{\bf #1} (19#2)\ }
\def\jd#1,#2,{{\sl J.\ Diff.\ Geom.\/ }{\bf #1} (19#2)\ }
\def\invm#1,#2,{{\sl Invent.\ Math.\/ }{\bf #1} (19#2)\ }
\def\cq#1,#2,{{\sl Class.\ Quantum Grav.\/ }{\bf #1} (19#2)\ }
\def\cqg#1,#2,{{\sl Class.\ Quantum Grav.\/ }{\bf #1} (19#2)\ }
\def\ijmp#1,#2,{{\sl Int.\ J.\ Mod.\ Phys.\/ }{\bf A#1} (19#2)\ }
\def\jmphy#1,#2,{{\sl J.\ Geom.\ Phys.\/ }{\bf #1} (19#2)\ }
\def\jams#1,#2,{{\sl J.\ Amer.\ Math.\ Soc.\/ }{\bf #1} (19#2)\ }
\def\grg#1,#2,{{\sl Gen.\ Rel.\ Grav.\/ }{\bf #1} (19#2)\ }
\def\mpl#1,#2,{{\sl Mod.\ Phys.\ Lett.\/ }{\bf A#1} (19#2)\ }
\def\nc#1,#2,{{\sl Nuovo Cim.\/ }{\bf #1} (19#2)\ }
\def\np#1,#2,{{\sl Nucl.\ Phys.\/ }{\bf B#1} (19#2)\ }
\def\pl#1,#2,{{\sl Phys.\ Lett.\/ }{\bf #1B} (19#2)\ }
\def\pla#1,#2,{{\sl Phys.\ Lett.\/ }{\bf #1A} (19#2)\ }
\def\pr#1,#2,{{\sl Phys.\ Rev.\/ }{\bf #1} (19#2)\ }
\def\prd#1,#2,{{\sl Phys.\ Rev.\/ }{\bf D#1} (19#2)\ }
\def\prl#1,#2,{{\sl Phys.\ Rev.\ Lett.\/ }{\bf #1} (19#2)\ }
\def\prp#1,#2,{{\sl Phys.\ Rept.\/ }{\bf #1C} (19#2)\ }
\def\ptp#1,#2,{{\sl Prog.\ Theor.\ Phys.\/ }{\bf #1} (19#2)\ }
\def\ptpsup#1,#2,{{\sl Prog.\ Theor.\ Phys.\/ Suppl.\/ }{\bf #1} (19#2)\ }
\def\rmp#1,#2,{{\sl Rev.\ Mod.\ Phys.\/ }{\bf #1} (19#2)\ }
\def\yadfiz#1,#2,#3[#4,#5]{{\sl Yad.\ Fiz.\/ }{\bf #1} (19#2) #3%
\ [{\sl Sov.\ J.\ Nucl.\ Phys.\/ }{\bf #4} (19#2) #5]}
\def\zh#1,#2,#3[#4,#5]{{\sl Zh.\ Exp.\ Theor.\ Fiz.\/ }{\bf #1} (19#2) #3%
\ [{\sl Sov.\ Phys.\ JETP\/ }{\bf #4} (19#2) #5]}

\def\beq{\begin{equation}}
\def\eeq{\end{equation}}
\def\beqar{\begin{eqnarray}}
\def\eeqar{\end{eqnarray}}

\newcommand{\be}{\begin{equation}}
\newcommand{\ee}{\end{equation}}
\newcommand{\bea}{\begin{eqnarray}}
\newcommand{\eea}{\end{eqnarray}}

\def\nfrac#1#2{{\displaystyle{\vphantom1\smash{\lower.5ex\hbox{\small$#1$}}%
        \over\vphantom1\smash{\raise.25ex\hbox{\small$#2$}}}}}

\def\n#1{\mskip-#1mu}

\def\to{\rightarrow}

\def\lae{\mathrel{\mathop{\smash{\lower .5 ex \hbox{$\stackrel<\sim$}}}}}
\def\lae{\mathrel{\mathop{\smash{\lower .5 ex \hbox{$\stackrel>\sim$}}}}}

\def\Tr{{\rm Tr}}
\def\l:{\mathopen{:}\,}
\def\r:{\,\mathclose{:}}

\catcode`\@=11
\def\theequation{\arabic{equation}}
\catcode`\@=12

\nblack
\bcites

\nblack

\catcode`\@=11
\def\theequation{\thesection.\arabic{equation}}
\@addtoreset{equation}{section}
\@addtoreset{footnote}{section}
\@addtoreset{footnote}{subsection}
\catcode`\@=12

\newcommand{\beqn}{\begin{equation}}
\newcommand{\eeqn}{\end{equation}}
\newcommand{\beqnarray}{\begin{eqnarray}}
\newcommand{\eeqnarray}{\end{eqnarray}}
\newcommand {\bear} [1] {\begin {array} {#1}}
\newcommand {\ear} {\end {array}}

\newcommand{\CP}{{\bf C}{\rm P}}

\newcommand {\beqarn} {\begin{eqnarray*}}
\newcommand {\eeqarn} {\end{eqnarray*}}

\newcommand{\PP}{\mbox{I}\!\mbox{P}}

\begin{document}

\begin{titlepage}

\begin{center}
\today
\hfill HUTP-00/A017\\
\hfill CTP-MIT-2941\\
\hfill                  hep-th/0005247

\vskip 1.5 cm
{\large \bf D-Branes And Mirror Symmetry}
\vskip 1 cm 
{Kentaro Hori${}^1$, Amer Iqbal${}^2$ and Cumrun Vafa${}^1$}\\
\vskip 0.5cm
${}^1${\sl Jefferson Physical Laboratory,
Harvard University\\
Cambridge, MA 02138, U.S.A.}\\

\medskip

${}^2${\sl Center for Theoretical Physics, MIT\\
Cambridge, MA 02139, U.S.A.\\}

\end{center}

\vskip 0.5 cm
\begin{abstract}

We study $(2,2)$ supersymmetric field theories
on two-dimensional worldsheet with boundaries.
We determine D-branes (boundary conditions and boundary interactions)
that preserve half of the bulk supercharges
in non-linear sigma models,
gauged linear sigma models, and Landau-Ginzburg models.
 We identify a mechanism for brane creation in LG theories
and provide a new derivation of a link between soliton numbers
of the massive theories
and R-charges of vacua at the UV fixed point.
  Moreover we identify Lagrangian submanifolds that arise as
the mirror of certain
 D-branes wrapped around holomorphic cycles of K\"ahler manifolds.
 In the case of Fano varieties this leads to the explanation
of Helix structure of the collection of exceptional bundles
and soliton numbers, through
Picard-Lefshetz theory applied to the mirror LG theory.
Furthermore using the LG realization of minimal models
we find a purely geometric realization of Verlinde
Algebra for $SU(2)$ level $k$ as intersection numbers
of D-branes.  This also leads to a direct computation of
modular transformation matrix and provides a geometric interpretation
for its role in diagonalizing the Fusion algebra.
 
\end{abstract}

\end{titlepage}

\thispagestyle{empty}

{\scriptsize

\tableofcontents

}

\pagenumbering{arabic}

\vspace{1cm}

\section{Introduction}

Two-dimensional quantum field theories formulated on
a surface with boundaries
are important arena in various fields of study.
Among others they provide the starting point for open string theories,
in particular for the study of D-branes.
D-branes have proven to be indispensable elements
in string theories.
The interplay between the target space properties of D-branes,
as sources for RR fields \cite{Dbranes}, and
how they couple to string worldsheet has been very important.
In most of the recent studies of D-branes
from the worldsheet point of view,
boundary conformal field theories, namely conformal field theories
with conformally invariant boundary conditions, have been the focus of
attention.
However, we believe that one can learn more,
especially on the off-shell properties of string theories,
 by studying quantum field theories
on a worldsheet with boundaries
where the bulk theories and/or the boundary conditions
are not necessarily conformally invariant.
We also expect that such a study may open a way to
reveal the geometric principle
of string theories.

One of the main aims of this work is to initiate the study of
D-brane associated with quantum field theories
with and without conformal invariance,
employing supersymmetry as the basic constraint.
In particular we study D-branes
in supersymmetric sigma models
on K\"ahler manifolds and their mirror description \cite{HV}
in terms of D-branes in Landau-Ginzburg theories.
We will consider both the conformal case (where target
space has vanishing $c_1$ and is a CY manifold) as
well as asymptotically free theories (where $c_1\geq 0$)
which have mass gaps in many cases. 
We will mainly consider D-branes corresponding
to holomorphic cycles on the K\"ahler manifold which are mirror,
as we will discuss,
to Lagrangian submanifolds on the Landau-Ginzburg side.

Along the way we find some interesting similarities
and differences between various aspects of D-branes
for the massive sigma models and
the conformal one.  In particular we see how Brane
creation also occurs for massive theories as we change
the parameters in the sigma model.  We also define
a notion of intersection between two Lagrangian
D-branes in the massive theory which is a refined
version of the classical intersection of the cycles in the Calabi-Yau
realization of it (in particular
the inner product in the massive case
is neither symmetric nor anti-symmetric). 
Also we apply the machinery of D-branes that we develop for Landau-Ginzburg
theories to the LG realization of ${\cal N}=2$ minimal models.
In this way we find a purely geometric realization of Verlinde
algebra for bosonic $SU(2)$ WZW model at level $k$, as well
as the modular transformation matrix.  
Also we are able to shed light on an old observation of
Kontsevich connecting ``helices of exceptional bundles'' on Fano
varieties with soliton numbers of certain Landau-Ginzburg
theories, as a consequence of mirror symmetry acting on D-branes.

The organization of this paper is as follow.  In section 2 we review
aspects of LG solitons in ${\cal N}=2$ theories \cite{CV,cfiv}.
 In section 3 we discuss
D-branes for supersymmetric sigma models and LG theories.
In this section we will consider both holomorphic and
Lagrangian D-branes.
For the most of this paper
we will mainly concentrate on holomorphic D-branes
(``B-type'') in the context
of supersymmetric sigma model and Lagrangian D-branes (``A-type'')
in the context of LG models. 
We define and study boundary states corresponding to such D-branes
following the study in string theory
and conformal field theory \cite{boundaryst,Ishibashi,cardy}.
In section 4 we discuss the phenomenon of D-brane creation
of massive LG theories, and show how these results give a
reinterpretation of the connection between 
R-charges of chiral fields at the ultra-violet
fixed point and the soliton numbers
of its massive deformation discovered in \cite{CV} .
In section 5 we apply these results to the study of
${\cal N}=2$ minimal models and show how aspects of conformal theory,
including certain properties of Cardy states,
and its relations with Verlinde algebra, as well as
its overlap with Ishibashi states in terms of modular
transformation matrix, can be derived in a purely geometric way.
In section 6 we derive,
using the results of \cite{HV} the mirror of certain D-branes
on Fano varieties in terms of D-branes in the mirror LG models.
In section 7 we apply the study of D-branes to the
LG mirrors of Fano varieties and uncover beautiful mirror interpretation
for helices of exceptional bundles on Fano
varieties in terms of D-branes of the LG mirror.
In section 8 we discuss connecting the LG mirror
for the case of non-compact
geometries in Calabi-Yau (such as $\PP^2$ inside a CY threefold)
discussed in \cite{HV}
to a local non-compact geometric mirror as was
used in \cite{geometriceng,kmv}.  Moreover we show how this
is related in the case of local non-compact threefolds
to the probe description in F-theory and its BPS states.

While completing this work,
a paper \cite{GJ}, which has some overlap with our discussions
in sections 3 and 5, appeared.

\newcommand{\onefigure}[2]{\begin{figure}[htbp]
         \caption{\small #2\label{#1}(#1)}
         \end{figure}}
\newcommand{\onefigurenocap}[1]{\begin{figure}[h]
         \begin{center}\leavevmode\epsfbox{#1.eps}\end{center}
         \end{figure}}
\renewcommand{\onefigure}[2]{\begin{figure}[htbp]
         \begin{center}\leavevmode\epsfbox{#1.eps}\end{center}
         \caption{\small #2\label{#1}}
         \end{figure}}
\newcommand{\comment}[1]{}
\newcommand{\myref}[1]{(\ref{#1})}
\newcommand{\secref}[1]{sec.~\protect\ref{#1}}
\newcommand{\figref}[1]{Fig.~\protect\ref{#1}}
\def\z{{\sf Z\!\!\!Z}}
\def\c{{\sf {\bf C}\!\!\!\,\!I}}
\def\bbbs{{\sf {\bf C}\!\!\!\,\!I}^{\,\,*}}
\def\sl2z{SL(2,\z)}
\newcommand{\q}{I\!\!Q}
\def\r{{\sf R\!\!\!\!\!I}}
\def\fracs#1#2{\textstyle\frac #1#2}
\newcommand{\nn}{\nonumber}
\newcommand{\unit}{1\!\!1}
\newcommand{\half}{\frac{1}{2}}
\newcommand{\shalf}{\mbox{$\half$}}
\newcommand{\transform}[1]{
   \stackrel{#1}{-\hspace{-1.2ex}-\hspace{-1.2ex}\longrightarrow}}
\newcommand{\inter}[2]{\null^{\#}(#1\cdot#2)}
\newcommand{\lprod}[2]{\vec{#1}\cdot\vec{#2}}
\newcommand{\mult}[1]{{\cal N}(#1)}
\newcommand{\Bn}{{\cal B}_N}
\newcommand{\B}{{\cal B}}
\newcommand{\Beight}{{\cal B}_8}
\newcommand{\Bnine}{{\cal B}_9}

\section{BPS Solitons in ${\cal N}$=2 Landau Ginzburg Theories}

In this section, we review some basic facts on Landau-Ginzburg models,
especially on the spectrum of BPS solitons and the relation to
Picard-Lefshetz theory of vanishing cycles.
The action for a Landau Ginzburg model of $n$ chiral superfields
$\Phi_i$ ($i=1,\ldots,n$) with superpotential $W(\Phi)$  
is given by
\be
S=\int \dd^{\,2}x\,\left[
\int\dd^{4}\theta \, K(\Phi_{i},\overline{\Phi}_{i})+
\fracs{{1}}{{2}}\left(\int\dd^{2}\theta\, W(\Phi_{i})
+\int \dd^{2}\bar{\theta}\,
\overline{W}(\overline{\Phi}_{i})\,\right)\,\right]\,.
\label{actio}
\ee
Here
$K(\Phi_{i},\bar{\Phi}_{i})$ is the K\"ahler potential which defines the
K\"ahler metric
$g_{i\bar{j}}=\partial_{i}\partial_{\bar{j}}K(\Phi_{i},\bar{\Phi}_{i})$.
If the superpotential $W(\Phi)$ is a quasi-homogeneous function with
an isolated critical point (which means $dW=0$ can only occur
at $\Phi_i=0$) then
the above action for a particular choice of $K(\Phi,\bar{\Phi})$ is believed
to define a superconformal theory \cite{vafa-1,mart}. 
For a general superpotential the vacua
are labeled by critical points of $W$, i.e., where  
\be
\phi^{i}(x)=\phi^{i}_{*}\,,\,\,\,
\,\,\partial _iW |_{\phi^{i}_{*}} =0\,.
\ee  
The theory is purely massive if all the critical points are isolated
and non-degenerate,
which means that near the critical points $W$ is quadratic in fields.
We assume this and label the non-degenerate critical points as
$\{\phi_{a}\,|\,a=1,\cdots, N\}$.  In such a case the number of vacua of the
theory is equal to the dimension of the local ring of $W(\Phi)$,
${\cal R}=\frac{\c~[\Phi]}{\partial_{\phi^{i}}W}$.
When we have more than one vacuum we can have solitonic
states in which the boundary conditions of the fields at the left
spatial infinity $x^{1}=-\infty$ is at one vacuum and is
different from the one at right
infinity $x^{1}=+\infty$ which is in another vacuum.  The geometry
of solitons and their degeneracies have been extensively
studied in \cite{CV,cfiv}  which we will now review.

Consider a massive Landau Ginzburg theory with superpotential 
$W(\Phi_{i})$.
Solitons are static
solutions, $\phi^{i}(x^{1})$, of the equations of motion interpolating
between {\it different} vacua i.e., $\phi^{i}(-\infty)=\phi^{i}_{a}$ and
$\phi^{i}(+\infty)=
\phi^{i}_{b}$, $a\neq b$.
The energy of a static field configuration interpolating between two
vacua is given by \cite{fimvw}
\begin{eqnarray}
E_{ab}&=&\int_{-\infty}^{+\infty} dx^{1}\left\{
g_{i\bar{j}}\frac{d\phi^{i}}{dx^{1}}\frac{d\bar{\phi}^{\bar{i}}}{dx^{1}}+
\fracs{{1}}{{4}}g^{i\bar{j}}\partial_{i}W\partial_{\bar{j}}\bar{W}\right\} \\ \nn
&=&\int_{-\infty}^{+\infty} dx^{1}
\left|\frac{d\phi^{i}}{dx^{1}}
-\fracs{{\alpha}}{{2}} g^{i\bar{j}}\partial_{\bar{j}}\bar{W}\right|^{2}
\,\,+\,\mbox{Re}\Bigl((\bar{\alpha}(W(b)-W(a))\Bigr)\,.
\end{eqnarray}
Where $g_{i\bar{j}}=\partial_{i}\partial_{\bar{j}}K$ is the K\"ahler
metric and $\alpha$ is an arbitrary phase. By choosing an appropriate
$\alpha$ we can maximize the second term. Since $\alpha$ is a phase it
is clear that the second term is maximum when phase of $W(b)-W(a)$ is
equal to $\alpha$. This implies a lower bound on the energy of the
configuration,
\be
E_{ab}\ \geq |W(b)-W(a)|.
\ee
In fact the central charge in the supersymmetry
algebra in this sector is $(W(b)-W(a))$.
BPS solitons saturate this bound and therefore satisfy the equation,
\be
\frac{d \phi^{i}}{dx^{1}}={\alpha \over 2}
g^{i\bar{j}}\partial_{\bar{j}}\bar{W}
\,,\,\,\,\,\,\,\alpha=\frac{W(b)-W(a)}{|W(b)-W(a)|}\,.
\label{solitoneq}
\ee
An important consequence of the above equation of motion of a BPS
soliton is that along the trajectory of the soliton the superpotential
satisfies the equation,
\be
\partial_{x^{1}}W={\alpha \over 2}
g^{i\bar{j}}\partial_{i}W\partial_{\bar{j}}\bar{W}\,.
\label{solitoneqW}
\ee
Now since the metric $g^{i\bar{j}}$ is positive definite, we know
$g^{i\bar{j}}\partial_{i}W\partial_{\bar{j}}\bar{W}$ is real,
and therefore
the image of the BPS soliton in the W-plane is a straight
line connecting the corresponding critical values $W(a)$ and $W(b)$.

The number of solitons between two vacua is equal to the number of
solutions of eq.\,(\ref{solitoneq}) satisfying the appropriate boundary
conditions.  The general way to count the number of solitons
has been determined in \cite{CV} and we will review it in the
next subsection.  Here we note that for the case of a single chiral
superfield the
number of solitons between two vacua can also be
determined using eq.\,(\ref{solitoneqW}). Since the image of the soliton
trajectory is a straight line in the $W$-plane therefore by looking at
the pre-image of the straight line connecting the corresponding
critical values in the $W$-plane we can determine the number of solitons
between the two vacua. But since the map to the $W$-plane is many to one,
not every pre-image of a straight line in the $W$-plane is a soliton. It
is possible for the trajectory to start at a critical point follow a
path whose image is a straight line in the $W$-plane and end on a point
which is not a critical point but whose image in the $W$-plane is a
critical value. The BPS solitons are those pre-images of the straight
line in the $W$-plane which start and end on the critical points.

\subsection{Vanishing cycles}

It was shown in \cite{CV} that the soliton
numbers also have a topological description in terms of intersection
numbers of vanishing cycles. The basic idea is to solve
the soliton equation (\ref{solitoneq}) along all possible directions
emanating from one of the critical points.
In other words the idea is to study
the ``wave front'' of all possible solutions to the $(\ref{solitoneq})$.

With no loss of generality we may assume $\alpha =1$.
Near a critical point $\phi^{i}_{a}$ we can 
choose coordinates $u_{a}^{i}$ such that, 
\be
W(\phi)=W(\phi_{a})+\sum_{i=1}^{n}(u^{i}_{a})^{2}\,.
\ee
In this case it is easy to see that the solutions to (\ref{solitoneq})
will have an image in the $W$-plane which is on a positive real line
starting from $W(\phi_a)$.  Consider a point $w$ on this line.
Then the space of solutions to (\ref{solitoneq}) emanating from
$u^{i}_{a}=0$ over this $w$ is a real $(n-1)$ dimensional sphere defined by
\be
\sum_{i=1}^{n}(\mbox{Re}(u^{i}_{a}))^{2}=w-w_a\,,\,
\,\,\,\,\mbox{Im}(u^{i}_{a})=0\,.
\label{vanishingeq}
\ee
where $w_{a}=W(\phi_{a})$.
Note that as we take $w\mapsto w_{a}$ the sphere vanishes.
This is the reason for calling these spheres ``vanishing cycles''.  
 As we move away,
the wavefront will no longer
be as simple as near the critical point, but nevertheless over
each point $w$ on the positive real line emanating from $w_{a}=W(\phi_{a})$
the pre-image is a real $(n-1)$
dimensional homology cycle $\Delta_{a}$ in the $n-1$ complex
 manifold defined by
$W^{-1}(w)$. Similarly as we move from $w_{b}$ toward
$w_{a}$ there is a cycle $\Delta_{b}$ evolving according to the
soliton equation eq.\,(\ref{solitoneq}) (this would
correspond to $\alpha =-1$). For a fixed value of $w$ we
can compare $\Delta_{a}$ and $\Delta_{b}$. Solitons originating from
$\phi_{a}$ and traveling all the way to $\phi_{b}$ correspond to the
points in the intersection $\Delta_{a}\cap \Delta_{b}$. This number,
counted with appropriate signs is the
intersection number of the cycles, $\Delta_{a}\circ \Delta_{b}$. The
intersection number counts the number of solitons weighed with
$(-1)^F$ for the lowest component of each soliton multiplet \cite{CV}.
This is independent of deformation of the D-terms.  In particular
this measures the net number of solitons that cannot disappear by
deformations in the D-terms.  We will denote this number by
$A_{ab}$ and sometimes loosely refer to it as the number of
solitons between $a$ and $b$.  We thus have
\be
A_{ab}=\Delta_a \circ \Delta_b.
\ee

Note that to calculate the intersection numbers we
have to consider the two cycles $\Delta_{a}$ and $\Delta_{b}$ in the
same manifold $W^{-1}(w)$.  Since the intersection number is topological,
a continuous deformation does not
 change them and hence we can actually
calculate them using some deformed path in the $W$-plane (rather than
the straight line) as long as the path we are choosing is homotopic to
the straight line.  How we transport the cycle along the path
will not change the intersection numbers as that is topological
and nothing is discontinuous, as long as the paths have
the same homotopy class in the $W$-plane with the critical values
deleted.  One way, but not the only way, to transport vanishing cycles along
arbitrary paths, is to use the soliton equation $(\ref{solitoneq})$
but instead of having a fixed $\alpha$, as would be the case
for a straight line, choose $\alpha$ to be $e^{i\theta}$
where $\theta$ denotes the varying slope
of the path.

\onefigure{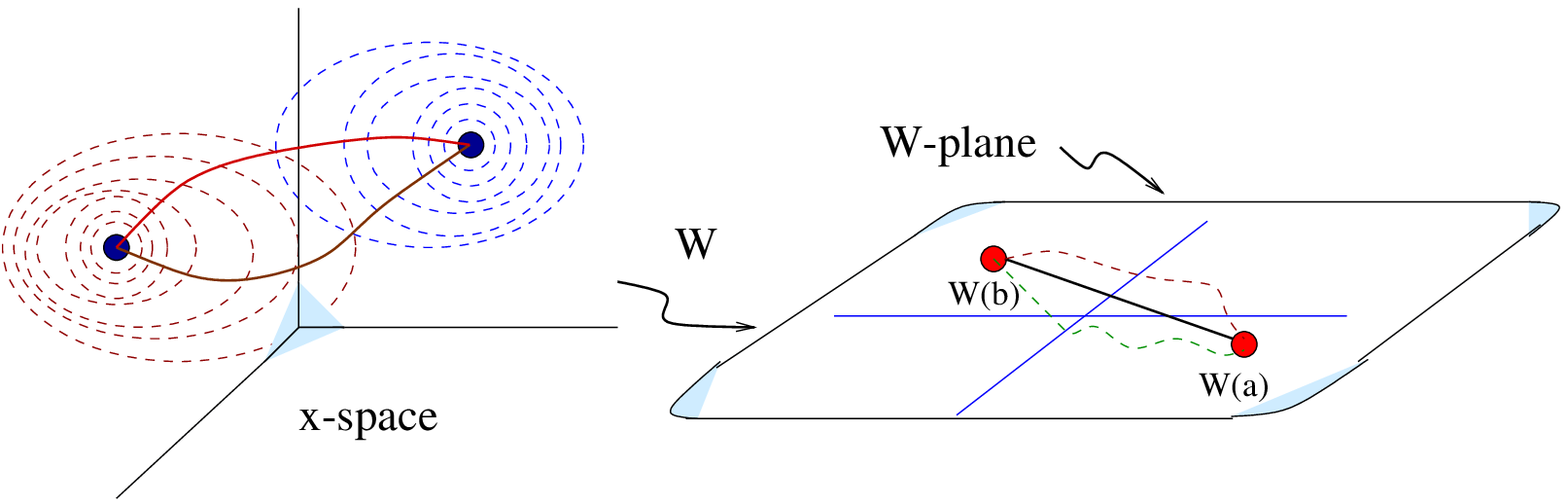}{BPS soliton map to straight line in the
 $W$-plane. Soliton solutions exist for each intersection point of
 vanishing cycles. Lines in the $W$-plane which are homotopic to the
 straight line (doted lines) can also be used to calculate soliton numbers.}

Let us fix a point $w$ in the $W$-plane.
For each critical point $a$ of $W$,
we choose an arbitrary path in the $W$-plane
emanating from $W(a)$ and ending on $w$,
but not passing through other critical values.
This yields $N$ cycles $\Delta_a$ over $W^{-1}(w)$
and it is known \cite{Arnold}
that these form a complete basis for the middle-dimensional homology
cycles of $W^{-1}(w)$.
Moreover, 
if we choose different paths the vanishing
cycle we get is a linear combination of the above and the
relation between them is known through the Picard Lefshetz theory as we
will now review.

\subsection{Picard-Lefshetz monodromy}

As we have discussed the basis for the vanishing cycle
over each point $w$ in the $W$-plane depends on the choice
of paths connecting it to the critical point.  Picard-Lefshetz
monodromy relates how the basis changes if we change
paths connecting $w$ to the critical values. This is
quite important for the study of solitons and leads to a jump
in the soliton numbers.  To explain the physical motivation
for the question, consider three critical values $W(a),W(b)$
and $W(c)$ depicted in \figref{monodromy-1}(a), with no other
critical values nearby.  Suppose we wish to compute
the number of solitons between them.  According to our
discussion above we need to connect the critical values by straight
lines in the $W$-plane and ask about the intersection
numbers of the corresponding cycles.  As discussed above
this is the same, because of invariance 
of intersection numbers under deformation, as the intersection
numbers of the vanishing cycles over the point $w$ connecting
to the three critical values as shown in \figref{monodromy-1}(a).
Thus the soliton number is $A_{ij}=\Delta_i \circ \Delta_j$.
 However suppose now that we change
the superpotential $W$ so that the critical values change
according to what is depicted in \figref{monodromy-1}(b),
and that the $W(b)$ passes through the straight line connecting
$W(a)$ and $W(c)$.
In this case to find the soliton numbers between the $a$ vacuum
and the $c$ vacuum we have to change the homotopy class of the
path connecting $w$ to the critical
value $W(a)$ as depicted by \figref{monodromy-1}(b).

\onefigure{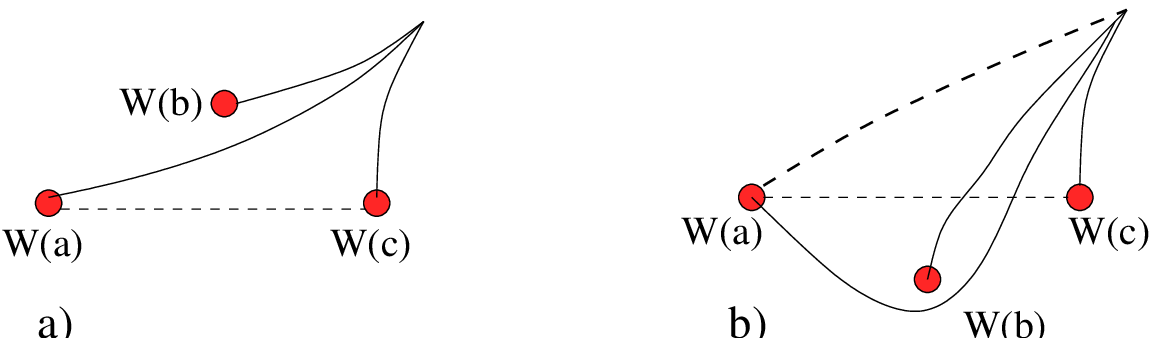}{As the positions
of critical values change in the $W$-plane, the 
choice of the vanishing cycles relevant for computing
the soliton numbers change.}

  In particular
the homology element corresponding to
vanishing cycle $a$ changes $\Delta_a\rightarrow \Delta_a^{\,\prime}$
and we need to find out how it changes.
Picard-Lefshetz theory gives a simple formula for this change.
In particular it states that
\bea 
\Delta_a^{\,\prime}=\Delta_a\pm (\Delta_a \circ \Delta_b) \Delta_b \,.
\eea
The sign in the above formula is determined once the orientation of
the cycles are fixed and will depend on the handedness of the crossing
geometry (see \cite{CV}).  This is perhaps most familiar to string
theorists in the context of moduli space of Riemann surfaces, where if
we consider a point on the moduli space of Riemann surfaces where a
1-cycle shrinks to zero, as we go around this point, all the other
cycles intersecting it will pick up a monodromy in the class of the
vanishing cycle (the case of the torus and the $\tau \rightarrow \tau
+1$ is the most familiar case, where the $b$ cycle undergoes a
monodromy $b\rightarrow b+a$).

As a consequence of the above formula we can now
find how the number of solitons between the $a$
and the $c$ vacuum change.  We simply have
to take the inner product $\Delta_a^{\,\prime}\circ \Delta_c$
and we find
\bea \nn
A_{ac}^{\,\prime}=A_{ac}\pm A_{ab}A_{bc}\,.
\eea

\subsection{Non-compact $n$ Cycles}

\newcommand{\Ggamma}{\mbox{\large $\gamma$}}

An equivalent description which will be important for later discussion
involves defining soliton numbers in terms of the intersection numbers
of $n$ real dimensional non-compact cycles which are 
closely related to the $n-1$
dimensional
vanishing cycles we have discussed. The idea is to consider the basis
for the vanishing cycles in the limit
where the point $w\rightarrow e^{i\theta }\infty$.  Let us
consider the case where $\theta =0$.  In this case we are taking
$w$ to go to infinity along the positive real axis.  Let us assume
that the imaginary part of the critical values are all distinct.
In this case a canonical choice of paths to connect the critical
points to $w$ is along straight lines starting from the critical values
$W(a)$ stretched
along the positive real axis.
We denote 
the corresponding non-compact $n$ dimensional cycles
by $\Ggamma_a$.  Then we have

\be
W(\Ggamma_{a})= I_{a}\,,\,\,\quad \mbox{and} \quad \partial 
\Ggamma_{a}\cong \Delta_{a}\Bigr|_{w \mapsto \,+\infty}\,,
\ee
where
\be
I_{a} \equiv \{w_{a}+t~|~t\in [0, \infty)\}\,.   
\ee
Two such cycles are shown in Figure \ref{vanishing-cycles-1}.
\onefigure{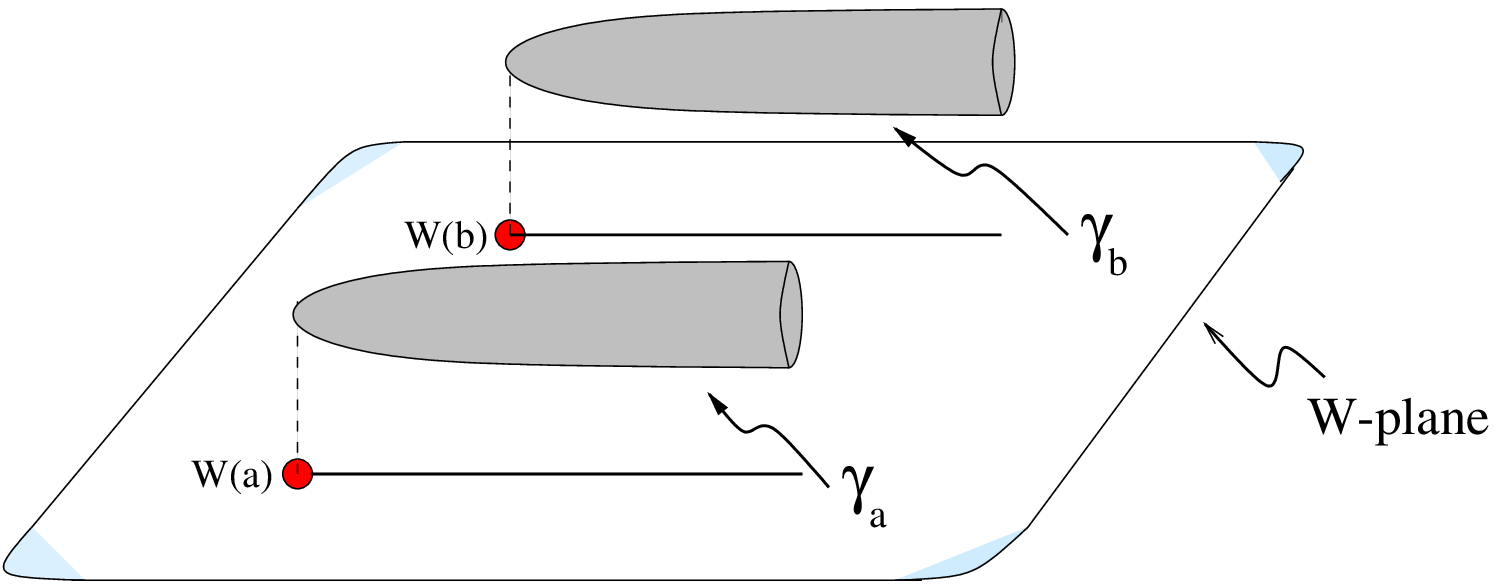}{The cycles emanating from the critical
points. The images in the $W$-plane
are the straight lines emanating from the
critical values and extending to the infinity
in the real positive direction.}

Let $B$ be the region of $\C^n$ where ${\rm Re}W$
 is larger than a fixed value which is chosen sufficiently large.
The non-compact cycles $\Ggamma_a$ can be viewed
as elements of the homology group $H_n({\bf C}^n, B)$
corresponding to $n$-cycles with boundary in $B$,
and again it can be shown \cite{Arnold} that they provide
a complete basis for such cycles.

For a pair of distinct critical points, $a$ and $b$,
the non-compact cycles $\Ggamma_a$ and $\Ggamma_b$ do
not intersect with each other,
since their images in the $W$-plane are parallel
to each other (and are separate from each other in the present situation).
In this situation
we consider deforming the second cycle $\Ggamma_b$
so that its image in the
$W$-plane is rotated with an infinitesimally small positive angle
$\epsilon$ against the real axis.
We denote this deformed cycle by $\Ggamma_b^{\prime}$.
We define the ``intersection number'' of $\Ggamma_a$ and $\Ggamma_b$
as the geometric intersection number
of $\Ggamma_a$ and $\Ggamma_b^{\prime}$.
Depending on whether ${\rm Im}W(a)$ is smaller or larger than 
${\rm Im}W(b)$, the images of
$\Ggamma_a$ and $\Ggamma_b^{\prime}$
in the $W$-plane do not or do intersect with
each other.
\onefigure{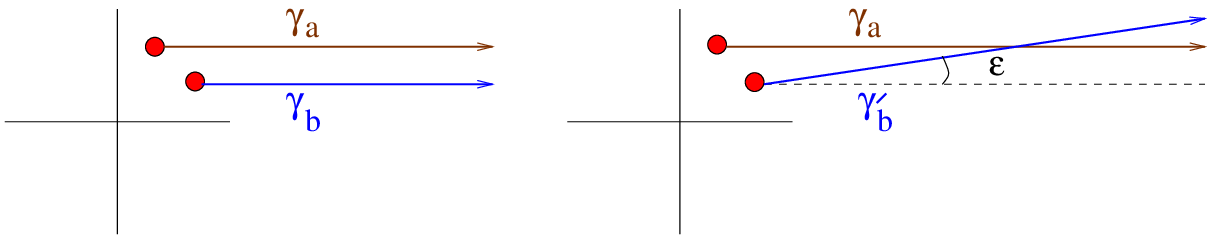}{The images in the $W$-plane of
$\Ggamma_a$ and $\Ggamma_b$ (left); and $\Ggamma_a$ and $\Ggamma_b^{\prime}$
(right).
The second will give
rise to``intersection number''.
As we will see in the next section,
this contains a certain information on D-branes in the LG model.}
In the former case the ``intersection number'' is of course zero.
In the latter case as shown in \figref{vanishing-cycles-2},
the intersection number
$\Ggamma_a \circ \Ggamma_b^{\prime}$
is counted by going to the point on the
$W$-plane where their images intersect and asking what is the intersection
of the corresponding vanishing cycles $\Delta_a\circ \Delta_b$.
Thus the intersection of these $n$-dimensional cycles has the
information about the soliton numbers. In particular if there
are no extra critical values between the $I_a$ and $I_b$ we will have
\begin{eqnarray}
\Ggamma_{a}\circ \Ggamma_b^{\,\prime}={ A}_{ab},\,\,\,\,\,a \neq b\,.
\end{eqnarray}
If there are extra critical values between $I_a$ and 
$I_b$ then these intersection numbers
are related to the soliton numbers by the Picard-Lefshetz
action as discussed before.

In the next section we will see that the cycles $\Ggamma_{a}$ 
defined through parallel transport by the soliton equation
(\ref{solitoneq}) can be viewed as D-branes for LG models that
preserve half of the supersymmetries on the worldsheet.  There we will also
see that the ``intersection number''
of $\Ggamma_a$ and $\Ggamma_b$
as defined above can be interpreted as
the supersymmetric index for the worldsheet theory
of open strings stretched between these cycles.

\subsection{Examples}

In this section we are going to discuss some examples
for the soliton numbers in the case of LG models.  We will concentrate
on LG models representing ${\cal N}=2$ minimal models as well
as the LG models mirror to $\PP^{N}$ sigma models.

\subsubsection{Deformed ${\cal N}=2$ Minimal models}

~~~~
${\cal N}=2$ minimal models are realized as
the infra-red fixed point of LG models \cite{vafa-1,mart} .
The soliton numbers for the deformed version of these
theories has been studied in detail for the massive deformations
of the A-series minimal models \cite{fimvw,CV}, which we will
now review.

The $k$-th minimal model is described by an LG theory with
one chiral superfield $X$ with superpotential
\be 
W(X)=\fracs{{1}}{{k+2}}X^{k+2}\,.
\ee
If we add generic relevant operators to the superpotential
we can deform this theory to a purely massive theory.  In this
case we will get $k+1$ vacua and we can ask how
many solitons we get between each pair.  For example if we consider the
integrable deformation,
\be
W(X)=\fracs{{1}}{{k+2}}X^{k+2}-X,
\ee
then there are $k+1$ vacua which are solutions of $dW=0$ given by 
$X=e^{\frac{2\pi i n}{k+1}}\,,n=0,\cdots,k$. In this case one
can count \cite{fimvw}\ the preimage of the
straight lines in the W-plane and ask which ones connect
critical points and in this way compute the number of solitons.
It turns out that in this case there is exactly
one soliton connecting each pair of critical points.  If we
deform $W$ the number of solitons will in general change
as reviewed above.  In this case one can show (by
taking proper care of the relevant signs in the 
 soliton number jump) that there is always at most one soliton
 between vacua. The precise number can be
determined starting from the above symmetric configuration
(see \cite{CV}).  The analog of the non-compact 1-cycles 
$\Ggamma_i$ in this case
will be discussed in more detail in section 4 after we discuss
their relevance as D-branes in section 3.
They are cycles in the $X$-plane
which asymptote an $(k+2)$-th root of unity as $X\rightarrow \infty$.
That there are $k+1$ inequivalent such homology classes
for $H_1({\bf C}, Re W=\infty )$ is related
to the fact
that there are
$k+1$ such classes defined by $\Ggamma$'s up to linear combinations.

\subsubsection{$\PP^{N-1}$}

~~~~
We next consider the $\PP^{N-1}$ sigma model.
The soliton matrix of the non-linear sigma model with target space
$\PP^{N-1}$ can be computed directly by studying the $tt^{*}$ equations
\cite{CV}\ and their relations to soliton numbers \cite{cfiv}.
This has been done in
\cite{cecvaf,zaslow}. The $tt^{*}$ equations are, however, 
very difficult to
solve for more non-trivial spaces such as toric del Pezzos. 
The mirror LG theory obtained in \cite{HV} provides
a simple way of calculating the soliton matrix. 
We start with the case $N=2$ where we can present explicit solutions
to the soliton equation.

The Landau Ginzburg theory which is mirror to 
the non-linear sigma model with $\PP^{1}$ target space is
the ${\cal N}=2$ sine-Gordon model with the superpotential,
\be
W(x)=x+\frac{\lambda}{x}\,.
\ee
Here $x=\e^{-y}$ is the single valued coordinate of the cylinder
$\c^{\,\times}$ 
and $-\log\lambda$ corresponds to the K\"ahler parameter of
$\PP^{1}$. 
The critical points are $x^{\pm}_{*}=\pm \sqrt{\lambda}$ with critical
values $w_{*}^{\pm}=\pm 2\sqrt{\lambda}$. As mentioned in the previous
section the BPS solitons are trajectories, $x(t)$,  starting and ending on
the critical points such that their image in the W-plane is a straight
line,
\be
x(t)+\frac{\lambda}{x(t)}=2\sqrt{\lambda}(2t-1)\,,\,\,\,t\in[0,1]\,.
\ee
This is a quadratic equation with two solutions given by,
\be
x(t)_{\pm}=\sqrt{\lambda}(2t-1)\pm 2i
\sqrt{\lambda}\sqrt{t-t^{2}} = \sqrt{\lambda}e^{\pm i \mbox{tan}^{-1} \frac{2\sqrt{t-t^2}}{2t-1}}.
\ee
\onefigure{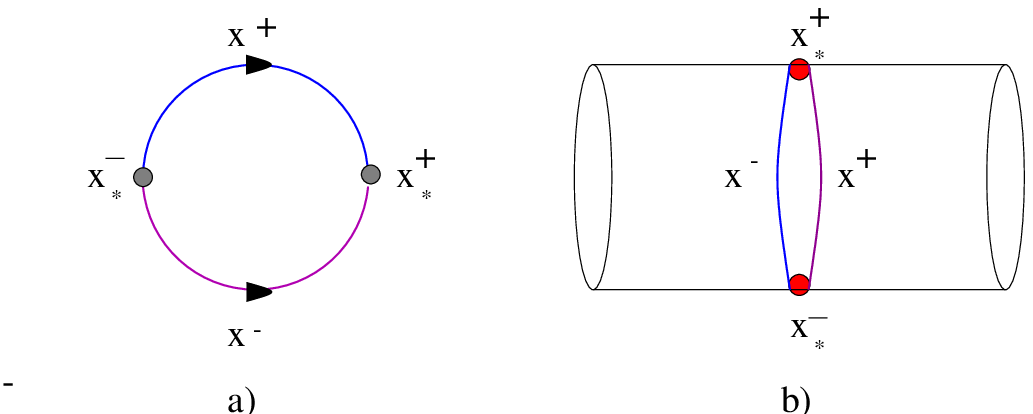}{The two solitons of the
$\PP^{1}$ model.}
Since $x_{+}(t)=x_{-}(t)^{*}$ and
$|x_{+}(t)|=|\sqrt{\lambda}|$, there are two solitons between the
two vacua such that their trajectories in the x-plane lie on two
half-circles as shown in \figref{cp1-a}(a).  Since $x$ is a $\c^{\,\times}$
coordinate we can consider the x-plane as a cylinder. Soliton
trajectories on the cylinder are shown in \figref{cp1-a}(b).
This description is useful in determining the intersection numbers of
middle dimensional cycles. As described in the previous section the
number of solitons between two critical points is given by the
intersection number of middle dimensional cycles starting from the
critical points. In our case there are two such cycles which are the
preimages of two semi-infinite lines in the $W$-plane starting at the
critical values as shown in \figref{cp1-d}(a). The preimage of these cycles on
the cylinder is shown in \figref{cp1-d}(b). 
\onefigure{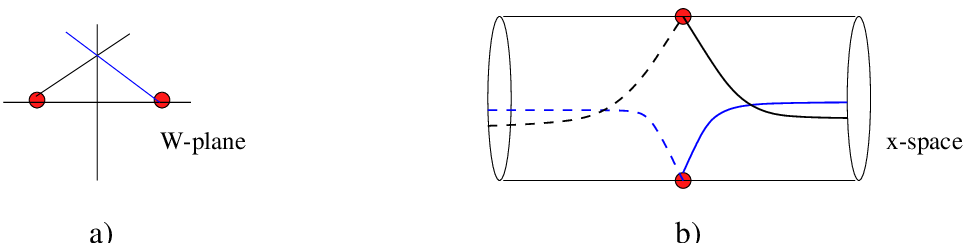}{Intersecting lines in the W-plane and the 
corresponding intersecting cycles in the x-space.}
The cycles in the x-space intersect only if the lines in the $W$-plane
intersect each other and the intersection number in this case is
two.

We now turn to the study of solitons
of the $\PP^{N-1}$ sigma model.
The LG theory mirror to the non-linear sigma model with
$\PP^{N-1}$ target space
has superpotential \cite{HV}
\be
W(X)=\sum_{k=1}^{N-1}X_{k}+\frac{\lambda}{X_{1}\cdots X_{N-1}}\,.
\label{cpnsuperpotential}
\ee
This superpotential has $N$ critical points given by
\be
X^{(a)}_{\,i}=\e^{\frac{2\pi i a}{N}}\,\,\,\,\, i=1,\cdots,N-1
\,\,; \quad a=0,\cdots, N-1\,,
\ee
with the critical values
\be
w_{a}\equiv W(\vec{X}^{(a)})=N\e^{\frac{2\pi i a}{N}}\,.
\ee
Here unlike the previous case of $\PP^{1}$, to be able to solve
for the preimage of a straight line, we will make an
 assumption
about the soliton solution (For the case of $\PP^{2}$ and
its blowups we will also find another way to count
the soliton numbers as will be discussed in section 8).
Even though we will not justify this
ansatz, the results we find are consistent with what is known
based on $tt^*$ equations.
We assume that the soliton trajectory is
determined by a function $f(t)$ such that
\be
X_{1}=X_{2}=\cdots=X_{k}=f(t)^{N-k}\,,\,\,X_{k+1}=X_{k+2}=
\cdots =X_{N}=f(t)^{-k}\,.
\label{ansatz}
\ee
This parameterization of the solution satisfies the constraint
$\prod_{i=1}^{N}X_{i}=1$ by construction.
With this ansatz the
straight line equation in the $W$-plane becomes (for $\lambda$=1)
\be
P(f):= kf^{N-k}+(N-k)f^{-k}
=N(1-t+t e^{ \frac{2\pi i k}{N}})\,,
\label{straightlineeq}
\ee
where the right hand side is the straight line $w(t)$
starting from $w(0)=N$ and ending on $w(1)=N\e^{2\pi i k\over N}$.
Here we have chosen the parameter $t$ running in the range $[0,1]$
that is linear in the $W$-plane.
We are interested in the solutions which start at $t=0$ from
$X_{i}^{(0)}$ and end at $t=1$ on $X_{i}^{(k)}$.
This implies that $f(0)^{N-k}=f(0)^{-k}=1$ and
$f(1)^{N-k}=f(1)^{-k}=e^{\frac{2\pi i k}{N}}$.  Thus the number of
solitons which satisfy eq.\,(\ref{ansatz}) is given by the number of
solutions to eq.\,(\ref{straightlineeq}) such that $f(0)=1$ and
$f(1)=e^{-\frac{2\pi i }{N}}$.  We will show that there is only a
single solution which satisfies these conditions.

Since $P'(1)=0$ and $P''(1)\neq 0$, where prime denotes a
differentiation with respect to $f$, only two trajectories
start from $f=1$. Thus it follows that the number of solutions is less
than or equal to two. From eq.\,(\ref{straightlineeq}) it is clear that $f$
can be real only at $t=0$. Thus a trajectory cannot cross the real
axis for $t > 0$. For $t$ very close to zero one of the trajectories
move into the upper half plane. Since the
trajectory in the upper half plane cannot cross the real axis it
cannot end on $e^{-\frac{2\pi i k}{N}}$ \footnote{Very close to $f=1$
the two solutions are given by $f_{\pm}=1\pm
\sqrt{\frac{2t(e^{\frac{2\pi i k}{N}}-1)}{k(N-k)}}$}. Thus there can be
at most one solution.

To show that there actually exists a solution we will construct a
solution whose image in the $W$-plane is homotopic to the straight line
$w(t)$. Consider the function $f_{*}(t)=e^{-\frac{2\pi i }{N}t}$
where $t\in [0,1]$.  Since
\be
|P(f_{*}(t))| = |k e^{-2\pi i t}+(N-k)|
\leq |ke^{-2\pi i t}|+(N-k) = N\,, 
\ee
the image of $f_{*}(t)$ in the W-plane always lies inside
the circle of radius $N$ and only intersects the circle for $t=0$
and $t=1$ at $w=w_{0}$ and $w=w_{k}$ respectively. Thus the image is
homotopic to the straight line $w(t)$ and therefore there exists a
solution $f_{0}(t)$ homotopic to $f_{*}(t)$ with the required properties.

Since permuting the $N$ coordinates among themselves does not
change the superpotential, it follows that we can choose any
$k$ coordinates to be equal to $f^{N-k}$ and the remaining $(N-k)$
coordinates equal to $f^{-k}$. Thus we see that there are
${N\choose k}$ solitons between the critical points $X^{(0)}_{i}$ and
$X^{(k)}_{i}$ consistent with the ansatz of eq.\,(\ref{ansatz}).
In fact this is the same number anticipated by the
study of tt* equations \cite{CV}.  Note that if the
$\PP^{N-1}$ has a round metric having $SU(N)$ symmetry, then
the solitons should form representations of this group.
In fact the permutations of $X_i$ can be viewed
as the Weyl group of the $SU(N)$, as is clear
from the derivation of the mirror in this case \cite{HV}.
It thus follows, given how the permutation acts on the solutions
we have found, that in this case the solitons connecting
vacua $k$ units apart correspond to $k$ fold anti-symmetric tensor
product of the fundamental representation of $SU(N)$, a result
which was derived from the large $N$ analysis of this
theory \cite{wittencp}.


\section{D-Branes in ${\cal N}=2$ Supersymmetric Field Theories}

\newcommand{\bQ}{\overline{Q}}
\newcommand{\bD}{\overline{D}}
\newcommand{\bchi}{\overline{\chi}}
\newcommand{\bphi}{\overline{\phi}}
\newcommand{\bpsi}{\overline{\psi}}
\newcommand{\btheta}{\overline{\theta}}
\newcommand{\bepsilon}{\overline{\epsilon}}
\newcommand{\blambda}{\overline{\lambda}}
\newcommand{\bsigma}{\overline{\sigma}}
\newcommand{\bSigma}{\overline{\Sigma}}
\newcommand{\bPhi}{\overline{\Phi}}
\newcommand{\bj}{{\bar\jmath}}
\newcommand{\bi}{{\bar\imath}}
\newcommand{\bk}{{\bar k}}
\newcommand{\bz}{\overline{z}}
\newcommand{\bTheta}{\overline{\Theta}}
\newcommand{\bartial}{\overline{\partial}}
\newcommand{\lrD}{\overleftarrow{D}\!\!\!\!\!
\overrightarrow{\vbox to 8.2pt{}}}
\newcommand{\lrd}{\overleftarrow{\partial}\!\!\!\!\!
\overrightarrow{\vbox to 8.35pt{}}}
\newcommand{\lrdd}{\overleftarrow{\dd}\!\!\!\!\!
\overrightarrow{\vbox to 8.335pt{}}}
\newcommand{\cycle}{\Ggamma}
\newcommand{\bG}{\overline{G}}

In this section, we study the $N=(2,2)$ supersymmetric field theory formulated
on a 1+1 dimensional worldsheet with boundaries.
We mainly consider supersymmetric sigma models and Landau-Ginzburg models.
We find boundary conditions that preserve half of the (worldsheet)
supersymmetry. (See \cite{GJS,Warner} for earlier works.)
We define and compute the supersymmetric index
of a theory on an interval.
We also analyze the ${\cal N}=2$
boundary entropy defined as the pairing of the
boundary states and
 the supersymmetric
ground states.

\subsection{The Supersymmetric Boundary Conditions}

\begin{figure}[htb]
\begin{center}
\epsfxsize=1.5in\leavevmode\epsfbox{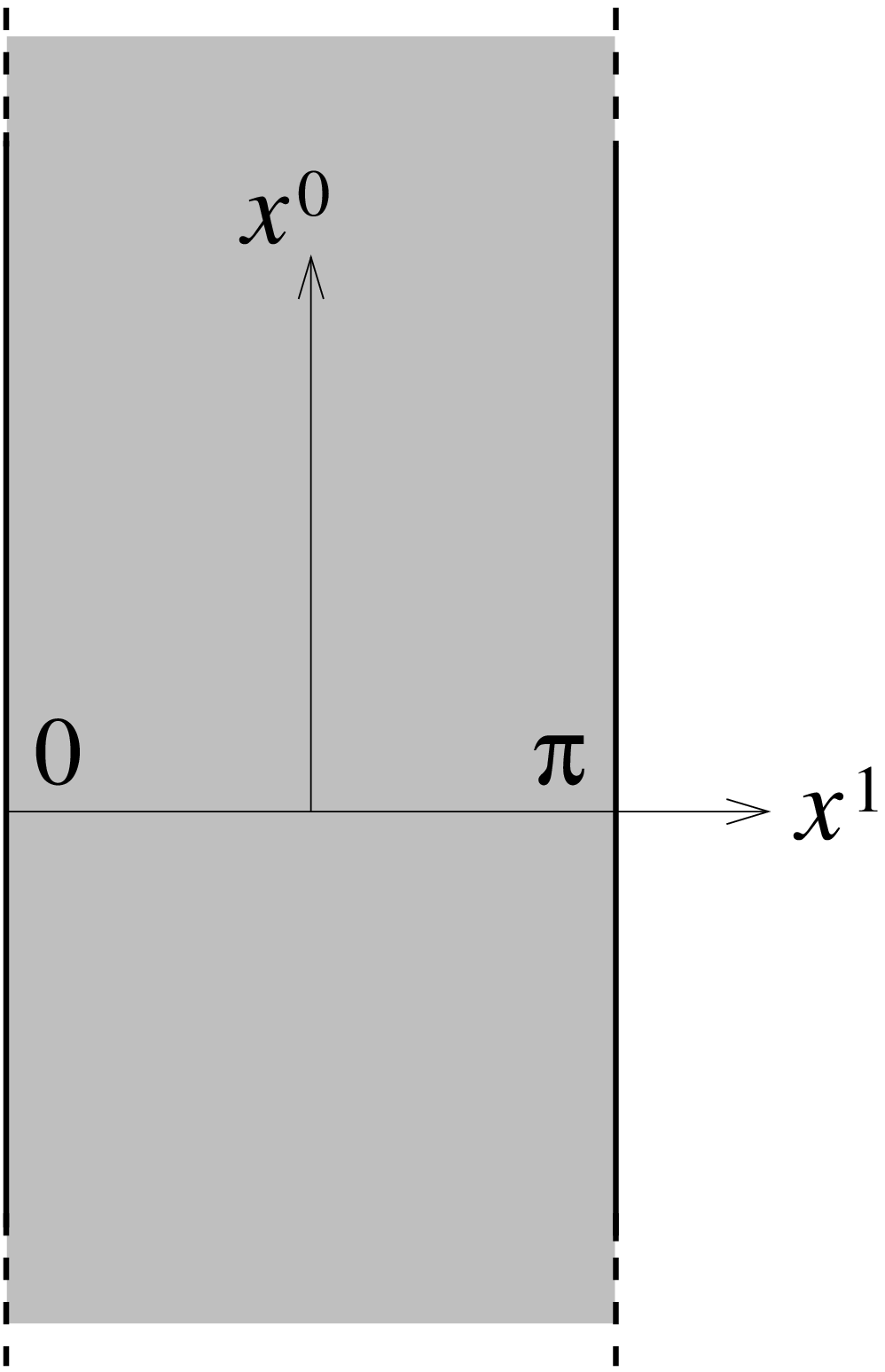}
\end{center}
\caption{The strip $\R\times I$}
\label{strip}
\end{figure}
Let us consider a supersymmetric sigma model on a K\"ahler manifold
$X$ of dimension $n$ with a superpotential $W$.
We denote the K\"ahler metric with respect to local complex
coordinates $z^i$ as $g_{i\bj}$.
We formulate the theory on the strip $\Sigma=\R\times I$
where $I$ is an interval $0 \leq x^1\leq \pi$
and $\R$ is parametrized by the time coordinate $x^0$.
Here, without any
loss in generality we have fixed the size of the interval to a fixed length.
Changing the length of the interval
is equivalent to changing the parameters in the
action according to the RG flow.

The action of the system is given by
\\
\beqa
S&=&\int\limits_{\Sigma}\dd^2x
\Bigl\{\,-g_{i\bj}\partial^{\mu}\phi^i\partial_{\mu}\bphi^{\bj}
+{i\over 2}g_{i\bj}\,\bpsi_-^{\bj}(\lrD_{\!\! 0}+\lrD_{\!\! 1})\psi^i_-
+{i\over 2}g_{i\bj}\,\bpsi_+^{\bj}(\lrD_{\!\! 0}-\lrD_{\!\! 1})\psi^i_+
\nonumber\\
&&~~~~~~~~~
-{1\over 4}g^{\bj i}\partial_{\bj}\overline{W}\partial_i W
-{1\over 2}(D_i\partial_jW)\psi_+^i\psi_-^j
-{1\over 2}(D_{\bi}\partial_{\bj}\overline{W})\bpsi_-^{\bi}\bpsi_+^{\bj}
\nonumber\\[0.2cm]
&&~~~~~~~~~~
+R_{i\bar kj\bar l}\,\psi_+^i\psi_-^j\bpsi_-^{\bar k}\bpsi_+^{\bar l}
~\Bigr\},
\label{action}
\eeqa
\\
where
$\bpsi^{\bj}\lrD_{\!\! \mu}\psi^i=
\bpsi^{\bj} (D_{\mu}\psi)^i-(D_{\mu}\bpsi)^{\bj}\psi^i$.
See \cite{HV} for other notations.
The above action is the same as the component expression of
(\ref{actio}) up to a boundary term. 
We require the equations of motion for the fields $\phi^i, \psi_{\pm}^i$
to be local. This yields the following conditions on the boundary
$\partial\Sigma$
\beqa
&&g_{IJ}\delta\phi^I\partial_1\phi^J=0,
\label{loc1}\\
&&g_{IJ}(\psi_-^I\delta\psi_-^J-\psi_+^I\delta\psi_+^J)=0,
\label{loc2}
\eeqa
where $\phi^I,\psi_{\pm}^I$ and $g_{IJ}$ are the components of the fields
and the metric with respect to the real coordinates of the target space.

Under the supersymmetry transformation
\beqa
&&\delta\phi^i=\epsilon_+\psi_-^i-\epsilon_-\psi_+^i,\\
&&\delta\psi_+^i=i\bepsilon_-(\partial_0+\partial_1)\phi^i+\epsilon_+F^i,\\
&&\delta\psi_-^i=-i\bepsilon_+(\partial_0-\partial_1)\phi^i+\epsilon_-F^i,
\eeqa
where
\beq
F^i=-{1\over 2} g^{i\bj}\partial_{\bj}\overline{W}
+\Gamma^i_{jk}\psi_+^j\psi_-^k,
\label{Fdef}
\eeq
the action varies as
\\
\beqa
\delta S\!\!\!&=&\!\!{1\over 2}\int\limits_{\partial \Sigma}\!\!\dd x^0
\Bigl\{
\epsilon_+\Bigl(-g_{i\bj}(\partial_0+\partial_1)\bphi^{\bj}\psi_-^i
+{i\over 2}\bpsi_+^{\bi}\partial_{\bi}\overline{W}\Bigr)
+\epsilon_-\Bigl(-g_{i\bj}(\partial_0-\partial_1)\bphi^{\bj}\psi_+^i
-{i\over 2}\bpsi_-^{\bi}\partial_{\bi}\overline{W}\Bigr)
\nonumber\\
&&~~~~
+\bepsilon_+\Bigl(g_{i\bj}\bpsi_-^{\bj}(\partial_0+\partial_1)\phi^i
+{i\over 2}\psi_+^i\partial_iW\Bigr)
+\bepsilon_-\Bigl(g_{i\bj}\bpsi_+^{\bj}(\partial_0-\partial_1)\phi^i
-{i\over 2}\psi_-^i\partial_iW\Bigr)\Bigr\}.
\nonumber\\
&&
\eeqa
If the boundary were absent, the action would be invariant under the
full $(2,2)$ supersymmetry and
the following four supercurrents would be conserved.
\beqa
&&G_{\pm}^0=g_{i\bj}(\partial_0\pm\partial_1)\bphi^{\bj}\psi_{\pm}^i
\mp {i\over 2}\bpsi_{\mp}^{\bi}\partial_{\bi}\overline{W},~~
G_{\pm}^1=\mp g_{i\bj}(\partial_0\pm\partial_1)\bphi^{\bj}\psi_{\pm}^i
-{i\over 2}\bpsi_{\mp}^{\bi}\partial_{\bi}\overline{W},~~
\nonumber\\
&&\bG_{\pm}^0=g_{i\bj}\bpsi_{\pm}^{\bj}
(\partial_0\pm\partial_1)\phi^{i}
\pm {i\over 2}\psi_{\mp}^{i}\partial_{i}W,~~~
\bG_{\pm}^1=\mp g_{i\bj}\bpsi_{\pm}^{\bj}
(\partial_0\pm\partial_1)\phi^{i}
+{i\over 2}\psi_{\mp}^{i}\partial_{i}W.~~
\nonumber
\eeqa

In what follows, we determine the boundary conditions on the fields
$\phi^i, \psi_{\pm}^i$ that preserve half of the supersymmetry.
We also wish to maintain the translation symmetry that maps
the worldsheet boundary to itself, which is the
time translation in the present set-up.
There are essentially two possibilities for the unbroken supercharges
\cite{OOY};

\medskip
(A)~ $Q=\bQ_++\e^{i\alpha}Q_-$ and $Q^{\dag}=Q_++\e^{-i\alpha}\bQ_-$,

(B)~ $Q=\bQ_++\e^{i\beta}\bQ_-$ and $Q^{\dag}=Q_++\e^{-i\beta}Q_-$.

\medskip

\noindent
Here $\e^{i\alpha}$ and $\e^{i\beta}$ are arbitrary phases.
In both cases, the supercharges satisfy
$\{Q,Q^{\dag}\}=2H$, up to a possible central term.
The variation parameters for these supersymmetries are
$\epsilon_-=\e^{i\alpha}\bepsilon_+$
for (A)
while $\epsilon_-=-\e^{i\beta}\epsilon_+$
for (B).
Conservation of the charges $Q$ and $Q^{\dag}$
requires that the spatial component of the corresponding
currents vanish at the boundary $\partial\Sigma$:
$\bG_+^1+\e^{i\alpha}G_-^1=G_+^1+\e^{-i\alpha}\bG_-^1=0$
for (A),
and $\bG_+^1+\e^{i\beta}\bG_-^1=G_+^1+\e^{-i\beta}G_-^1=0$
for (B).

The conditions we are interested in are the ones associated with
{\it D-branes} wrapped on a submanifold $\cycle$ of $X$.
Namely, we require the worldsheet boundary
to be mapped to $\cycle$.
In such a case, 
the derivative along the boundary $\partial_0\phi^I$ as well as
an allowed variation $\delta\phi^I$ at the boundary must be tangent
to $\cycle$.
The locality condition (\ref{loc1}) then tells that $\partial_1\phi^I$
must be normal to $\cycle$.
The other condition (\ref{loc2}) is satisfied if
$\psi_-^I$ and $\psi_+^I$ are related by an orthonormal transformation
$\psi_-^I=M^I_{\,\,J}\psi_+^J$, $g_{IJ}M^I_{\,\,K}M^J_{\,\,L}=g_{KL}$.
In fact, supersymmetry requires this and determines the matrix
$M^I_{\,\,J}$, as we now show.
For simplicity, we set the phases $\e^{i\alpha}$ and $\e^{i\beta}$
to be equal to $1$; the general case
can be easily recovered by
$U(1)_V$ and $U(1)_A$ rotations.
Then, both (A) and (B) contains an $N=1$ subalgebra generated by
the variations with  parameter
$\epsilon_+=i\epsilon$ and $\epsilon_-=-i\epsilon$ 
where $\epsilon$ is real.
Expressed in the real coordinates,
the action varies as
\beq
\delta S=
{i\epsilon\over 2}\int\limits_{\partial\Sigma}
\dd x^0\Bigl\{-g_{IJ}\partial_0\phi^I(\psi_-^J-\psi_+^J)
-g_{IJ}\partial_1\phi^I(\psi_-^J+\psi_+^J)
-{i\over 2}(\psi_-^I+\psi_+^I)\partial_I(W-\overline{W})
\Bigr\}.
\eeq
Since $\partial_0\phi^I$ and $\partial_1\phi^i$ are tangent and normal to
$\cycle$, the invariance of the action requires
$i\epsilon(\psi_-^I-\psi_+^I)$ and $i\epsilon(\psi^I_-+\psi_+^I)$
to be normal and tangent to $\cycle$ respectively.
This means that
\beq
\psi_-^I=\left\{
\begin{array}{ll}
\psi_+^I&I:~\mbox{tangent}\\[0.15cm]
-\psi_+^I&I:~\mbox{normal}
\end{array}
\right.
\label{bcfermi}
\eeq
for a choice of coordinates that separates
the tangent and normal directions.
Furthermore, invariance of $S$ requires $W-\overline{W}$
to be a constant along $\cycle$.

As we will see, A-type supersymmetry requires $\cycle$ to be a middle
dimensional Lagrangian submanifold whose image in the $W$-plane is
a straight line, while B-type supersymmetry requires $\cycle$
to be a holomorphic submanifold on which $W$ is a constant.

\subsection*{A-Type Supersymmetry}

We first consider A-type supersymmetry with the trivial
phase $\e^{i\alpha}=1$, which is generated by the variations with
parameters $\epsilon_-=\bepsilon_+$ and $\bepsilon_-=\epsilon_+$.
The bosonic fields $\phi^i$ transform as
\beqa
\delta\phi^i&=&\epsilon_+\psi_-^i-\bepsilon_+\psi_+^i
\nonumber\\
&=&\epsilon_1(\psi_-^i-\psi_+^i)
+i\epsilon_2(\psi_-^i+\psi_+^i),
\eeqa
where $\epsilon_1$ and $\epsilon_2$ are the real and the imaginary parts
of
$\epsilon_+$; $\epsilon_+=\epsilon_1+i\epsilon_2$.
This shows that, for a real parameter $\epsilon$,
$\epsilon(\psi_-^i-\psi_+^i)$
and $i\epsilon(\psi_-^i+\psi_+^i)$ are
the holomorphic components of
tangent vectors of $\cycle$.
On the other hand,
$N=1$ supersymmetry requires
$i\epsilon(\psi_-^i-\psi_+^i)$
and $i\epsilon(\psi_-^i+\psi_+^i)$ are
the holomorphic components of
normal and tangent vectors of $\cycle$ respectively.
Thus, multiplication by $i=\sqrt{-1}$ on the holomorphic components
sends tangent vectors to normal vectors and vice versa.
Namely, the cycle $\cycle$ must be a middle dimensional Lagrangian
submanifold of $X$ (where $X$ is considered as a symplectic manifold
defined by the K\"ahler form).
The supersymmetry transformation of the tangent vector
$\epsilon(\psi^i_--\psi_+^i)$ is
\beq
\delta[\epsilon(\psi^i_--\psi_+^i)]
=2i\epsilon\epsilon_1(\partial_0\phi^i)
+2i\epsilon\epsilon_2(i\partial_1\phi^i
+F^i).
\label{vartan}
\eeq
This must again be tangent to $\cycle$
when $\delta\phi^i=0$ (i.e. $\psi_{\pm}^i=0$ for which
$F^i=-{1\over 2}g^{i\bj}\partial_{\bj}\overline{W}$).
We note that $i\epsilon\epsilon_1$ and $i\epsilon\epsilon_2$
are real parameters, $(i\epsilon\epsilon_i)^{\dag}=
-i\epsilon_i\epsilon=i\epsilon\epsilon_i$.
Since $\partial_0\phi^i$, $i\partial_1\phi^i$
are both tangent to $\cycle$, the vector
$v^i=g^{i\bj}\partial_{\bj}\overline{W}$ must be tangent to $\cycle$.
This is consistent with the requirement of $N=1$ supersymmetry
that $W-\overline{W}$ must be a constant along $\cycle$,
since the vector $v^I$ annihilates $W-\overline{W}$
\beq
v^I\partial_I(W-\overline{W})=|\partial W|^2-|\partial W|^2=0.
\eeq
It is easy to see that under these conditions the action
$S$ is invariant under the full A-type supersymmetry with
$\e^{i\alpha}=1$ and also that no other condition is required.

It is easy to recover the phase $\e^{i\alpha}$ using the
$U(1)_V$ symmetry which rotates the fermions as
$\psi_{\pm}^i\to\e^{i\alpha/2}\psi_{\pm}^i$ and 
$\bpsi_{\pm}^{\bj}\to\e^{-i\alpha/2}\bpsi_{\pm}^{\bj}$
and the superpotential as
$W\to \e^{-i\alpha}W$.
The boundary conditions on the fermions
(\ref{bcfermi}) are rotated accordingly.
The condition on the cycle is also rotated:
{\it the cycle $\cycle$ is a middle dimensional Lagrangian submanifold of
$X$ with respect to the K\"ahler form, whose image
in the $W$-plane is a straight line in the 
$\e^{i\alpha}$-direction.}

The axial $U(1)$ R-symmetry is not broken by the boundary condition.
Indeed, $\epsilon(\psi_-^i-\psi_+^i)$
and $i\epsilon(\psi_-^i+\psi_+^i)$, which are holomorphic component of
tangent vectors to $\cycle$ (for $\e^{i\alpha}=1$),
are rotated within themselves
by the axial rotation $\psi_{\mp}^i\to\e^{\pm i\theta_A}\psi_{\mp}^i$.
If not anomalous (i.e. if $c_1(X)=0$),
the axial R-charge is conserved and
the spatial component of the current must vanish at the boundary,
$J_A^1=0$ at $\partial\Sigma$.
We note that the conserved supercharges $Q$ and $Q^{\dag}$
have axial charge $1$ and $-1$ respectively.

The basic example of a cycle satisfying this condition is the
wave-front trajectory emanating from a critical point of
the superpotential.
We consider here the case $\e^{i\alpha}=1$ for simplicity.
Let $p_*\in X$ be a non-degenerate critical point of $W$
and let us consider
a wave-front trajectory $\cycle_{p_*}$ emanating from $p_*$ in the
positive real direction.  As discussed in section 2 this
corresponds to the totality of all potential soliton
solutions starting at $p_*$ whose image in the $W$-plane is stretched
along the positive real axis.
We recall that the one parameter family of maps $f_t$
generated by the vector field $v^i=g^{i\bj}\partial_{\bj}\overline{W}$
acts on $\cycle_{p_*}$ and any point on it is mapped by $f_t$ to
$p_*$ in the limit $t\to -\infty$.
By definition, $\cycle_{p_*}$ is of middle dimension and
the image in the $W$-plane
is a straight-line in the real direction.
To see that $\cycle_{p_*}$ is a Lagrangian submanifold of $X$ with respect
to the K\"ahler form $\omega=ig_{i\bj}\dd z^i\wedge\dd \overline{z}^{\bj}$,
it is crucial to note that
\beq
i_v\omega=i(g_{i\bj}v^i\dd\overline{z}^{\bj}
-g_{i\bj}\dd z^i\overline{v}^{\bj})
=i\dd(\overline{W}-W),
\eeq
and hence
\beq
{\cal L}_v\omega=\dd i_v\omega+i_v\dd\omega=0.
\eeq
Thus,
$\omega$ is invariant under the diffeomorphisms $f_t$.
Let $V_1$ and $V_2$ be tangent vectors of $\cycle_{p_*}$
at any point.
Since the K\"ahler form is $f_t$-invariant,
$\omega(f_tV_1,f_tV_2)=(f_t^*\omega)(V_1,V_2)$ is independent of $t$.
However, in the limit
$t\to-\infty$, the vectors $f_tV_i$ become the zero vector at $p_*$.
Thus, we have shown $\omega(V_1,V_2)=0$.
Namely, $\cycle_{p_*}$ is Lagrangian.

\subsection*{B-Type Supersymmetry}

We next consider B-type supersymmetry with the phase $\e^{i\beta}=1$
which is generated by $\epsilon_-=-\epsilon_+$
and $\bepsilon_-=-\bepsilon_+$.
The bosonic fields $\phi^i$ transform as
\beq
\delta\phi^i=\epsilon_+(\psi_-^i+\psi_+^i).
\eeq
Since $\epsilon_+$ is a complex parameter, this shows that
the tangent space to $\cycle$ is invariant under
the multiplication by $i=\sqrt{-1}$ on the holomorphic components.
Namely, the cycle $\cycle$ must be a complex submanifold of $X$.
The supersymmetry transformation of
the tangent vector $\psi_-^i+\psi_+^i$ is
$\delta(\psi_-^i+\psi_+^i)=-2i\bepsilon_+\partial_0\phi^i$ which is indeed
tangent to $\cycle$.
On the other hand, the normal vector
$\psi_-^i-\psi_+^i$ transforms as
\beq
\delta(\psi_-^i-\psi_+^i)=2i\bepsilon_+\partial_1\phi^i
+\epsilon_+g^{i\bj}\partial_{\bj}\overline{W},
\eeq
at $\psi_{\pm}^i=0$ for which $\delta\phi^i=0$.
This must again be normal to $\cycle$.
Since $\partial_1\phi^i$ is normal to $\cycle$,
this requires that
$n^i=g^{i\bj}\partial_{\bj}\overline{W}$ is also
a normal vector to $\cycle$.
Namely, for a tangent vector $v^i$ we have
\beq
0=g_{i\bj}v^i\overline{n}^{\bj}
=v^i\partial_iW.
\eeq
Thus,
not only the imaginary part $W-\overline{W}$
but $W$ itself must be a constant on $\cycle$.
It is easy to see that under these conditions
the action is invariant under the full B-type
supersymmetry with $\e^{i\beta}=1$
and also that no other condition is required.

It is again easy to recover the phase $\e^{i\beta}$
using the $U(1)_A$ symmetry
which rotates the fermions as
$\psi_{\pm}^i\to \e^{\pm i\beta/2}\psi_{\pm}^i$
and $\bpsi_{\pm}^{\bj}\to\e^{\mp i\beta/2}\bpsi_{\pm}^{\bj}$.
The boundary conditions on the fermions (\ref{bcfermi}) are
rotated accordingly, but the condition on the cycle remains the same:
 {\it the cycle $\cycle$ is a complex submanifold of $X$
on which $W$ is a constant.}

The vector $U(1)$ R-symmetry is not broken by the boundary condition.
Indeed,
the tangent vector
$\epsilon(\psi_-^i+\psi_+^i)$ to $\cycle$ (for $\e^{i\beta}=1$)
is rotated by phase under the vector rotation
$\psi_{\mp}^i\to\e^{i\theta_V}\psi_{\mp}^i$
and hence remains tangent.
If not broken by the superpotential,
the vector R-charge is conserved and
the spatial component of the current must vanish at the boundary,
$J_V^1=0$ at $\partial\Sigma$.
We note that the conserved supercharges $Q$ and $Q^{\dag}$
has the vector charge $1$ and $-1$ respectively.

\subsubsection{Inclusion of the B-field}

~~~~We can deform the theory by adding the following term to the action
(\ref{action})
\beq
{1\over 2}\int\limits_{\Sigma}B_{IJ}\dd\phi^I\wedge\dd\phi^J,
\eeq
where $B={1\over 2}B_{IJ}\dd x^I\wedge\dd x^J$
is a closed two-form on the manifold $X$.
This term alters the condition (\ref{loc1}) of locality
for the bosonic equations of motion as
\beq
\delta\phi^I(g_{IJ}\partial_1\phi^J+B_{IJ}\partial_0\phi^J)=0,
\label{loc3}
\eeq
but the condition (\ref{loc2}) for the fermionic equations of motion
remains the same.

We look for the boundary conditions associated with the D-branes
wrapped on a cycle $\cycle$ in $X$ which preserves
A-type or B-type supersymmetry.
By definition and by the requirement (\ref{loc3}),
the bosonic fields must obey the boundary conditions
\beq
\begin{array}{ll}
g_{IJ}\partial_1\phi^J+B_{IJ}\partial_0\phi^J=0,
&I:~\mbox{tangent},
\\[0.2cm]
\partial_0\phi^I=0,
&I:~\mbox{normal},
\end{array}
\label{locB}
\eeq
where we have chosen the coordinates that separate
the tangent and the normal directions.
For invariance under the
$N=1$ supersymmetry generated by $\epsilon_+=i\epsilon$ and
$\epsilon_-=-i\epsilon$ with $\epsilon$ being real (which is contained
in both (A) and (B) supersymmetries with the trivial phases), the
following boundary conditions on the fermions are required:
\beq
\begin{array}{ll}
g_{IJ}(\psi_-^J-\psi_+^J)-B_{IJ}(\psi_-^J+\psi_+^J)=0,
&I:~\mbox{tangent},
\\[0.2cm]
\psi_-^I+\psi_+^I=0,
&I:~\mbox{normal}.
\end{array}
\label{N1B}
\eeq
This also guarantees the condition (\ref{loc2}).
We also obtain the condition that the imaginary part of $W$ is a constant
along $\cycle$.

Proceeding as in the case without $B$-field,
we obtain the following conditions on the cycle $\cycle$
for the A- and B-type
supersymmetry to be preserved.
We only state the conditions for the cases with the trivial phase
$\e^{i\alpha}=\e^{i\beta}=1$ since the generalization is clear.

\noindent
\underline{A-type Supersymmetry}

{\it $\cycle$ is a middle dimensional Lagrangian submanifold of $X$
on which (not only the K\"ahler form but also) the $B$-field
is annihilated, $B|_{\cycle}=0$.
The image in the $W$-plane must be a straight line
in the real direction.}

\noindent
\underline{B-type Supersymmetry}

{\it $\cycle$ is a complex submanifold of $X$.
$B$-field evaluated on the holomorphic tangent vectors to
$\cycle$ is zero, $(B|_{\cycle})^{(2,0)}=0$. Also, $W$ must be a
constant on $\cycle$.}

\subsubsection{Coupling to the Gauge Fields on the Branes}

~~~~We can couple the worldsheet boundaries to
the gauge fields on the branes. In the case of the strip
$\Sigma=\R\times I$, this corresponds to
adding to the action (\ref{action}) the terms
\beq
\int\limits_{\partial\Sigma}A_M\dd\phi^M
=\int\limits_{x^1=\pi}\dd x^0\partial_0\phi^{M_b}A^{(b)}_{M_b}
-\int\limits_{x^1=0}\dd x^0\partial_0\phi^{M_a}A^{(a)}_{M_a},
\label{Acoup}
\eeq
where
$A^{(a)}$ and $A^{(b)}$ are the
$U(1)$ gauge fields on the branes
$\cycle_a$ and $\cycle_b$
on which the left and the right boundaries end.
(We use $M,N,\ldots$ for coordinate indices on the branes.)
If the left and the right boundaries are coupled to the same gauge field
$A$ that extends to the whole target space $X$,
the boundary terms (\ref{Acoup}) can be written as
\beq
{1\over 2}\int\limits_{\Sigma}F_{IJ}\dd\phi^I\wedge\dd\phi^J
\eeq
where $F={1\over 2}F_{IJ}\dd x^I\wedge\dd x^J$
is the curvature of the gauge field, $F=\dd A$.
Thus, in this case, we can
treat the gauge field coupling in the same way as the
coupling to the $B$-field.
In particular, 
we have the local equations of motion and $N=1$ supersymmetry
by imposing the boundary conditions
 (\ref{locB}) and (\ref{N1B}) with $B\to F$.
When the cycle $\cycle$ is a middle dimensional Lagrangian submanifold
of $X$ whose image in the $W$-plane is a straight line,
the theory is invariant under
A-type supersymmetry if the gauge field is flat on the cycle,
$F|_{\cycle}=0$.
When the cycle $\cycle$ is a complex submanifold of $X$
on which $W$ is a constant, the theory is invariant under B-type
supersymmetry if the gauge field has a $(1,1)$ curvature on $\cycle$,
$(F|_{\cycle})^{(2,0)}=0$, namely, if $A|_{\cycle}$
determines a holomorphic line bundle on $\cycle$.

The conclusion obtained above remains valid
even if the left and the right boundary components are coupled to
different gauge fields that are defined only on the branes.
Thus, {\it one can deform the A-type supersymmetric theory 
by flat gauge fields on $\cycle$ while 
B-type supersymmetric theory can be deformed by
holomorphic line bundles on $\cycle$.}

\subsection*{\it An Alternative Formulation for B-type D-branes}

For A-type D-branes with a flat gauge field,
the boundary condition given by (\ref{locB}) and (\ref{N1B})
(with $B_{IJ}\to F_{IJ}$) is the same as the standard one
\beq
\begin{array}{ll}
\partial_1\phi^I=0,~~\psi_-^I-\psi_+^I=0,
&I:~\mbox{tangent},
\\[0.2cm]
\partial_0\phi^I=0,~~\psi_-^I+\psi_+^I=0
&I:~\mbox{normal}.
\end{array}
\label{stbc}
\eeq
However, it is in general different from (\ref{stbc})
for B-type D-branes where the gauge field is not necessarily flat.
There is actually an alternative formulation for B-type D-branes
where we still impose the standard boundary condition (\ref{stbc}).
It is easy to see that
(with $\psi^I:=(\psi_+^I+\psi_-^I)/2$)
\beq
\int\limits_{\partial\Sigma}
\dd x^0\,\left\{\,
\partial_0\phi^MA_M+iF_{MN}\psi^M\psi^N
\,\right\}
\label{Sbound}
\eeq
is invariant by itself under the B-type supersymmetry
if the gauge field is holomorphic, $F_{mn}=F_{\bar m\bar n}=0$.
Thus, instead of (\ref{Acoup}),
one can add the boundary term (\ref{Sbound})
without breaking the B-type supersymmetry of the bulk action
(\ref{action}) which holds under (\ref{stbc}).
We note however that the equations of motion for the fields
$\phi^I$, $\psi_{\pm}^I$
are modified by boundary terms.
This formulation was used in \cite{Fradkin,CLNY,abouel}
to study the fluctuation
of the target space gauge fields in string theory.

\subsection*{\it Non-Abelian Gauge Fields}

One can generalize the above analysis to 
non-abelian $U(k)$ gauge group \cite{CLNY,WCS}.
In this case, the path-integral weight $\exp (iS)$
is accompanied by the matrix factors
\beq
P_{\partial\Sigma}\exp\left(\,i\int_{\partial\Sigma}
\dd x^0\,\left\{\,
\partial_0\phi^MA_M+iF_{MN}\psi^M\psi^N
\,\right\}
\,\right)
\label{extwt}
\eeq
where $P_{\partial\Sigma}$ is (the product of) the path-ordering
along the boundary $\partial\Sigma$.
Under the standard boundary condition (\ref{stbc}), the weight
(\ref{extwt}) is invariant
under A-type supersymmetry if $A$ is flat, $F_{MN}=0$,
while it is invariant under B-type supersymmetry
if $A$ is holomorphic, $F_{mn}=F_{\bar m\bar n}=0$.

\subsection{Supersymmetric Ground States}

As in any supersymmetric field theory, in the theory
on the segment $I=[0,\pi]$ with the boundary condition that
preserves A or B-type supersymmetry, one can define the supersymmetric
index $\Tr(-1)^F$
which is invariant under deformations of the theory.
We denote this index as
\beq
I(a,b)=\Tr\,(-1)^F,
\label{Win}
\eeq
where $a$ and $b$ are the boundary conditions
at the left and the right boundaries.\footnote{The index (\ref{Win})
for supersymmetric
D-branes in Calabi-Yau manifolds was studied in \cite{Douglas-Fiol,BDL}.}
We shall compute this index in the two basic examples;
A-type D-branes in Landau-Ginzburg models
and B-type D-branes in non-linear sigma models.
Actually, in the LG models (not only the index but also)
the complete spectrum of supersymmetric ground states can be determined.
This can also be done for non-linear sigma models
under a certain condition on the cohomology of the gauge bundles.
For simplicity, we set the phases $\e^{i\alpha}=1$ and $\e^{i\beta}=1$.

\subsubsection{Landau-Ginzburg Models}

~~~~
Let us consider a LG model with superpotential $W$.
We assume that the bosonic potential
$U=|\partial W|^2$ diverges at 
infinity in the configuration space $X$.
We also assume that there is no non-trivial $B$ field and we will not
consider coupling to gauge field on the branes.
Let $a$ and $b$ be two non-degenerate critical points of $W$.
We consider the wave-front trajectories $\cycle_a$ and $\cycle_b$
emanating from $a$ and $b$ in the positive real direction
in the $W$-plane.
We assume for now that
the half-lines $W(\cycle_a)$ and $W(\cycle_b)$
are separated in the imaginary direction,
and there is no other critical values of $W$ between them.
We consider the theory on $[0,\pi]$ where the left boundary $x^1=0$ is
mapped to $\cycle_a$ and the right boundary $x^1=\pi$
is mapped to $\cycle_b$.
For the boundary condition described earlier, the theory is invariant under
A-type supersymmetry generated by the
supercharges $Q=\bQ_++Q_-$ and $Q^{\dag}=Q_++\bQ_-$, which
are expressed as
\beqa
Q\!\!\!&=&\!\!\!\!\sqrt{2}\int_0^{\pi}\dd x^1\left\{\,
\bpsi_{+}^{\bj}\Bigl(
g_{i\bj}(\partial_0+\partial_1)\phi^i
+{i\over 2}\partial_{\bj}\overline{W}\Bigr)
+\psi_{-}^i\Bigl(g_{i\bj}(\partial_0-\partial_1)\bphi^{\bj}
+{i\over 2}\partial_iW\Bigr)\,\right\},~~
\nonumber\\
Q^{\dag}\!\!\!&=&\!\!\!\!\sqrt{2}\int_0^{\pi}\dd x^1\left\{\,
\bpsi_{-}^{\bj}\Bigl(
g_{i\bj}(\partial_0-\partial_1)\phi^i
-{i\over 2}\partial_{\bj}\overline{W}\Bigr)
+\psi_{+}^i\Bigl(g_{i\bj}(\partial_0+\partial_1)\bphi^{\bj}
-{i\over 2}\partial_iW\Bigr)\,\right\}.~~
\nonumber\\
\label{QpmbQpm}
\eeqa
The supercharges $Q$ and $Q^{\dag}$
are nilpotent and satisfy the anti-commutation relation
$$
\{Q,Q^{\dag}\}=4(H+{\mit \Delta}{\rm Im}W),
$$
where ${\mit \Delta}{\rm Im}W={\rm Im}W(b)-{\rm Im}W(a)$
is the separation of the two half-lines in the imaginary direction.
We shift the definition of the Hamiltonian as
$\widetilde{H}=H+{\mit \Delta}{\rm Im}W$
so that the supersymmetry algebra takes
the standard form
$\{Q,Q^{\dag}\}=4\widetilde{H}$.
Since ${\mit \Delta}{\rm Im}W$ is a constant,
this is done simply by the shift of the action
\beq
\widetilde{S}=S-\int\limits_{x^1=\pi}\dd x^0\, {\rm Im}W
+\int\limits_{x^1=0}\dd x^0\,{\rm Im}W.
\label{shift}
\eeq
The index can be defined by
$I(a,b)=\Tr\,(-1)^F\e^{-\beta \widetilde{H}}$,
and only the ground states with energy $\widetilde{H}=0$
can contribute to this.
One can see from the expressions (\ref{QpmbQpm}) that
$Q=Q^{\dag}=0$ for a static configuration 
such that
\beq
\partial_1\phi^i=-{i\over 2}g^{i\bj}\partial_{\bj}\overline{W}.
\label{straightup}
\eeq
Namely, the supersymmetry is classically preserved 
for a static configuration that goes from $\cycle_a$ to $\cycle_b$,
straight down in the negative imaginary direction of the $W$-plane.
Such a configuration would indeed have $H=-{\mit \Delta}{\rm Im}W$
or $\widetilde{H}=0$
and satisfy the required boundary condition.
We note that there is no
such configuration if ${\rm Im}W(a)<{\rm Im}W(b)$.
In such a case $I(a,b)=0$.

\begin{figure}[htb]
\begin{center}
\epsfxsize=5in\leavevmode\epsfbox{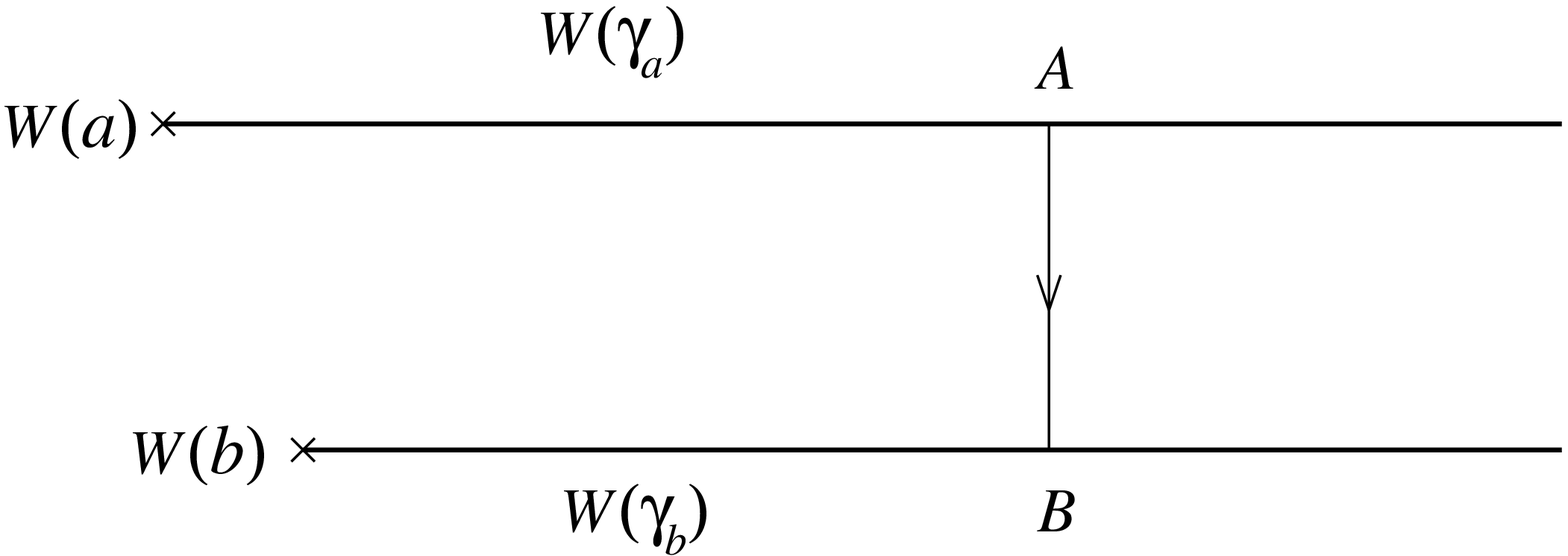}
\end{center}
\caption{The image of D-branes in the $W$-plane and a path between
them}
\label{paths}
\end{figure}
\begin{figure}[htb]
\begin{center}
\epsfxsize=2.7in\leavevmode\epsfbox{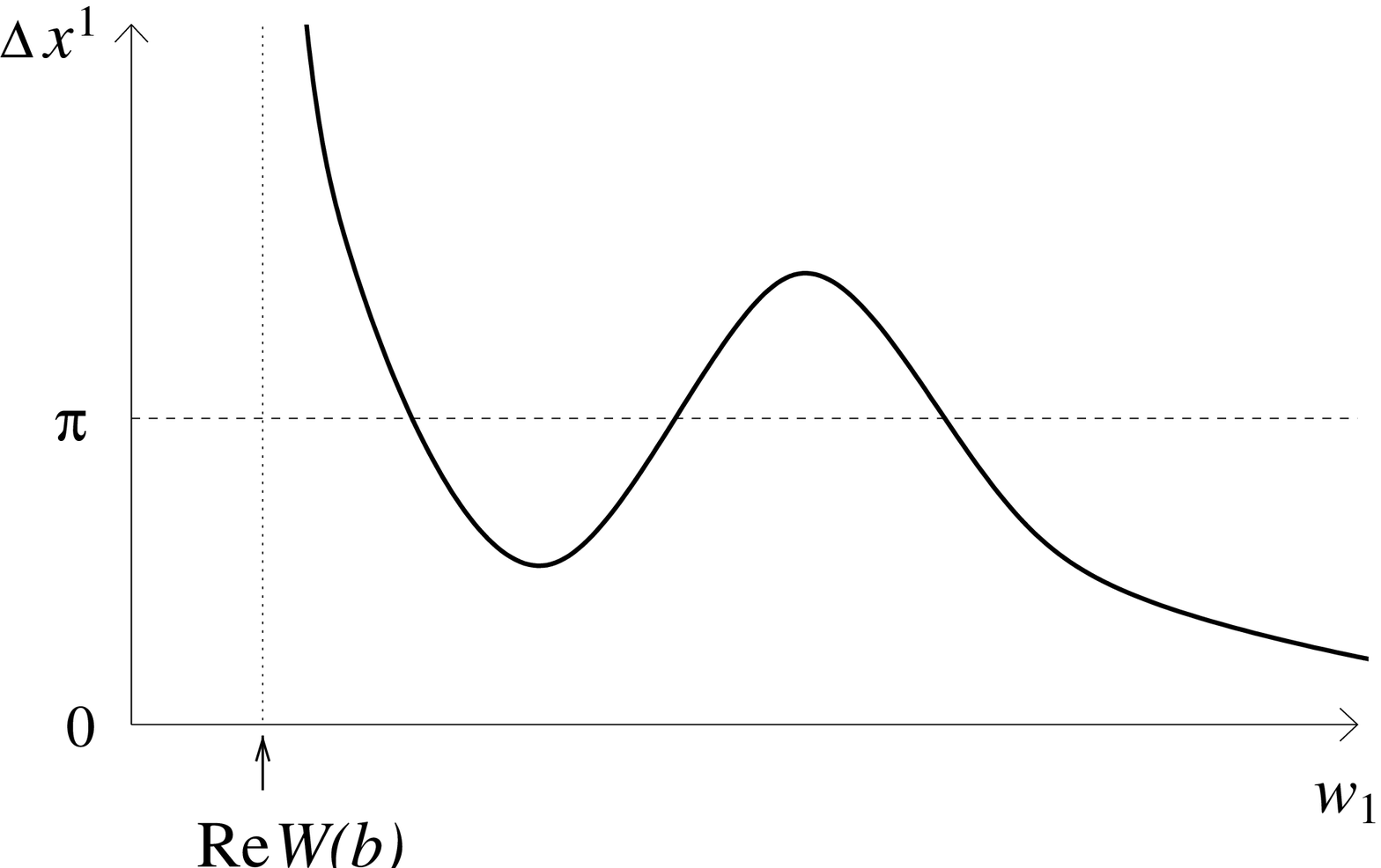}
\end{center}
\caption{A Graph of ${\mit\Delta}x^1$ as a function of
$w_1={\rm Re}A={\rm Re}B$.
Corresponding to the situation in Figure \ref{paths},
the left end is set at $w_1={\rm Re}W(b)$.}
\label{mdel}
\end{figure}

Now let us compute the index.
We are considering the situation as depicted in Figure \ref{paths}
where the arrowed line from $A$ to $B$ is a straight segment in the negative
imaginary direction of the $W$-plane.
We consider the wave-front at the point $B$
along the straight line from $W(b)$ 
and another wave-front at $B$ along the broken segment starting
from $W(a)$ and bending at the point $A$.
 From the general theory of singularities, the two wave-fronts
have intersection number $\Delta_a\circ\Delta_b$, the same as the soliton
number between $a$ and $b$.
This means that there are $\Delta_a\circ\Delta_b$ paths
from $\cycle_a$ to $\cycle_b$ that maps
to the straight segment from $A$ to $B$ in the $W$-plane.
Since this holds for any starting point $A$,
there are $\Delta_a\circ\Delta_b$
families
of such paths parametrized by $w_1:={\rm Re}A={\rm Re}B$.
It may appear that there are infinitely many
solutions to (\ref{straightup}) and therefore infinite degeneracy of
supersymmetric ground states.
However, we note that the length of $x^1$ that is required
to go from $\cycle_a$ to $\cycle_b$ depends on each path
and does not necessarily coincide with $\pi$.
The required length of $x^1$ for each path $P$ is given by
\beq
{\mit\Delta}x^1=\left|2\int\limits_{P}
{\dd{\rm Im}W\over |\partial W|^2}\right|.
\label{mitdel}
\eeq
Only the path with ${\mit\Delta}x^1=\pi$
defines a classical supersymmetric ground state.
If the starting point $A$ or the end point $B$ is
the critical value $W(a)$ or $W(b)$, the required length
is infinity ${\mit\Delta}x^1=+\infty$.
In the massive theory where the bosonic potential $U=|\partial W|^2$
diverges at infinity, ${\mit\Delta}x^1$ approaches zero when
$w_1={\rm Re}A$ goes to infinity.
Thus, for each of the $\Delta_a\circ\Delta_b$ families,
${\mit\Delta}x^1$ is roughly a decreasing function as a function of
$w_1$.
If it is a monotonic function,
the function (\ref{mitdel}) cut through ${\mit\Delta}x^1=\pi$
exactly once and hence the contribution to the index of that family is
$1$.
However, one may encounter a family where
it cuts through ${\mit\Delta}x^1=\pi$ more than once
as depicted in Figure \ref{mdel}.
In such a case, we make use of the fact that the index is invariant under
the deformation of the theory. In particular, we can rescale the
superpotential as $W\to \e^t W$. This changes the function
(\ref{mitdel}) as ${\mit\Delta}x^1\to\e^{-t}{\mit\Delta}x^1$.
For an appropriate choice of $\e^t$ one can make
$\e^{-t}{\mit\Delta}x^1$ to cut through $\pi$ exactly once.
Thus, in any case, the contribution to the Witten index is $1$
for each family.
Thus, the total index is
given by
\beq
I(a,b)=\left\{
\begin{array}{ll}
\Delta_a\circ\Delta_b&\mbox{if}~~{\rm Im}W(a)>{\rm Im}W(b),\\[0.2cm]
0&\mbox{if}~~{\rm Im}W(a)<{\rm Im}W(b).
\end{array}\right.
\label{Wind}
\eeq

In the case $a=b$ there is of course a unique classical supersymmetric
configuration: $\phi^I(x^1)=a^I$ (constant along $x^1$).
Thus, we have
\beq
I(a,a)=1.
\eeq
It is easy to generalize the above analysis to the case where there are
critical values between $W(\cycle_a)$ and $W(\cycle_b)$.
Let us consider the simplest case where there is one critical value
$W(c)$. Then, the number of paths from $\cycle_a$ to $\cycle_b$
depends on whether the image
in the $W$-plane is on the left or right of $W(c)$, i.e.
whether $w_1<{\rm Re}W(c)$ or $w_1>{\rm Re}W(c)$;
we denote these numbers as $(\Delta_a\circ\Delta_b)_{<}$
and $(\Delta_a\circ\Delta_b)_{>}$ respectively.
Some paths on the left smoothly continue to the right of $W(c)$.
However, some others hit the critical
point $c$ at $w_1={\rm Re}W(c)$ where ${\mit\Delta}x^1$ blows up to
infinity.
The length ${\mit\Delta}x^1$
is bounded from below
by a positive value for the paths on the left of $W(c)$.
 Thus, by rescaling the superpotential if necessary, we can have a
situation where the solution with ${\mit\Delta}x^1=\pi$
exists only on the right of $W(c)$. Thus, the index is
given by $I(a,b)=(\Delta_a\circ\Delta_b)_{>}$.
It is obvious how to generalize this argument to the case where
there are more than one critical values in the region between $W(\cycle_a)$
and $W(\cycle_b)$.
The index is still given by (\ref{Wind})
where it is understood that
$\Delta_a\circ\Delta_b$ stands for the intersection number
of the wavefronts corresponding to the paths in the $W$-plane
that meet with each other
on the right of the right-most critical value.
Note that this is the same as the ``intersection number''
of $\cycle_a$ and $\cycle_b$ defined in the previous section; namely,
the number $\#(\cycle_a\cap\cycle_b^{\prime})$
where $\cycle_b^{\prime}$ is obtained by tilting $\cycle_b$ with an
infinitesimal positive angle against the real axis in the $W$-plane.

The asymmetry in $I(a,b)$ under the interchange
of $a$ and $b$ has an interesting interpretation, as we will
discuss later in this paper.
The mirror version of the same
asymmetry is discussed in the next subsection, for holomorphic D-branes
on sigma models.

One might be interested in exactly how many supersymmetric ground states
are there.
If the function (\ref{mitdel}) for a family of paths
cut through ${\mit\Delta}x^1=\pi$ exactly once,
there is of course one ground state from that family.
However, one may find a family where the graph of
${\mit\Delta}x^1$ looks like Figure \ref{mdel}.
In such a case, extra pairs of states may potentially become supersymmetric
ground states (though do not contribute to the index).
To see whether this is possible or not,
we show that, under a certain assumption,
the system under consideration is nothing but
the supersymmetric quantum mechanics considered
in \cite{SUSYB,Morse}
applied to the infinite-dimensional space of paths.
In \cite{SUSYB}, the system with Hamiltonian
$H={1\over 2}p^2+{1\over 2}(h^{\prime}(x))^2
+{1\over 2}h^{\prime\prime}(x)(\psi\bpsi-\bpsi\psi)$
is considered where $h(x)$ is a real valued function.
This system possesses supersymmetry generated by
$Q=\sqrt{2}\bpsi(p+ih^{\prime}(x))$ and
$Q^{\dag}=\sqrt{2}\psi(p-ih^{\prime}(x))$ which satisfy
$\{Q,Q^{\dag}\}=4H$.
It is shown that there is a single supersymmetric ground state
as long as $h^{\prime}(x)$ cuts through $h^{\prime}(x)=0$
odd times (no matter how many), but there is none
if it cuts through $h^{\prime}(x)=0$ even times.
The analysis is extended in \cite{Morse} to the supersymmetric
quantum mechanics on a Riemannian manifold deformed by a Morse function $h$.
In particular, the supersymmetric ground states
are realized as cohomology classes of a cochain complex
constructed from the critical points of $h$ with the grading determined by
the Morse index.
Let $\Omega_{ab}X$ be the space of paths
$[0,\pi]\to X$ from $\cycle_a$ to $\cycle_b$,
with the boundary condition that the derivatives at
$x^1=0$ and $\pi$ are normal to $\cycle_a$ and $\cycle_b$ respectively.
{}From the inspection of the supercharges (\ref{QpmbQpm}),
we see that the present system is nothing but the supersymmetric quantum
mechanics on $\Omega_{ab}X$ deformed in the same way as
\cite{SUSYB,Morse} if there is a function $h$ on $\Omega_{ab}X$
such that
$\delta h/\delta\phi^i
=ig_{i\bj}\partial_1\bphi^{\bj}+{1\over 2}\partial_i W$,
and
$\delta h/\delta\bphi^{\bj}
=-ig_{i\bj}\partial_1\phi^i+{1\over 2}\partial_{\bj}\overline{W}$;
in other words
\beq
\delta h=\int_0^{\pi}\dd x^1\,\left\{\,
\omega_{IJ}\delta\phi^I\partial_1\phi^J+\delta\phi^I\partial_I{\rm Re}W
\,\right\},
\label{hvar}
\eeq
where $\omega$ is the K\"ahler form of $X$,
$\omega=ig_{i\bj}\dd z^i\wedge\dd \bz^{\bj}$.
If we choose a base point $\phi_0$ of (each connected component of)
$\Omega_{ab}X$,
one can ``define'' a function
\beq
h[\phi]={1\over 2}\int\limits_{[0,1]\times [0,\pi]}
\widehat{\phi}^*\omega~+~\int_0^{\pi}\dd x^1{\rm Re}W,
\label{hphi}
\eeq
where $\widehat{\phi}(s,x^1)$ is a homotopy
in $\Omega_{ab}X$
connecting $\phi$ and $\phi_0$, namely a map
$\widehat{\phi}:[0,1]\times [0,\pi]\to X$ such that
$\widehat{\phi}(0,x^1)=\phi_0(x^1)$,
$\widehat{\phi}(1,x^1)=\phi(x^1)$,
and obeying the Dirichlet/Neumann boundary condition
at $x^1=0$ and $x^1=\pi$.
Using the fact that $\cycle_a$ and $\cycle_b$ are Lagrangian
and recalling the boundary condition that
$\partial_1\phi^I|_{\partial\Sigma}$
is normal to the brane,
it is easy to see that $h[\phi]$ is invariant under a small variation of
the homotopy and that it satisfies (\ref{hvar})
for the variation of $\phi$.
For a large change of homotopy,
using the fact that the cycles $\cycle_a$ and $\cycle_b$ 
are simply connected, we can show that
$h[\phi]$ changes by 
${1\over 2}\int_C\omega$ where $C$ is a two-cycle in $X$.
We discard this subtlety by focusing our attention only to those cases
where $\int_C\omega=0$ for all (compact) two-cycle $C$.
We also assume $c_1(X)=0$. These assumptions hold in
the class of models we consider later in this paper.
Then, (\ref{hphi}) is independent of the choice of homotopy
and becomes a well-defined function on $\Omega_{ab}X$.
Under the assumption $c_1(X)=0$,
$U(1)$ axial R-charge is conserved
and the supercharge $Q$ has charge $1$.
Thus, we can define the cochain complex as in \cite{Morse}
graded by the axial R-charge.\footnote{There is a subtlety
for defining the grading (Morse index) from the fact that the Hessian of
$h[\phi]$ has unbounded spectrum.
However, one can regularize it, up to an additive constant,
by the index of the corresponding
Dirac-type operator which is well-defined under the assumption that
$c_1(X)=0$ (this is related to the conservation of the axial R-charge).
We set the ground of the grading so that
it is $0$ for the critical path which is unique in the family
(unlike Figure \ref{mdel}).}
The coboundary operator is defined by
counting the number of instantons connecting different critical points
of $h[\phi]$.
We note that an instanton connecting critical paths $\phi_1$ and $\phi_2$
is a configuration $\phi(\tau,x^1)$ such that
$\phi(-\infty,x^1)=\phi_1(x^1)$, 
$\phi(+\infty,x^1)=\phi_2(x^1)$ and satisfying the following equation
\beq
{\partial \phi^i\over\partial \tau}=-i{\partial \phi^i\over\partial x^1}
+{1\over 2}g^{i\bj}\partial_{\bj}\overline{W}.
\eeq
We denote the corresponding cohomology groups as
\beq
{\rm HF}_W^p(\cycle_a,\cycle_b),
\label{Floer}
\eeq
where the grading $p$ is given by the axial R-charge
of the ground states.
This is a Landau-Ginzburg generalization of the Floer homology group
\cite{Floer}.
Under the rescaling of the superpotential $W\to\e^t W$,
the supercharge simply changes as
$Q\to \e^{-\Delta h}Q\e^{\Delta h}$
where $\Delta h=(\e^t-1)\int_0^{\pi}\dd x^1 {\rm Re}W$.
Since multiplication by $\e^{-\Delta h}$ is well-defined,
the cohomology is invariant under this rescaling.\footnote{Alternatively,
as in \cite{Floer}
one may construct a cochain homotopy equivalence
of the reduced complexes introduced above.}
Then, we see that for ${\rm Im}W(a)>{\rm Im}W(b)$
\beq
{\rm HF}_W^p(\cycle_a,\cycle_b)=\left\{
\begin{array}{ll}
\R^{|\Delta_a\circ\Delta_b|}&p=0,\\[0.2cm]
0&p\ne 0.
\end{array}\right.
\label{HFr}
\eeq
In the situation as in Figure \ref{mdel},
we see that pairs of classical supersymmetric ground states are lifted 
by an instanton effect.
Those states have very small (but positive) energies.

It would be interesting to generalize the above consideration to the case
where there are two-cycles with $\int_C\omega\ne 0$ and
also to the case where $c_1(X)\ne 0$.
In some of the latter cases, we expect that the cohomology
(\ref{Floer})
is not graded by integers, but by some cyclic group.

\subsubsection{Sigma Models}

~~~~
The other example we consider is the supersymmetric sigma model
on $X$ with trivial superpotential $W=0$
where the D-branes are wrapped totally on $X$.
We couple the left and the right boundaries to 
$U(1)$ gauge fields $A^{(a)}$ and $A^{(b)}$ respectively
that define holomorphic line bundles $E_a$ and $E_b$ on $X$.
We use the formulation where the boundary term is given by (\ref{Sbound})
and the boundary condition is the standard one (\ref{stbc})
(where there is no normal direction in the present case).
The theory is invariant under B-type supersymmetry
generated by
$Q=\bQ_++\bQ_-$ and $Q^{\dag}=Q_++Q_-$.
Since the boundary term (\ref{Sbound})
includes the time derivatives of the fields,
the Noether charges are modified.
Thus, the supercharge $Q$ is expressed as
\beqa
\lefteqn{Q~=~\sqrt{2}\Biggl(\,\int_0^{\pi}\dd x^1\left\{
g_{i\bj}(\bpsi_+^{\bj}+\bpsi_-^{\bj})\partial_0\phi^i
-g_{i\bj}(\bpsi_+^{\bj}-\bpsi_-^{\bj})\partial_1\phi^i
\right\}}
\nonumber\\
&&~~~~~~~~~~
+(\bpsi_+^{\bj}+\bpsi_-^{\bj})A^{(b)}_{\bj}\Bigr|_{x^1=\pi}
-(\bpsi_+^{\bj}+\bpsi_-^{\bj})A^{(a)}_{\bj}\Bigr|_{x^1=0}
\,\Biggr).
\eeqa
For the purpose of computing the index, we can focus on the zero modes
($x^1$-independent modes).
Then from the boundary condition, the left and the right
fermionic zero modes are related as
$\psi^i_{-0}=\psi_{+0}^i$ and $\bpsi_{-0}^{\bi}=\bpsi_{+0}^{\bi}$.
We can identify the quantum mechanical Hilbert space as
the space of sections on the bundle
\beq
\left(\bigwedge T^{*(0,1)}X\right)
\otimes E_a^*\otimes E_b,
\label{bund}
\eeq
on which the fermionic zero modes
act as\footnote{
Unlike in the closed string case \cite{Wtop},
we do not have the factor $\bigwedge TX^{(1,0)}$ in (\ref{bund})
nor
$${1\over\sqrt{2}}g_{i\bj}(\bpsi_{+0}^{\bj}-\bpsi_{-0}^{\bj})
\leftrightarrow
(\partial/\partial z^i)\wedge,~~
{1\over\sqrt{2}}(\psi_{+0}^i-\psi_{-0}^i)
\leftrightarrow
i_{\displaystyle \dd z^i},$$
because
$\bpsi_{+0}^{\bj}-\bpsi_{-0}^{\bj}=0$
and $\psi_{+0}^i-\psi_{-0}^i=0$ from the boundary condition.}
\beqa
{1\over\sqrt{2}}(\bpsi_{+0}^{\bi}+\bpsi_{-0}^{\bi})
&\longleftrightarrow&
\dd \bz^{\bi}\wedge,\\
{1\over\sqrt{2}}g_{i\bj}(\psi_{+0}^i+\psi_{-0}^i)
&\longleftrightarrow&
i_{\displaystyle \partial/\partial \bz^{\bj}}.
\eeqa
Then the supercharge $Q$ corresponds to the Dolbeault operator
on the bundle $E_a^*\otimes E_b$:
\beq
Q~\leftrightarrow~
2\overline{\partial}_A=
2\dd\bz^{\bi}\left(\partial_{\bi}+A^{(b)}_{\bi}-A^{(a)}_{\bi}\right).
\label{Dolbeault}
\eeq
Thus, the Witten index, which is defined as the index of $Q$ operator,
is equal to the index of this Dolbeault operator.
By the standard index theorem, we obtain
\beq
I(a,b)
=\chi(E_a,E_b):=\int\limits_X{\rm ch}(E_a^*\otimes E_b){\rm Td}(X),
\label{chi}
\eeq
where ${\rm Td}(X)$ is the total Todd class of the tangent bundle of $X$
which are given by polynomials of the Chern classes (see e.g.
\cite{Hirtz}).
It is easy to extend this analysis to the case where the
bundles $E_a$ and $E_b$ have higher ranks.
The conclusion remains the same as (\ref{chi}).

In general,
the index (\ref{chi}) is not symmetric nor anti-symmetric
under the exchange of $a$ and $b$.  This is
related by mirror symmetry, as we will discuss later,
with the fact noted earlier, that supersymmetric index $I(a,b)$
for Lagrangian D-branes
in LG models is neither symmetric nor anti-symmetric. 
However, since odd Todd classes are divisible by the first Chern class
of $X$ \cite{Hirtz}, for a Calabi-Yau manifold
${\rm Td}(X)$ is a sum of $4k$-forms.
Under the exchange $E_a^*\otimes E_b\to E_b^*\otimes E_a$ the Chern
character changes by sign flip in the $(4k+2)$-form components.
Thus, for a Calabi-Yau manifold of dimension $n$, the index
$I(a,b)$ is symmetric for even $n$ and anti-symmetric for odd $n$
under the exchange of $a$ and $b$.

One can actually
obtain an upper bound on the number of supersymmetric ground
states, using a technique of section 3 of \cite{Morse}.
The cohomology of operator $Q$ is actually invariant under the rescaling
$\partial_1\phi^i\to\e^t\partial_1\phi^i$,
since it is done by conjugation
by $\e^{tP}$ where $P$ is an operator counting
the number of fermions of combination $\bpsi_+^{\bj}-\bpsi_-^{\bj}$.
This means that the cohomology group is independent of the
width of the strip in the $x^1$ direction, as long as it is finite.
A zero energy state should remain a zero energy state
as we make the strip thinner and thinner. 
In particular,
a ground state must correspond to a ground state of the quantum
mechanics of the zero modes.
In this way we obtain the upper bound on the number of ground states.
The ground states of the quantum mechanics are the cohomology classes
of the Dolbeault complex $\Omega^{0,p}(X,E_a^*\otimes E_b)$ 
with (\ref{Dolbeault}) as the coboundary operator.
Thus, the quantum mechanical ground states are given by
the Dolbeault cohomology
\beq
{\rm H}^{0,p}(X,E_a^*\otimes E_b).
\label{DolH}
\eeq
We note that
the vector R-symmetry, for which $Q$ has charge $1$,
is not broken in the bulk nor by the boundary condition.
Thus, the grading $p$ of the cohomology group (\ref{DolH}) is
the same as the vector R-charge.
The group (\ref{DolH}) gives us an upper bound on the number of
supersymmetric ground states, but
we do not have a lower bound in general
(the argument in section 3 of \cite{Morse} does
not apply here)\footnote{This is analogous to the situation
in the sigma model on a worldsheet without boundary.
B-type supercharge yields Dolbeault complex with
the coefficient $\bigwedge TX^{(1,0)}$ in the zero mode approximation.
This indeed has the correct Witten index
$$
\sum_{p,q} (-1)^{p+q}\dim {\rm H}^{0,p}(X,\wedge^qTX^{(1,0)})
=\pm \chi(X).
$$
However, the cohomology group
$\oplus_{p,q}{\rm H}^{0,p}(X,\bigwedge^qTX^{(1,0)})$
itself is in general larger than
the space of ground states
which we know to be $\oplus_{p,q} {\rm H}^{q,p}(X)$,
unless $X$ is a Calabi-Yau manifold.}.
However, if the cohomology is non-vanishing only for
even $p$ (or only for odd $p$),
the cohomology group (\ref{DolH}) is indeed the same as the space of
supersymmetric ground states.
Later in this paper, we will consider a certain set of bundles
such that the cohomology (\ref{DolH}) vanishes except $p=0$
and hence it can be identified as the space of ground states.

\subsection{The Boundary States}

Let us consider a Euclidean quantum field theory
formulated on a Riemann surface $\Sigma$ with boundary circles.
We choose an orientation of each component $S^1$
of the boundary
and we call it an {\it incoming} (resp. {\it outgoing})
component if the $90^{\circ}$ rotation
of the positive tangent vector of $S^1$ (with respect to
the orientation of $\Sigma$) is an inward (outward)
normal vector at the boundary.
We choose the metric on $\Sigma$ such that it is a
flat cylinder near each boundary component.
Suppose $\Sigma$ has a single outgoing boundary, $S^1=\partial\Sigma$.
The partition function on $\Sigma$
depend on the boundary condition $a$ on the fields at
$\partial\Sigma$ and we denote it
by $Z^a(\Sigma)$.
On the other hand, the path-integral over the fields on $\Sigma$
defines a state $|\Sigma\rangle$
that belongs to the quantum Hilbert space
${\cal H}_{S^1}$ at the boundary circle.
We define the {\it boundary state} $\langle a |$
corresponding to the boundary condition $a$
by the property
\beq
Z^{a}(\Sigma)
=\langle a|\Sigma\rangle.
\eeq
If $\Sigma$ has a single incoming boundary
$\partial\Sigma=S^1$,
we have a state $\langle\Sigma |$ that belongs to the dual space
${\cal H}_{S^1}^{\dag}$.
For a boundary condition $b$ at $S^1$, we define the boundary state
$|b\rangle$ by
\beq
Z_b(\Sigma)
=\langle\Sigma |b\rangle,
\eeq
where $Z_b(\Sigma)$ stands for the partition function on $\Sigma$
with the boundary condition $b$.
In general, the boundary state $\langle a|$
(resp. $|b\rangle$) does not belong to ${\cal H}_{S^1}^{\dag}$
(resp. ${\cal H}_{S^1}$)
but is a formal sum of elements therein.
If $\partial\Sigma$
consists of several incoming components $S^1_i$ and outgoing components
$S^1_j$, we have a map
$f_{\Sigma}:\otimes_i{\cal H}_{S_i^1}\to \otimes_j{\cal H}_{S_j^1}$.
The partition function on $\Sigma$ with
the boundary conditions $\{a_j\}$
and $\{b_i\}$ can be expressed using the boundary states as
\beq
Z^{\{a_j\}}_{\,\,\{b_i\}}(\Sigma)=
\Biggl(\bigotimes_j\langle a_j|\Biggr)
f_{\Sigma}
\Biggl(\bigotimes_i |b_i\rangle\Biggr).
\eeq
For instance, let us consider a flat finite size cylinder
$\Sigma$ of length $T$ and circumference $\beta$. With a choice of
orientation in the circle direction, we have one incoming
and one outgoing boundaries. We choose the boundary conditions
$b$ and $a$ there. Then, the partition function is given by
$Z^a_{\,b}(\Sigma)=\langle a |\e^{-TH(\beta)}| b\rangle$,
where $H(\beta)$ is the Hamiltonian of the theory on
the circle of circumference $\beta$.
\begin{figure}[htb]
\begin{center}
\epsfxsize=2.8in\leavevmode\epsfbox{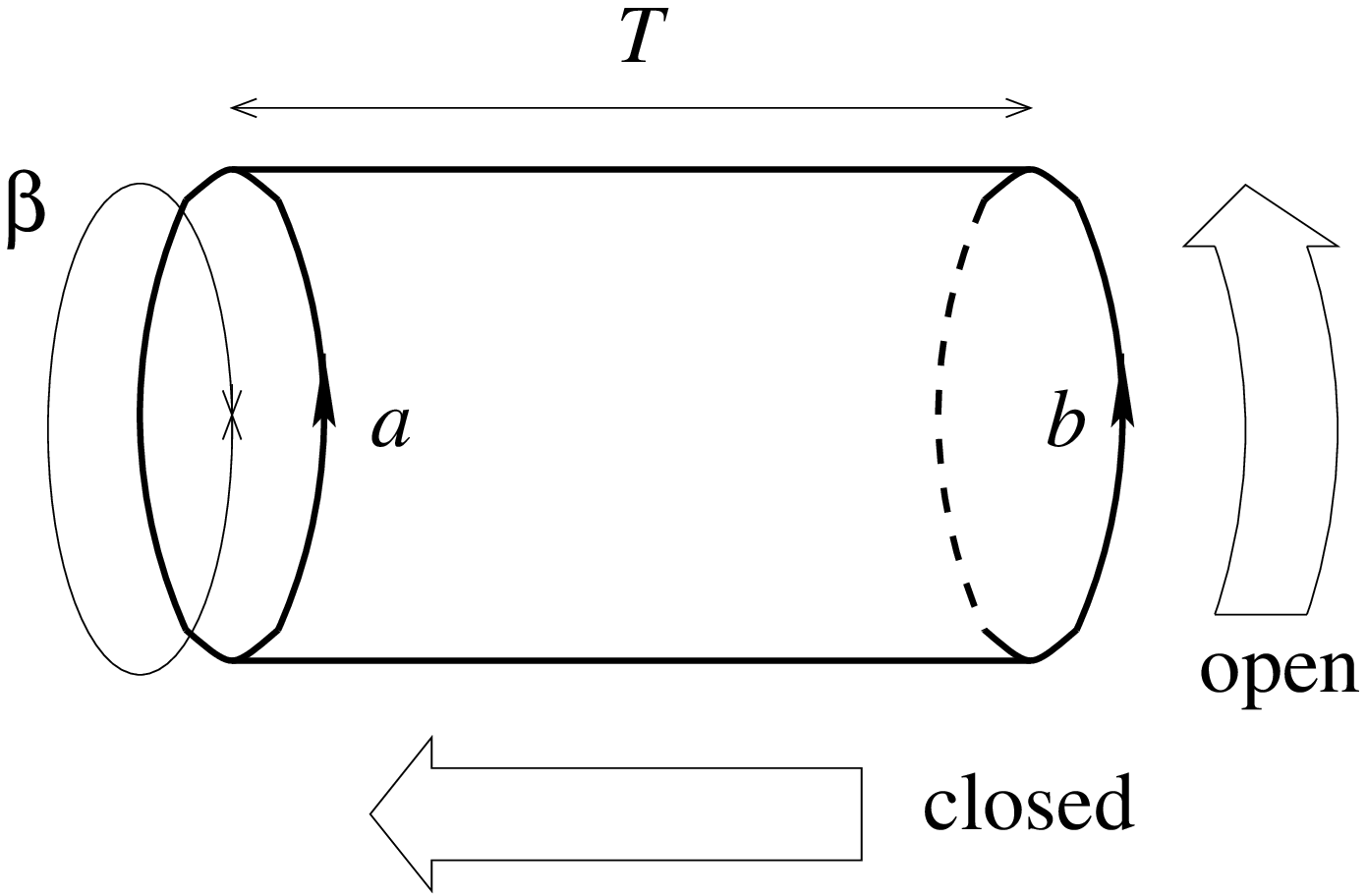}
\end{center}
\caption{Open and closed string channels}
\label{openclosed}
\end{figure}
This is the interpretation of
the partition function from the closed string view point.
On the other hand, one can interpret it
from the point view of open strings.
Let ${\cal H}_{ab}$ be the space of states on
the interval of length $T$ with $a$ and $b$
as the left and the right boundary conditions and let
$H(T)$ be the Hamiltonian generating the evolution in the
circle direction.
If the theory has spin half fermions and if the spin structure
is periodic (anti-periodic) along the circle direction,
the partition function is the trace of $(-1)^F\e^{-\beta H(T)}$
($\e^{-\beta H(T)}$) over ${\cal H}_{ab}$.
Thus, we have
\beqa
&&\Tr_{{\cal H}_{ab}}(-1)^F\e^{-\beta H(T)}
={}_{{}_{\rm RR}\!}\langle a |
\e^{-TH(\beta)}| b\rangle_{\!{}_{\rm RR}},\\
&&\Tr_{{\cal H}_{ab}}\e^{-\beta H(T)}
={}_{{}_{{\rm NS}^2}\!}\langle a |
\e^{-TH(\beta)}| b\rangle_{\!{}_{{\rm NS}^2}},
\eeqa
where RR (NS${}^2$) shows that the fermions on the circle
are periodic (anti-periodic).

Let us consider a $(2,2)$ supersymmetric field theory
formulated on the strip $\Sigma=\R\times I$ of Minkowski signature,
with the boundary conditions $a$ and $b$ that preserves A- or B-type
supersymmetry.
We recall that
the $x^1$ components of the supercurrents are required
to obey $\bG_+^1+G_-^1=G_+^1+\bG_-^1=0$ for A-type supersymmetry
and $\bG_+^1+\bG_-^1=G_+^1+G_-^1=0$ for B-type supersymmetry
(for the trivial phases $\e^{i\alpha}=\e^{i\beta}=1$).
Now let us compactify the time direction $\R$ to $S^1$
and continue the theory to Euclidean signature
by the Wick rotation $x^0=-ix^2$ where
we choose the orientation so that $z=x^1+ix^2$ is a complex coordinate.
The boundary conditions of the supercharges remains the same as
in the Minkowski theory. 
If we change the coordinates as
$(x^{1^{\prime}},x^{2^{\prime}})=(x^2,-x^1)$,
the conditions become
$\e^{\pi i\over 4}\bG_{+^{\prime}}^{2^{\prime}}
+\e^{-{\pi i\over 4}}G_{-^{\prime}}^{2^{\prime}}
=\e^{\pi i\over 4}G_{+^{\prime}}^{2^{\prime}}
+\e^{-{\pi i\over 4}}\bG_{-^{\prime}}^{2^{\prime}}=0$
for A-type supersymmetry
and $\e^{\pi i\over 4}\bG_{+^{\prime}}^{2^{\prime}}
+\e^{-{\pi i\over 4}}\bG_{-^{\prime}}^{2^{\prime}}
=\e^{\pi i\over 4}G_{+^{\prime}}^{2^{\prime}}
+\e^{-{\pi i\over 4}}G_{-^{\prime}}^{2^{\prime}}=0$
for B-type supersymmetry,
where the phases $\e^{\pm{\pi i\over 4}}$ come from the
spin of the supercurrent.
This means that the boundary states satisfy
\beqa
&&\Bigl(\bG_{+^{\prime}}^{2^{\prime}}
-iG_{-^{\prime}}^{2^{\prime}}\Bigr)|b\rangle
=\Bigl(G_{+^{\prime}}^{2^{\prime}}
-i\bG_{-^{\prime}}^{2^{\prime}}\Bigr)|b\rangle
=J_A^{2^{\prime}}|b\rangle
=0,\\
&&\langle a|\Bigl(\bG_{+^{\prime}}^{2^{\prime}}
-iG_{-^{\prime}}^{2^{\prime}}\Bigr)
=\langle a|\Bigl(G_{+^{\prime}}^{2^{\prime}}
-i\bG_{-^{\prime}}^{2^{\prime}}\Bigr)
=\langle a|J_A^{2^{\prime}}
=0
\label{conda1}
\eeqa
for A-type supersymmetry and
\beqa
&&\Bigl(\bG_{+^{\prime}}^{2^{\prime}}
-i\bG_{-^{\prime}}^{2^{\prime}}\Bigr)|b\rangle
=\Bigl(G_{+^{\prime}}^{2^{\prime}}
-iG_{-^{\prime}}^{2^{\prime}}\Bigr)|b\rangle
=J_V^{2^{\prime}}|b\rangle
=0,\\
&&\langle a|(\bG_{+^{\prime}}^{2^{\prime}}
-i\bG_{-^{\prime}}^{2^{\prime}}\Bigr)
=\langle a|\Bigl(G_{+^{\prime}}^{2^{\prime}}
-iG_{-^{\prime}}^{2^{\prime}}\Bigr)
=\langle a|J_V^{2^{\prime}}
=0,
\label{conda2}
\eeqa
for B-type supersymmetry. Here we have added
the conditions for conservation of the R-charge,
which applies when the R-symmetry is not broken in the bulk theory.
Note that, in the quantization of the closed strings, the Hermiticity
condition is imposed so that
$(G_{\pm^{\prime}}^{\mu^{\prime}})^{\dag}
=\bG_{\pm^{\prime}}^{\mu^{\prime}}$
(whereas the quantization of open strings would lead to
$(G_{\pm}^{\mu})^{\dag}
=\bG_{\pm}^{\mu}$).
Thus, the above conditions on the boundary states
are not invariant under Hermitian conjugation.
If 
$|b\rangle$ and $\langle a|$ correspond to
the boundary conditions preserving A- or B-type supersymmetry with the
phase $\e^{i\alpha}$ or $\e^{i\beta}$, 
the Hermitian conjugates $\langle\overline{b}|$ and
$|\overline{a}\rangle$
correspond to the boundary conditions preserving A- or B-type supersymmetry
with the phase $-\e^{i\alpha}$ or $-\e^{i\beta}$.
If the sign flip $(-1)^{F_L}$ of the left-moving worldsheet fermions
is a symmetry of the theory,
the states $\langle\overline{b}|(-1)^{F_L}$ and
$(-1)^{F_L}|\overline{a}\rangle$ correspond to
the boundary conditions preserving the A- or B-type supersymmetry
with the phase $\e^{i\alpha}$ or $\e^{i\beta}$,
which is the same as the original
supersymmetry.\footnote{Our convention differs from that in the reference
\cite{Douglas-Fiol,BDL} where
$\langle b|$ stands for the Hermitian conjugate of $|b\rangle$.
In particular, we do not need an extra $(-1)^{F_L}$ insertion in the
r.h.s. of eq.\,(\ref{Windex}) that is required in the notation of
\cite{Douglas-Fiol,BDL}.}

As above, let $a$ and $b$ be the boundary conditions
that preserve the same combinations of the supercharges
(A-type or B-type).
We can use the boundary states to
represent the supersymmetric index as
\beq
I(a,b)={}_{{}_{\rm RR}\!}\langle a|
\e^{-TH(\beta)}|b\rangle_{\!{}_{\rm RR}},
\label{Windex}
\eeq
where ${}_{{}_{\rm RR}\!}\langle a|$ and
$|b\rangle_{\!{}_{\rm RR}}$ are the
boundary states in the RR sector.
By the basic property of the index, it is independent of
the various parameters, such as $\beta$ and $T$.
It is an integer and therefore
must be invariant under the complex conjugation
that induces the replacement $(a,b)\to(\overline{b},\overline{a})$.
We note, however, that the latter preserves a
different combination of the
supercharges compared to the original one.

\subsubsection{Boundary Entropy}

~~~~
The boundary states
are in general a sum of infinitely many
eigenstates of the Hamiltonian.
An important information on the boundary states
can be obtained by looking at the contribution by the ground state.
For instance,
in boundary conformal field theory, the coefficient 
$g_b=\langle 0|b\rangle$ of the expansion is known to play a role
analogous to that of the central charge $c$ in the bulk theory
\cite{AL91} and is called {\it boundary entropy}.
In supersymmetric field theory, there are several
supersymmetric ground states $|i\rangle$ in the RR sector.
Thus the ${\cal N}=2$ analog of the boundary entropy would be the pairings
\beq
\Pi_{i}^a={}_{{}_{\rm RR}\!}\langle a|i\rangle.
\label{pairi}
\eeq
These overlaps were studied in \cite{HV}
especially on the relation to the period integrals,
which will be further elaborated here.
If the axial R-symmetry is unbroken in the bulk theory,
we see from (\ref{conda1}) that an A-type boundary state $\langle a|$
has zero axial charge. Thus for the pairing (\ref{pairi})
to be non-vanishing, the ground state $|i\rangle$ must also have
zero axial R-charge.
Likewise, if vector R-symmetry exists, the pairing (\ref{pairi})
for B-type boundary state is non-vanishing only for the ground state
$|i\rangle$ with zero vector R-charge.
If the theory has a mass gap, this selection rule is vacuous since 
all ground states have zero R-charges.
However, if there is a non-empty IR fixed point,
some of the ground states can have non-vanishing R-charges
and this selection rule is non-trivial.
For example, in LG models all ground states
have vanishing R-charges (even if it is quasihomogeneous and has
a non-trivial fixed point) and the selection rule is vacuous
for A-type boundary states, but in LG orbifold, there are usually
ground states of nonzero axial R-charges and the selection rule is
non-trivial.

If the vector (resp. axial) R-charge is conserved and
integral, there is a one to one correspondence between the
supersymmetric ground states and the elements of
the $ac$ ring (resp. $cc$ ring) \cite{tt}.
The state $|\phi_i\rangle$
corresponding to a chiral ring element $\phi_i$
is the one that appears at the boundary $S^1$ of the semi-infinite cigar
$\Sigma$ with the insertion of $\phi_i$ at the tip,
where
the theory is twisted to a topological field theory in the curved region.
\begin{figure}[htb]
\begin{center}
\epsfxsize=5.1in\leavevmode\epsfbox{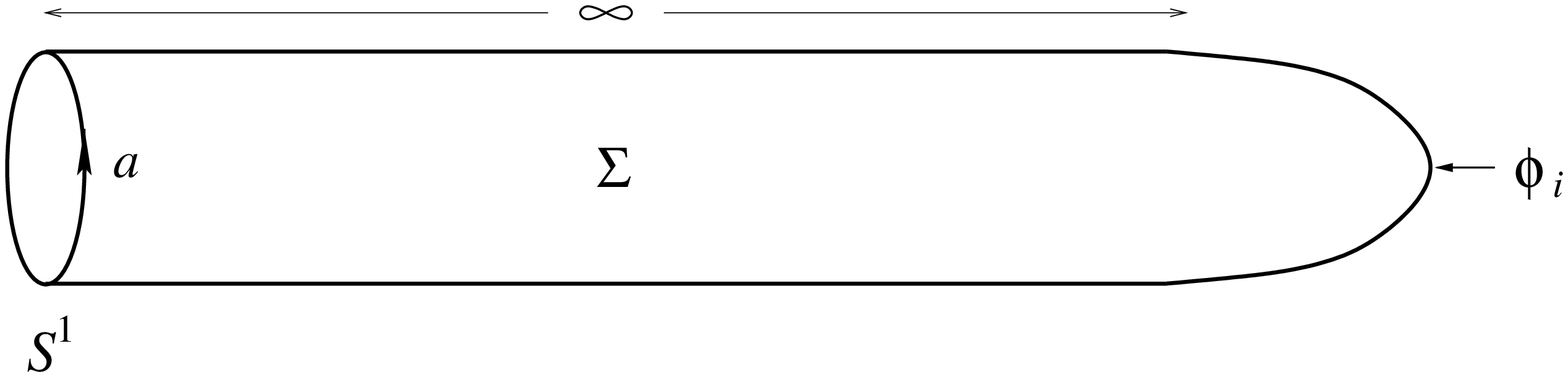}
\end{center}
\caption{The semi-infinite cigar leading to
$\Pi_{i}^a={}_{{}_{\rm RR}\!}\langle a|\phi_i\rangle$}
\label{hemisphere}
\end{figure}
Thus, for those states, the pairings
${}_{{}_{\rm RR}\!}\langle a|\phi_i\rangle$
can be identified as the
path-integral on the semi-infinite cigar where the
boundary condition $a$ is imposed at the outgoing
boundary and the operator $\phi_i$ is inserted at the tip
(see Figure \ref{hemisphere}).

For concreteness, let us consider a theory with conserved and integral
axial R-charge where B-twist is possible.
The operators $\phi_i$ we use to define the supersymmetric ground states
are the $cc$ ring elements.
We will be interested in A-type boundary conditions $a$
and the corresponding
boundary states $\langle a|$ obeying (\ref{conda1}) which in particular
yields
\beq
\langle a|\Bigl(\oint_{S^1}\bG_{+}
-i\oint_{S^1} G_{-}\Bigr)
=\langle a|\Bigl(\oint_{S^1} G_{+}
-i\oint_{S^1}\bG_{-}\Bigr)=0.
\label{Abc}
\eeq
Here we use the closed string coordinates as in
(\ref{conda1}) but omit the primes.
Also, we use the current notation
$G_{\pm}=\dd x^1 G_{\pm}^2-\dd x^2 G_{\pm}^1$
which are one forms with values in the spinor bundles of $\Sigma$.
After B-twisting, the currents $\bG_{\pm}$ become ordinary one forms
but the current $G_-$ (resp. $G_+$) becomes a one form
with values in the bundle of holomorphic (resp. antiholomorphic)
one forms of $\Sigma$.

The pairings $\Pi_{a,i}={}_{{}_{\rm RR}\!}\langle a|\phi_i\rangle$
are invariant under the twisted F-term deformations of the theory.
\beq
{\partial \Pi^a_i\over\partial t_{ac}}=0,~~
{\partial \Pi^a_i\over\partial \overline{t}_{ac}}=0.
\label{acinv}
\eeq
As for the F-term deformations generated by
$cc$ ring elements, they satisfy the following equation
\beq
(\nabla_i\Pi^a)_j=(D_i\delta^k_j+i\beta C_{ij}^k)\Pi^a_k=0,
~~
(\nabla_{\bi}\Pi^a)_{\bj}=(D_{\bi}\delta^{\bk}_{\bj}
-i\beta C_{\bi \bj}^{\bk})\Pi^a_{\bk}=0,
\label{parallel}
\eeq
where $\beta$ is the circumference of the boundary circle $S^1$.
Here $D_i$ is the covariant derivative defined in \cite{tt} and
$C_{ij}^k$ is the structure constant of the chiral ring.
These can be shown as follows by the standard gymnastics
in $tt^*$ equation.

\begin{figure}[htb]
\begin{center}
\epsfxsize=5.2in\leavevmode\epsfbox{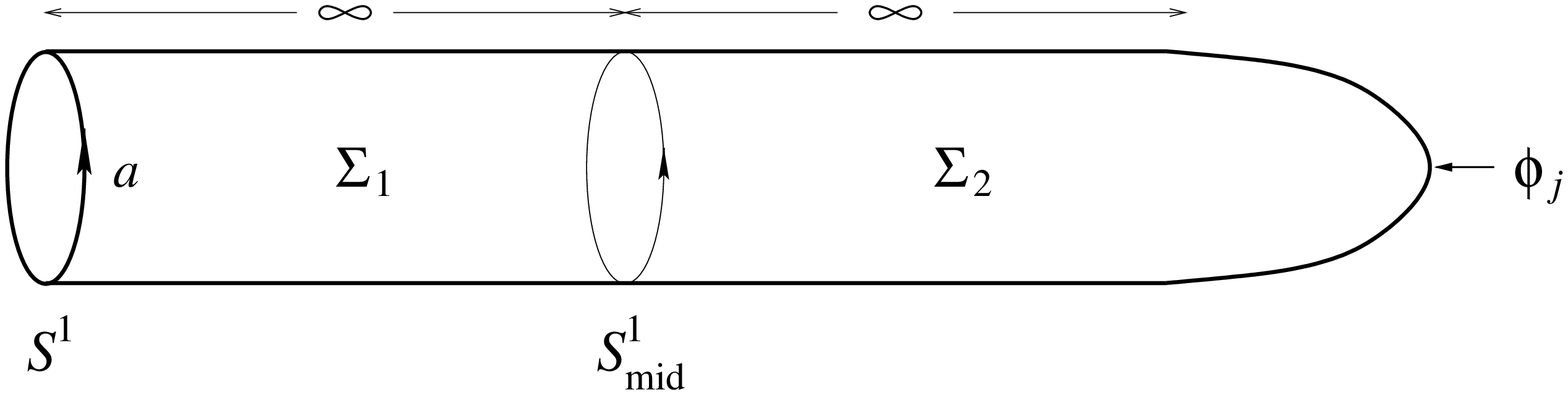}
\end{center}
\caption{Separation into two regions}
\label{separation}
\end{figure}

We start with the twisted F-term deformation which can be written as
\footnote{Note that 
$(Q_+\phi_{ac})(x)=\oint_x G_+\phi_{ac}(x)$ where the contour integral
is along a small circle around the point $x$.}
\beq
{1\over 2}\int_{\Sigma} \bQ_-Q_+\phi_{ac} \sqrt{h}\dd^2x
\label{acdeform}
\eeq
plus its complex conjugate.
Here $\phi_{ac}$ is a twisted chiral operator of
axial R-charge $2$ (and therefore $\bQ_-Q_+\phi_{ac}$
has spin zero even in the twisted theory;
the spin of $Q_+$ cancels that of $\phi_{ac}$).
Now, let us divide the semi-infinite cigar into two infinite regions
$\Sigma_1$ and $\Sigma_2$ separated by a circle $S^1_{\rm mid}$
as shown in Figure \ref{separation}.
We first consider the integral (\ref{acdeform}) in the region $\Sigma_1$.
We recall that
$(Q_+\phi_{ac})(x)=\oint_x G_+\phi_{ac}(x)$ etc,
where the contour integral
is along a small circle around the point $x$.
We can deform the contour of the $\bG_-$ integral
from the small circle around $x$
to two boundary circles of $\Sigma_1$; $S^1$ and $S^1_{\rm mid}$.
The one on $S^1_{\rm mid}$ can be considered as the supercharge
acting on the state at the boundary of $\Sigma_2$.
Since $\Sigma_2$ is infinitely long, the state that appears at
the boundary is a ground state. Thus, the contour integral along
$S^1_{\rm mid}$ vanishes.
The one on $S^1$ turns into $i\oint_{S^1}G_+$
by the boundary condition (\ref{Abc}). By deforming the contour back
into the interior, it becomes the sum of the integral
on a small circle around $x$ and the one on $S^1_{\rm mid}$.
The latter vanishes for the same reason as before.
The former becomes $iQ_+Q_+\phi_{ac}$ which is zero from the nilpotency of
$Q_+$.
We next consider the region $\Sigma_2$. This has a curved region but one can
still deform the contour of the $\bG_-$ integral since it is an ordinary
one-form on $\Sigma$. The contour can be deformed to the sum of
$S^1_{\rm mid}$ and a small circle around the tip $x_0$ at which
$\phi_j$ is inserted.
The integral on $S^1_{\rm mid}$ can be considered as the supercharge
acting on the (dual) state that appears on the incoming boundary
of $\Sigma_1$. Since $\Sigma_1$ is infinite, the state is a ground state
and thus the supercharge vanishes.
The contour integral around $x_0$
yields $\bQ_-\phi_j$ but this vanishes since
$\phi_j$ is a chiral operator.
To summarize, the pairing $\Pi_{i,a}$ is independent of the twisted F-term
deformations.

Next, we consider the F-term deformation generated by the chiral operator
$\phi_i$, which can be written as
\beq
{1\over 2}\int_{\Sigma}Q_-Q_+\phi_i
+{1\over 2}\int_{\Sigma}\bQ_+\bQ_-\bphi_{\bi}\sqrt{h}\dd^2 x
-i\oint_{S^1}\dd x^1 (\phi_i-\bphi_{\bi}).
\label{Fdeform}
\eeq
Here we have included the (constant) boundary term
which is required to
set the ground state energy at zero, as found before in this section
(see (\ref{shift})).
We separate the integral into two regions as before.
We first consider the region $\Sigma_1$.
Since $\Sigma_1$ is flat, one can treat the currents $G_{\pm}$
as ordinary one forms.
The contour of the $G_-$ integral can be deformed to
the one on $S^1$ and the one on $S^1_{\rm mid}$.
The latter vanishes by the same reason as before.
By the boundary condition the former turns to $-i\oint_{S^1}\bG_+$
which can be deformed to an integral around the insertion point
of $\phi_i$ (plus an integral over $S^1_{\rm mid}$ that vanishes).
This yields the term $-{i\over 2}\bQ_+Q_+\phi_i$
which is the same as $-{i\over 2}\{\bQ_+,Q_+\}\phi_i$
since $\phi_i$ is a chiral operator.
On the other hand, the same manipulation for $G_+$ rather than $G_-$
leads to $-{i\over 2}\{Q_-,\bQ_-\}\phi_i$.
By taking the average of the two we
obtain
\beqa
{1\over 2}\int_{\Sigma_1}Q_-Q_+\phi_i
&=&\int_{\Sigma_1}\left(
-{i\over 4}\{\bQ_+,Q_+\}\phi_i-{i\over 4}\{Q_-,\bQ_-\}\phi_i\right)
\dd^2 x
=-{i}\int_{\Sigma_1}H\phi_i\,\dd^2 x
\nonumber\\
&=&i\int_{\Sigma_1}{\partial\over\partial x^2}\phi_i\,\dd^2 x
=i\oint_{S^1} \phi_i\,\dd x^1-i\oint_{S_{\rm mid}^1}\phi_i\,\dd x^1,
\eeqa
where we have used the supersymmetry algebra
$\{Q_{\pm},\bQ_{\pm}\}=2(H\mp P)$.
The first term on the right hand side cancels the boundary term in
(\ref{Fdeform}). The integrand of the last term is a constant along the
circle and hence we obtain $-i\beta\phi_i$. One can move the
operator $\phi_i$ toward the tip $x_0$ where $\phi_j$ is inserted,
and this will yield the term
$-i\beta C_{ij}^k\phi_k$.
Next we consider the term ${1\over 2}\int_{\Sigma_2}Q_-Q_+\phi_i$.
Since we have an infinite cylinder to the left of $\Sigma_2$, by
definition, this yields the term $A_{ij}^k\phi_k$ where
$-A_{ij}^k\dd t^i$ is the connection form defining the covariant derivative
$D_i$. Thus, we obtain
\beq
\partial_i\Pi^a_j
=A_{ij}^k\Pi^a_k-i\beta C_{ij}^k\Pi^a_k.
\eeq
This is nothing but the first equation in (\ref{parallel}).
The derivation of the second equation is similar.

\subsubsection{Period Integral as the Boundary Entropy for LG Models}

~~~~
We study the pairings $\Pi^a_i$ in a LG model on a non-compact
Calabi-Yau manifold $X$,
where the axial R-charge is conserved and integral.
We consider the D-brane wrapped on a wave-front trajectory $\cycle_a$
emanating from a critical point in the positive real direction.
The corresponding boundary condition $a$ preserves A-type supersymmetry.
As we have shown,
the pairings $\Pi^a_i=\langle a|\phi_i\rangle$
are invariant under twisted F-term deformations (\ref{acinv}).
In particular they are invariant under the K\"ahler deformation and can be
studied by taking the large volume limit
where the contribution of constant maps dominates.
We thus expect the quantum mechanical expression
$\Pi^a_i=\int_{\cycle_a}\omega_i$ where $\omega_i$ are
the vacuum wave functions
corresponding to the chiral fields $\phi_i$.
It is known that $\omega_i$
are middle dimensional differential forms on $X$ \cite{Klimek}
which have the right dimension to be integrated over the middle dimensional
cycles $\cycle_a$.
However, we recall that, in addition to the ordinary path integral
with the boundary condition $a$,
we have the following boundary term in the Euclidean action
\beq
{i\over 2}\oint\dd x^1 \Bigl(\,W-\overline{W}\,\Bigr)
\eeq
which comes from the shift (\ref{shift}).
This is simply
$i\beta (W-\overline{W})/2$ since the integrand is a constant.
Thus, the pairing is given by
\beq
\Pi^a_i=\int\limits_{\cycle_a}\e^{-i\beta(W-\overline{W})/2}
\omega_i.
\label{ove}
\eeq

The vacuum wave forms $\omega_i$ are in general difficult to determine.
However, a simplification is expected when we make a replacement
$(W,\overline{W})\to(\lambda W,\overline{\lambda}\overline{W})$
and take the limit $\overline{\lambda}\to 0$
but keeping $\lambda$ finite.
In this limit (and in the quantum mechanical approximation),
the supercharges $Q_{\pm}$, $\bQ_{\pm}$
correspond to the operators
\beq
\begin{array}{ll}
\bQ_+\propto\bartial-{i\over 2}\lambda\partial W\wedge,&
Q_+\propto\bartial^{\dag},\\[0.2cm]
\bQ_-\propto \partial^{\dag}
-{i\over 2}\lambda(\bartial\overline{W}\wedge)^{\dag},&
Q_-\propto \partial.
\end{array}
\label{Qp}
\eeq
The vacuum wave forms $\omega_i$ in the $\blambda\to 0$ limit
must be annihilated by these operators.
Reference \cite{Klimek} studies
the cohomology of the Dolbeault operator
deformed as $\bQ_+$ in (\ref{Qp}).
Under suitable assumption about $X$\footnote{
The assumption is that $X$ be a Stein space, where
ordinary Dolbeault cohomology $H^{p,q}(X)$ vanishes except $q=0$ where
it is isomorphic to the space of holomorphic $p$-forms.},
it was shown that the cohomology of
$\bartial-{i\over 2}\lambda\partial W\wedge$ is isomorphic to
the cohomology of Kozsul complex given by the operator
$\partial W\wedge$ acting on the holomorphic forms.
Furthermore, under the assumption that $W$ has a finite number of critical
points, the latter cohomology group is
non-zero only at middle dimension and is isomorphic to
the underlying group of the local ring of $W$
which is nothing but the chiral ring
of the LG model.
Here we use this fact and the arguments in \cite{Klimek}
to study the overlap integral
(\ref{ove}) in the $\blambda\to 0$ limit.
The vacuum wave form $\omega=\omega_i$,
which in particular defines a $\bQ_+$ cohomology class,
can be written as
\beq
\omega=\Omega+(\bartial-{i\over 2}\lambda\partial W\wedge)\eta,
\eeq
where $\Omega$ is a holomorphic $n$-form (where $n$ is the complex
dimension of $X$)
and $\eta$ is an $(n-1)$-form.
It is clear that a holomorphic $n$-form $\Omega$
is annihilated by all operators
$Q_{\pm}$, $\bQ_{\pm}$ in (\ref{Qp}).
Now, let us evaluate the overlap integral
\beqa
\lim_{\blambda\to 0}
\Pi^a&=&
\int\limits_{\cycle_a}\e^{-i\beta\lambda W/2}
\omega
=\int\limits_{\cycle_a}\e^{-i\beta\lambda W/2}
\Bigl(\Omega+(\bartial-{i\over 2}\lambda\partial W\wedge)\eta\Bigr)
\nonumber\\
&=&\int\limits_{\cycle_a}\e^{-i\beta\lambda W/2}\Omega
+\int\limits_{\cycle_a}\left\{
\dd\Bigl(\e^{-i\beta\lambda W/2}\eta\Bigr)
-\e^{-i\beta\lambda W/2}\partial\eta\right\}.
\label{perlim}
\eeqa
The total derivative term vanishes under the assumption that
the integrand vanishes at infinity of $\cycle_a$.
Let us focus on the term involving $\partial\eta$.
Here we use the fact that the vacuum wave form $\omega$
must be annihilated by all
supercharges, not just by $\bQ_+$.
In particular it must be annihilated by
$Q_-\propto \partial$.
Since $\Omega$ is trivially annihilated by $\partial$,
this leads to the condition
$\partial \bQ_+\eta=0$. Since $\partial$ and
$\bQ_+$ anti-commute with each other, this means that $\partial\eta$
is annihilated by $\bQ_+$.
In particular, it can be written as
\beq
-\partial\eta=\Omega_1+(\bartial-{i\over 2}\lambda\partial W\wedge)\eta_1,
\eeq
where $\Omega_1$ is a holomorphic $n$-form and $\eta_1$
is an $(n-1)$-form.
Inserting this expression to (\ref{perlim}),
we obtain $\int_{\cycle_a}\e^{-i\beta\lambda W/2}(\Omega+\Omega_1
-\partial\eta_1)$,
where again we assumed that
the total derivative term vanishes.
Since $\partial\eta$ has no $(0,n)$ component, we can choose
$\eta_1$ to have no $(0,n-1)$ component.
Continuing this procedure, we finally obtain an expression
of $\Pi^a$ as an integral over $\cycle_a$ of
$\e^{-i\beta\lambda W/2}$ times a holomorphic $n$-form only.
The vanishing of the total derivative terms
is assured by taking $\lambda=-i$,
since the exponential factor becomes $\e^{-\beta W/2}$
which quickly vanishes at infinity of $\cycle_a$
which extends to real positive directions in the
$W$-plane.
Thus, we obtain
\beq
\lim_{\lambda\to -i\atop\blambda\to 0}
\Pi^a_i
=\int\limits_{\cycle_a}\e^{-\beta W/2}\Omega_i.
\eeq
where $\Omega_i$ is a holomorphic $n$-form.\footnote{
In this derivation we have assumed that $X$ is a Stein space.
In the cases of interest in this paper we will be dealing this is
the case for some examples.  But we will
also also consider cases where $X$ has a non-trivial $\pi_1$.  In such
a case one can repeat the arguments above for the covering
space and obtain the same results.}
Even though we considered the limit ${\overline \lambda}\rightarrow 0$
in finding this overlap between ground states and D-brane
boundary states, in the conformal limit this is unnecessary
(as the conformal limit corresponds to taking $\lambda, {\overline
\lambda }\rightarrow 0$).  We will use this result
in section 5 in the context of the LG realization of minimal
models.

The result obtained here was anticipated in part in 
\cite{tt} and can be viewed
as an interpretation of some of the observations
there.  The argument presented there shows that 
for flat coordinates, i.e. for
a special choice of chiral fields,  the period integrals $\Pi_i^a$
given above
satisfy the holomorphic
part of the flatness equations given in eq.\,(\ref{parallel}).
The anti-holomorphic part trivializes in the
limit ${\overline \lambda}\rightarrow 0$, and thus
we obtain the above result in this limit.

\section{Brane Creation}

As we have already discussed the D-branes preserving
the A-type supersymmetry in a LG theory are
Lagrangian submanifolds, and their image
in the $W$-plane correspond to straight lines.  The slope
of the straight lines depend on which phase combination of
A-type supercharges one preserves.  In particular for
$Q_A^{\alpha}={\overline Q}_++{\rm exp}(i\alpha){ Q}_-$
the image in the $W$-plane forms an angle $\alpha$
relative to the real axis.  Moreover
D-branes which lead to boundary states with finite
overlap with Ramond ground states correspond to D-branes
whose image in the $W$-plane correspond to straight lines
emanating from a critical point.

We consider a LG model of $n$ variables which are coordinates of
$\C^n$.
Let us assume that the superpotential
$W$ has $N$ isolated critical points $x_1,\ldots,x_N$.
We denote by $B_{\alpha}$ the region in $\C^n$ on which
 ${\rm Re} [e^{-i\alpha}W]$ is larger than a fixed large value.
Let $\Ggamma_i$ be the
wavefront trajectory emanating from the critical point $x_i$
along the straight line in the $W$-plane with the angle $\alpha$ against
the real axis.
As discussed before $\Ggamma_i$ are the cycles on which
the D-branes can wrap without breaking the supersymmetry $Q_A^{\alpha}$.
It is known \cite{Arnold} that the cycles $\Ggamma_i$ form a basis of
the middle-dimensional homology group
$H_n(\C^n, B_{\alpha})$ relative to the boundary $B_{\alpha}$.
In other words $H_n(\C^n, B_{\alpha})$
can be viewed as the lattice of charge for the D$n$ branes. 

Now, let us consider a one parameter family $\Ggamma_1(t)$ of D-branes
emanating from a critical point $x_1$.  Here $t$ is
a deformation parameter either in the couplings in the superpotential
$W$, or the angle $\alpha$ in the combination of
supercharges the D-brane preserves. In such a situation
a special thing may happen:  The image of the $\Ggamma_1(t)$ brane
in the W-plane may pass through a critical value $W(x_2)$ at
some $t=t_0$, so that as we go from $t_0-\epsilon$ to $t_0+\epsilon$
the position of the critical value relative to the
image of the $D$-brane on the W-plane, goes from
one side to the other.  In such a case $\Ggamma_1(t_0-\epsilon)$
and $\Ggamma_1(t_0+\epsilon)$ will label different elements
of $H_n(\C^n,B_\alpha)$, i.e. they will have different
D-brane charges.  In particular as discussed before,
\be 
[\Ggamma_1(t_0-\epsilon)]
=[\Ggamma_1(t_0+\epsilon )]-(\Delta_1 \circ \Delta_2)\,
[\Ggamma_2(t_0+\epsilon )]\,.
\ee
In the context of string theory, charge
conservation would imply that we have to have
created $+\Delta_1 \circ\Delta_2$ of $\Ggamma_2$ branes in order
to guarantee charge conservation during this process.
\onefigure{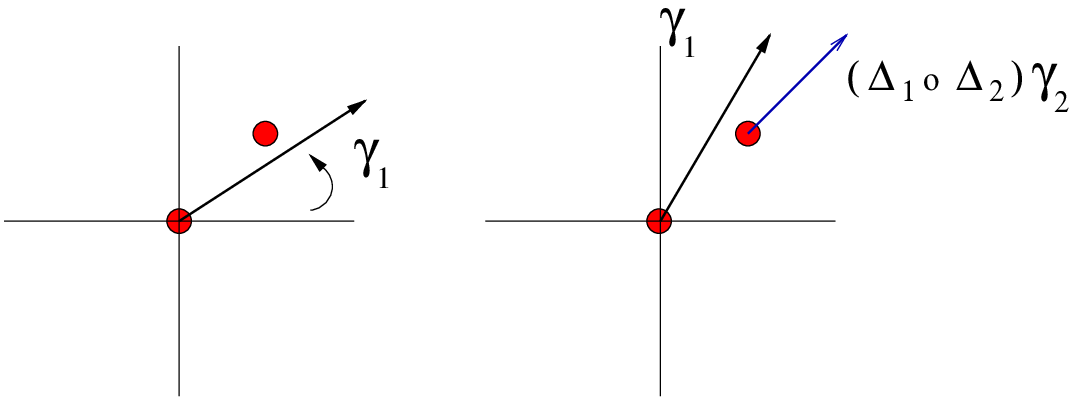}{As the image of D-brane in
the W-plane passes through a critical value, new
D-branes are created whose image start from the
crossed critical value.}
In the present context the same can be said if
we demand {\it continuity} of the correlation functions
of the 2-dimensional theory. For example if we consider
the 2-dimensional theory on a cylinder with one boundary ending
on the $\Ggamma_1(t)$, then the continuity of the correlation function
with this boundary condition as a function of $t$ demands
that as we change $t$ from $t_0-\epsilon$ to $t_0+\epsilon$
we would have to also add to the correlation function
the correlator involving boundaries on the $\Ggamma_2$ brane
with multiplicity factor $+\Delta_1 \circ\Delta_2$.
Also, continuity of the
overlap with the Ramond ground states in the closed channel
will already imply this.  Note that $\Delta_1 \circ\Delta_2$ 
possibly being
negative simply means we have an opposite orientation for the 
$\Ggamma_2$-brane (i.e. the rules of Grassmann integration
over the fermions has picked an extra minus sign).

In string theory a similar process was discovered
in \cite{HW} where again
charge conservation leads to creation of new branes.

\subsection{Monodromy and R-Charges}

Now we revisit 
a result obtained in \cite{CV}
which relates the number of BPS solitons
in 2d theories with $(2,2)$ supersymmetry
to the R-charges of the Ramond ground
states at the conformal point.  In particular
we show how this result follows very naturally
from the realization of D-branes in LG theories, together
with the Brane creation discussed above.
Our proof will be based on the case of LG theories,
though the generalization to arbitrary
massive theories should hold, as already shown in \cite{CV}.

Consider an LG theory with $N$ isolated massive vacua.
As discussed before we can associate $N$ natural D-branes
to these vacua, one for each critical point.  The image
in the $W$-plane is a straight line emanating from the critical
point and going to infinity along a line whose slope depends
on the combination of A-type supercharges we are preserving.
In particular for $Q_A^{\alpha}=\bQ_++{\rm exp}(i\alpha) Q_-$
they make an angle $\alpha$ relative to the real axis.

Let us start with $\alpha=0$ and order the $N$ D-branes 
according to the lower value for $Im(W)$, as depicted
in \figref{creation-1}.  
\onefigure{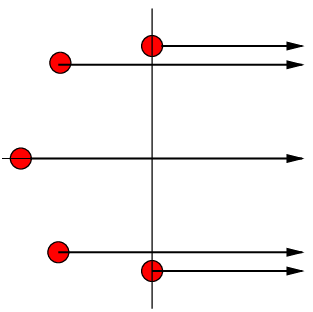}{A convex arrangement of critical
values in the $W$-plane is the most convenient
one for deriving the monodromy action by $2\pi$ rotation.}
Let us further
assume that the critical values have a convexity
compatible  with the ordering of $Im(W)$ as shown in the figure.
This can be done, by deforming the coefficients of $W$ if necessary.
 As we increase $\alpha$ from $0$
to $\pi$  we rotate the
image of branes in the W-plane counter clockwise.
As discussed in the previous section, during this process
we create new branes.  In particular the action of
brane creation in the basis of branes emanating
from the critical points $\Ggamma_i$ is rather simple:  The
rotation of branes by $\pi$ in the W-plane causes the
$\Ggamma_i$ brane to cross all the other $\Ggamma_j$ branes with $j>i$
exactly once.  Moreover during this crossover it creates
$(\Delta_i \circ \Delta_j)$ new $\Ggamma_j$ branes. This action 
of rotation of branes by $\pi$ is thus realized by
an $N\times N$ upper triangular matrix with 1 on the diagonal
and $\Delta_i \circ \Delta_j$ for each $i<j$.  This is denoted by
\be 
S=1+A\,,
\ee
where $A$ is the upper
triangular matrix of inner product of $\Delta$'s.

Now consider instead going from $\alpha=0$ to $\alpha=-\pi$.
In this case for each $i>j$ we get $\Delta_i \circ\Delta_j$ brane
creation of $\Ggamma_j$.  Thus this action is realized as
\be 
S^t=1+A^t\,.
\ee
Now we consider going around from $\alpha=0$ to
$\alpha=2\pi$.  In this case the action on the
$\Ggamma_i$ brane basis is given by 
\be 
H=SS^{-t}\,,
\ee
where we used here the fact that going from $\alpha=\pi$
to $\alpha=2\pi$ is the inverse of the action of 
changing $\alpha$ from $0$ to $-\pi$.

Now consider rescaling the superpotential $W\rightarrow \lambda W$
as we send $\lambda \rightarrow 0$.  In this limit we approach
a conformal point.  For any $\lambda$ the monodromy operator
$H$ we have obtained is the same, because the rescaling of $W$
does not affect the relative location of D-branes or their
intersection numbers.

Consider the boundary states $|\Ggamma_i\rangle $ corresponding
to the $i$-th D-brane at $\alpha=0$.
  At the conformal point we obtain a
new
conserved R-charge, which is the fermion number of the
right-moving fermions.  In particular we have
\be
Q_A^{\alpha}={\rm exp}(-i \alpha R) Q_A^0 {\rm exp}(i\alpha R)\,,
\ee
where $R$ denotes the right moving fermion number charge.
Thus the $H$ monodromy is realized in the conformal limit as
\be
H|\Ggamma_i\rangle ={\rm exp}(2\pi i R )|\Ggamma_i\rangle\,.
\ee
On the other hand we can go to a basis where the action of
$R$ is diagonal. Note that since the $|\Ggamma_i\rangle $ have invertible
overlaps with the Ramond ground states, we can choose
linear combination of Ishibashi type states associated
to Ramond ground states to represent $|\Ggamma_i \rangle $.
  We thus learn that,
\be
{\rm Eigenvalues}[SS^{-t}]={\rm Spectrum}[{\rm exp}
(2\pi i R)] \ {\rm on} \ {\rm Ramond}\ {\rm Ground}\ {\rm states}\,,
\ee
which is a result of \cite{CV} rederived in a purely D-brane
language.  Note that the choice we have made in the convexity
of the critical values is irrelevant for the final result,
in that the brane creation was derived precisely based
on the continuity of physical correlation functions.
The operator ${\rm exp}(2\pi i R)$ is a physical observable
and the structure of brane creation guarantees that for any
distribution of critical values, going around the $W$ plane by 
$2\pi$ will yield the same operator on the $\Ggamma_i$ brane states.

In the next section, after we discuss minimal models
we show that we can make a slightly stronger statement
than just equating the eigenvalues of $SS^{-t}$ with
the spectrum of ${\rm exp}(2\pi iR )$. Namely we can actually
find the change of basis which diagonalizes $SS^{-t}$ by considering
the overlap of chiral fields with definite $R$
charges with the corresponding boundary
states.  That this should be possible is clear, because
the chiral fields provide a basis where $R$ acts diagonally.

\section{D-Branes in ${\cal N}=2$ Minimal Models}

In this section, we study D-branes of ${\cal N}=2$
minimal models using their realizations as the infra-red fixed points
of Landau-Ginzburg models
\cite{mart,vafa-1}.
We will see that the D-branes of the LG models naturally
gives rise to the Cardy states of the minimal models and we will be able to
study their properties using purely geometric method.
In particular,
we will find a beautiful geometric realization of the Verlinde ring for
$SU(2)$ level $k$ Wess-Zumino-Witten models
as well as a simple understanding of
the $\tau\rightarrow -\frac{1}{\tau} $ modular transformation matrix $S^{\,j}_{i}$.
We first review the construction of the D-branes in the minimal
models and then see how they
are realized as the D-branes in the LG models.

\subsection{Cardy States, Ishibashi States and $N=2$ Minimal models}

${\cal N}=2$ minimal models are unitary
$(2,2)$ superconformal field theories in two dimensions with
central
charge $c=\frac{3k}{k+2}$, where $k$ is a positive
integer. They can be viewed as an $SU(2)/U(1)$ super-GKO
construction at level $k$.
 The superconformal
primary fields are
labeled by three integers $(l,m,s)$ such that
\bea
l&=&0,\cdots, k\,,\\ \nn
m&=&-(k+1),\cdots,(k+2)\,\,(\mbox{mod}\,\,2k+4)\,\,,\\ \nn
s&=&-1,0,1,2~(\mbox{mod}\,\,4)\,,
\eea
with the constraint $l+m+s\equiv 0~(\mbox{mod}~2)$ and field
identification $(l,m,s)=(k-l,m+k+2,s+2)$.
$ s=0,2$ in the NS sector
and s=$\pm 1$ in the Ramond sector. The two
different values of $s$ denote the GSO parity
of various states in the Ramond or NS sector.
The
conformal weights and the $U(1)$ charges of the primary fields are
(mod integer),
\be
h^{l}_{m,s}= \frac{l(l+2)-m^{2}}{4(k
+2)}+\frac{s^{2}}{8}\,,\,\,\,\,\,q^{l}_{m,s}=\frac{m}{k+2}-\frac{s}{2}\,.
\ee
The ${\cal N}=2$ chiral primary states are $(l, l,0)$ in the NS
sector. The related Ramond states $(l, l+1,1)$ can be reached by
spectral flow. These models can also be described by the IR fixed
point of the LG model with a single chiral superfield $X$ with 
superpotential $W=X^{k+2}$ \cite{mart,vafa-1}.
The 
chiral primary fields $X^{l}$ correspond to the states 
$(l,l,0)$ and provide a representation of the chiral ring. 
Note that there are only $k+1$ chiral primary fields 
(as $l$ ranges from $0$ to $k$), which
correspond to the $k+1$ ground states in the Ramond
sector.  However there are
a total of $(k+1)(k+2)$ primary states in the Ramond sector
(up to a choice of GSO action $(-1)^F$).


An A-type boundary state satisfies the following boundary conditions,
\be
(L_{n}-\bar{L}_{-n})|B\rangle=
0\, ,\,(J_{n}-\bar{J}_{-n})|B\rangle =0\,,\,(G^{\pm}_{r}+ 
i\,\bar{G}^{\mp}_{r})|B\rangle =0\,.
\ee
 For a rational 
conformal field theory it was shown in \cite{Ishibashi} that the boundary 
states are linear combinations of ``Ishibashi states'' on which the left and
the right generators of the superconformal algebra are linearly related. 
Ishibashi state corresponding to 
the primary state $(l,m,s)$ is given by \cite{Ishibashi}
\be
|l,m,s \rangle \!\rangle=\sum_{N}|l,m,s;N\rangle \otimes U\Omega
\overline{|l,m,s;N\rangle}\,.
\ee
Where $U$ is an anti-linear operator acting only on the right moving
sector as $U\bar{{\cal O}}_{n}U^{-1}=(-1)^{h_{{\cal O}}}\bar{{\cal
O}}_{n}$, $\Omega$ is the mirror automorphism of the ${\cal N}=2$
algebra and the states $|l,m,s;N\rangle$ form an orthonormal basis of
${\cal H}_{l,m,s}$. The boundary states are particular linear
combination of Ishibashi states \cite{cardy}
\be
|l,m,s\rangle _{BS}=\alpha \sum_{(l',m',s')}\frac{S^{~l',m',s'}_{l,m,s}}
{\sqrt{S^{~l',m',s'}_{0,0,0}}}|l',m',s'\rangle \!\rangle
\,.
\label{boundary-state}
\ee
Where the constant, $\alpha$, is fixed
by the condition that the partition function in the open sting channel
is integral linear combination of the characters.  The summation in
the above equation is only over allowed states modulo the field
identification and $S^{i}_{j}$ is the matrix representation of the
modular transformation $\tau\mapsto -\frac{1}{\tau}$ for the
characters $\chi_{l,m,s}(\tau)=
\mbox{Tr}_{{\cal H}_{l,m,s}}q^{L_{0}-\frac{c}{24}}$,
\be
\chi_{l,m,s}(-\fracs{1}{\tau})=
\sum_{(l',m',s')}S^{~l',m',s'}_{l,m,s}\chi_{(l',m',s')}(\tau)\,,
\ee
and is given by,
\be
S^{~l',m',s'}_{l,m,s}=\fracs{{1}}{{\sqrt{2}\,(k+2)}}\,
\,\,\mbox{Sin}(\pi \fracs{{(l+1)(l'+1)}}{{k+2}})
e^{\frac{i \pi m m'}{k+2}}e^{-\frac{i \pi s s'}{2}}\,.
\ee
The above identification of boundary state is motivated mainly by
demanding integral expansion in the characters of $Tr_{\alpha,\beta}
q^{L_0}$ corresponding to open strings ending on $\alpha$ and $\beta $
D-branes.  The integrality of characters in this sector follows from
properties of the Verlinde algebra.  We are interested in the Ramond
part of the boundary state which can be obtained by restricting the
sum in eq.\,(\ref{boundary-state}) to Ramond states only. The properly
normalized Ramond part of the boundary state is,
\be
|l,m,s\rangle_{{}_{\rm RR}} =\sqrt{2\sqrt{2}}\,\sum_{(l',m',s')_R}\frac{S^{~l',m',s'}_{l,m,s}}
{\sqrt{S^{~l',m',s'}_{0,0,0}}}|l',m',s'\rangle \!\rangle \,.
\label{boundast}
\ee

Consider an open string in the $(a,b)$ sector.  As we have discussed
in section 3, the index $I(a,b)=\Tr_{a,b} (-1)^F \e^{-\beta H}$
corresponds in the closed string channel to an overlap in the Ramond
sector boundary states $I(a,b)=\Tr_{a,b}(-1)^{F}={}_{{}_{\rm
RR}}\langle a|b\rangle_{{}_{\rm RR}}$.  Using the expression
(\ref{boundast}), it is straightforward to compute the index in the
$(a,b)=((l_{1},m_{1},s_{1}),(l_{2},m_{2},s_{2}))$ sector. Since the
index gets contribution from the Ramond ground states only we have,
\bea \nn
I(a,b)&=&_{{}_{\rm RR}}\langle
l_{1},m_{1},s_{1}|l_{2},m_{2},s_{2}\rangle_{{}_{\rm RR}} \\ \nn 
&=&2\sqrt{2}
\sum_{l=0}^{k}
\frac{(S_{l_{1},m_{1},s_{1}}^{l,l+1,1})^{*}\,S_{l_{2},m_{2},s_{2}}^{l,l+1,1}}{S_{0,0,0}^{l,l+1,1}}\,\langle
l,l+1,1| l,l+1,1\rangle \\ \nn
&=& \fracs{{-2\sqrt{2}i}}{{\sqrt{2}(k+2)}}\sum_{l=0}^{k}
\frac{\mbox{Sin}(\pi \fracs{{(l+1)(l_{1}+1)}}{{k+2}})\mbox{Sin}(\pi \fracs{{(l+1)(l_{2}+1)}}{{k+2})}}{\mbox{Sin}(\pi \fracs{{l+1}}{{k+2})}}
e^{\frac{i\pi (l+1)(m_{2}-m_{1}+1)}{k+2}}\,e^{-\frac{i \pi (s_{2}-s_{1})}{k+2}}\, \\ \nn
&=&\fracs{{2\,e^{-\frac{i \pi (s_{2}-s_{1})}{2}}}}{{k+2}}
\sum_{l=0}^{k}
\frac{\mbox{Sin}(\pi \fracs{{(l+1)(l_{1}+1)}}{{k+2}}) 
      \mbox{Sin}(\pi \fracs{{(l+1)(l_{2}+1)}}{{k+2}})
      \mbox{Sin}(\pi \fracs{{(l+1)(m_{2}-m_{1}+1)}}{{k+2}})}
     {\mbox{Sin}(\pi \fracs{{l+1}}{{k+2}})}\,. 
\eea
Finally using the fact that $s_{2}-s_{1}$ is an even integer in the Ramond sector we get 
\cite{RS,BDL,GJ},
\be 
I((l_{1},m_{1},s_{1}),(l_{2},m_{2},s_{2}))= (-1)^{\frac{s_{2}-s_{1}}{2}} \,\,N^{\,m_{2}-m_{1}}_{l_{1},\,l_{2}}\,.
\label{verl}
\ee
Where 
\bea \nn
N^{\,l_{1}}_{l_{2},\,l_{3}}=\fracs{{2}}{{k+2}}\sum_{l=0}^{k}
\frac{\mbox{Sin}(\pi \fracs{{(l+1)(l_{1}+1)}}{{k+2}}) 
      \mbox{Sin}(\pi \fracs{{(l+1)(l_{2}+1)}}{{k+2}})
      \mbox{Sin}(\pi \fracs{{(l+1)(l_{3}+1)}}{{k+2}})}
     {\mbox{Sin}(\pi \fracs{{l+1}}{{k+2}})}\,
\eea
are the $SU(2)_{k}$ fusion coefficients, 
\beq
N^{~m_{2}-m_{1}}_{{l_{1},\,l_{2}}}\,=\,
\left\{
\begin{array}{ll}
1&\mbox{if}\,\,\,\,\,|l_{1}-l_{2}|\leq m_{2}-m_{1} \leq \mbox{min}
\{l_{1}+l_{2},2k-l_{1}-l_{2}\}\,, \\[0.2cm]
0&\,\,\, \mbox{otherwise}\,. 
\end{array}\right.
\eeq
As we have studied in section 3, the index for a pair of
D-branes in the LG model can be identified as the ``intersection number''
 of the corresponding cycles.
In the LG realization of the minimal model D-branes, as we will see below,
the index (\ref{verl}) can indeed be considered as such an
``intersection number''.

\subsection{
 D-branes in LG description}

The Landau Ginzburg description of
$A_{k+1}$ minimal model consists of
a single chiral superfield $X$ with
superpotential $W=X^{k+2}$ \cite{mart,vafa-1}. In section 3 we saw that the
D-branes in the
LG description are  preimages of the straight lines
in W-plane starting from
the critical values which correspond to vanishing cycles in
the x-space fibered over the W-plane.

The superpotential $W=X^{k+2}$ has a single critical point
$X=0$ of
multiplicity $k+1$ with critical value $w_{*}=0$.  If we consider
deforming the superpotential by lower powers of $X$ we
will generically obtain $k+1$ isolated and non-degenerate critical points
with distinct critical values $w_i$.
We assume that ${\rm Im}(w_i)$ are separate from one another. 
Then as we discussed
before we would get $k+1$ D-branes, one associated to each
of the critical points.  Moreover the image of these D-branes
are straight lines in the W-plane going to $+\infty$ in the
real positive direction.  For large values of $X$ the
lower order terms which deform $W$ are irrelevant and
the D-branes approach the preimages of
the positive real axis $X^{k+2}\in \R_{\geq 0}\subset \C$,
namely
\be
X= \mbox{\large $r$}\cdot
\exp\left({2\pi ni\over k+2}\right),\,\,\,n=0,\cdots,k+1\,,\,~~\,
\mbox{\large $r$}\in [0,\infty).
\ee 
Thus we see that the $X$-plane is divided up into $k+2$ wedge shaped
regions by the $k+2$ lines going from the origin to infinity making an
angle of $\frac{2\pi n}{k+2}$ with the positive real axis, we will
denote such a line by ${\cal L}_{n}$.

Any D-brane of the deformed theory is a curve
in the $X$-plane that will asymptote to a pair of such lines,
say ${\cal L}_{n_1}$ and ${\cal L}_{n_2}$ with $n_1\ne n_2$. 
To see this, we note that the deformed superpotential $W$ is
approximately quadratic
around any (non-degenerate) critical point $a$ and the preimage of
the straight line emanating from $W(a)$ in the $W$-plane
splits to trajectories of two points (wavefronts) starting from $a$.
 The two wavefronts approach the lines 
${\cal L}_{n_1}$ and ${\cal L}_{n_2}$ as they move away from the critical
point.
To see that $n_1\ne n_2$ it is sufficient to note that
the two wavefronts can merge only at a critical point (but
the $(k+1)$ critical values are assumed
to be separate in the imaginary direction).
Note that the homology
class of the D-branes is completely specified
by the choices of the combinations of $k+2$ wedges in the x-plane, and
that the $k+1$
D-branes will be enough to provide a linear basis for the
non-trivial cycles (since the sum of all wedges is homologically
a trivial contractible cycle). Precisely which combination of
$k+1$ pairs of asymptote we obtain will depend
on which deformation of $W$ away from criticality
we are considering.

  In general the $k+1$ D-branes
we obtain in this way will not intersect with 
each other (as their images in the W-plane do not intersect
one another).  Nevertheless, as discussed in section 3
the index $I(a,b)=\Tr_{a,b}(-1)^F $ is
not in general zero and will depend on
the number of solutions to (\ref{straightup}), that is,
how many orthogonal
gradient trajectories
there are from $a$ to $b$ D-branes, with a fixed length $x^1\in [0,\pi]$.
This in turn is given 
by the ``intersection number'' of the D-branes
which is defined as the geometric intersection number
where the $b$-brane is tilted with a small positive
angle in the $W$-plane.

 We thus see that away from the
conformal point there are $k+1$ distinct
pairs of D-branes, each labeled by an ordered pair of integers
($n_1,n_2$) which label the asymptotes that it makes
(taking into account the orientation of the D-brane).
In particular $n_1$ and $n_2$ are well defined modulo $k+2$
and $n_1\not= n_2$. We will label such a D-brane by 
$\gamma_{n_1n_2}$.  However there are only
$k+1$ such pairs for a generically deformed $W$.
In particular we do not have
both branes of the type $\gamma_{n_1n_2}$ and $\gamma_{r_1r_2}$
with $n_1<r_1<n_2<r_2$ for generically deformed $W$
as that would have required them to geometrically intersect.

Let us consider two allowed
branes $\gamma_{n_1n_2}$ and $\gamma_{r_1r_2}$.
We are interested in computing the
Witten index
in the oriented open string sector starting from the
$\gamma_{n_1n_2}$ brane and ending on the $\gamma_{r_1r_2}$ brane.
Let us denote this by the overlap of the corresponding
boundary states, namely $\langle \gamma_{n_1n_2}|\gamma_{r_1r_2}\rangle$.
If none of the $n_i$ and $r_i$ are equal the branes do not
intersect even when one of them is slightly tilted
(as noted before in the massive theory the case $n_1<r_1<n_2<r_2$
is not allowed) and thus the index is zero.
The more subtle case is when one of the $n_i$ is equal to
one of the $r_i$.  If they are both equal then we
get the Witten index to be $1$ as discussed before.
Without loss of generality we can order the branes so
that $n_1<n_2$ and $r_1<r_2$ (otherwise the intersection
number gets multiplied by a minus sign for each
switch of order).  Thus there are only four more cases
to discuss:  $n_i=r_j$, for some choice of pair of $i,j=1,2$.
Let us also assume that $r_1+r_2>n_1+n_2$ (by $r_2=n_1$
we mean equality mod $k+2$, i.e. this is $r_2=n_1+k+2$).
It turns out that in such cases
\be
\langle \gamma_{n_1n_2}|\gamma_{r_1r_2}\rangle =1 \quad \mbox{iff} \quad
n_i=r_i \,\,\, \mbox{for some \,} i
\label{equalcase}
\ee
and zero otherwise.  To see this, as discussed in section 2 and 3
it suffices to consider tilting the slope of the image in the
$W$-plane of the D-brane corresponding to $\gamma_{r_1r_2}$ in the 
positive direction and seeing if they intersect or not.
Tilting the slope
in the $W$-plane in this case will also correspond to tilting the
asymptotes $r_1,r_2$ in the positive direction and seeing if they
intersect the $n_1, n_2$ brane. This leads to the above formula.
The case of $r_1=n_1$ which leads to intersection number 1 and
$r_1=n_2$ which leads to intersection number zero is
depicted in the \figref{ade-4}.
\onefigure{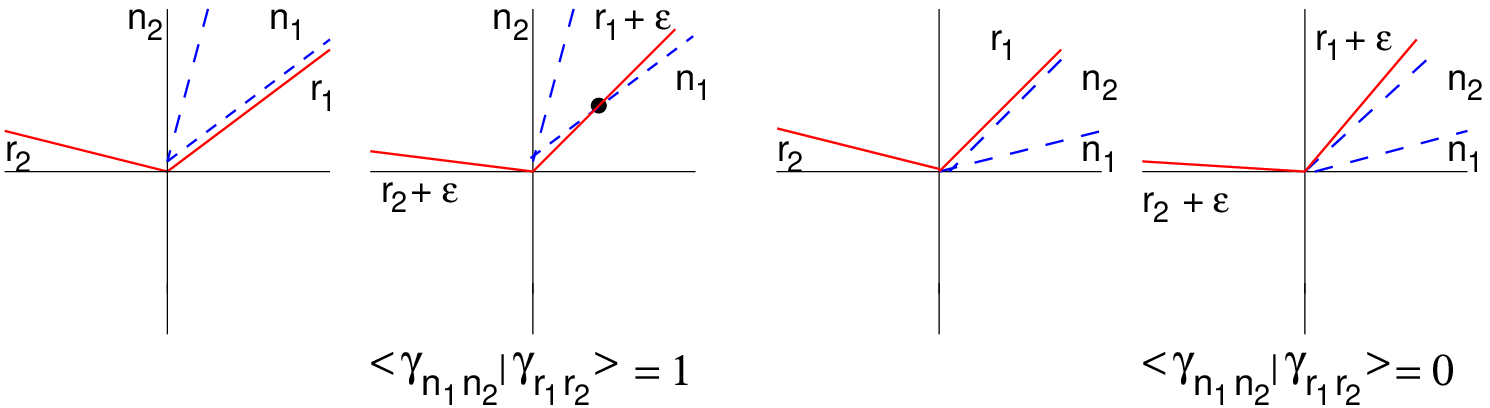}{To get the index in the open
string sector stretched between $\gamma_{n_1n_2}$ and
$\gamma_{r_1r_2}$ we have to rotate the image in the
$W$-plane of the $\gamma_{r_1r_2}$ brane
in the positive direction and compute the corresponding
intersection number.} 

Now we come to the D-branes at the conformal point.
Since the $k+1$ D-branes make sense arbitrary close
to the conformal point, they survive in the limit of
conformal point as well.  But here since we have different
allowed D-branes at the massive theory, depending on the choice
of the deformation polynomials, we learn that {\it all}
of them survive at the conformal point.  Since all pairs
$(n_1,n_2)$ are realized in terms of a D-brane for
{\it some} deformation of $W$ (which follows from Picard-Lefshetz
action discussed before) we learn that all D-branes
$\gamma_{n_1n_2}$ exist for arbitrary unequal integers
$n_1,n_2$ defined mod $k+2$, which now correspond
to exact straight lines in the $x$ space along
the half-lines given by $n_1$ and $n_2$, passing through
the origin.
  This gives us a total
of $(k+2)(k+1)$ D-branes, which are pairwise the same
upto orientation at the conformal point.  Here we are
encountering an interesting effect: {\it  The number
of D-branes jump as we go from the conformal point
to the massive theory}.

The fact that we have obtained $(k+2)(k+1)$ of such
branes at the conformal point
 is very encouraging as that is exactly the
same as the predicted number of Cardy states, as already
discussed. Moreover, if we consider the range of parameters
where $0\le n_1<n_2\leq k+1$
we see that 
$|n_{1}-n_{2}|\in \{1,\cdots
,k+1\}$ and $n_{1}+n_{2}\in \{0,\cdots 2k+2\}$. The range of these
parameters exactly corresponds to the quantum numbers $(l,m)$ labelling
the boundary states.
Note also that $s=\pm 1$
for the Ramond sector boundary
states which we are considering.
Thus we claim the following identification
\be
|n_{2}-n_{1}|=l+1,\qquad n_{1}+n_{2}=m,\qquad
s={\rm sign}(n_2-n_1).\, 
\label{identification}
\ee
It follows from (\ref{identification})
that $m+l+s=2n_1=0 \, (\mbox{mod} \, 2)$ as is required.  
 The field identification $(l,m,s) =(k-l,m+k+2,s+2)$ also
has a natural identification as shown in \figref{ade-1}
and relates to the statement that if we change $n_1\rightarrow
n_2$ and $n_2\rightarrow n_1+k+2$ we get the same brane back
up 
to a flip in the orientation (reflected in the shift in $s$).
\onefigure{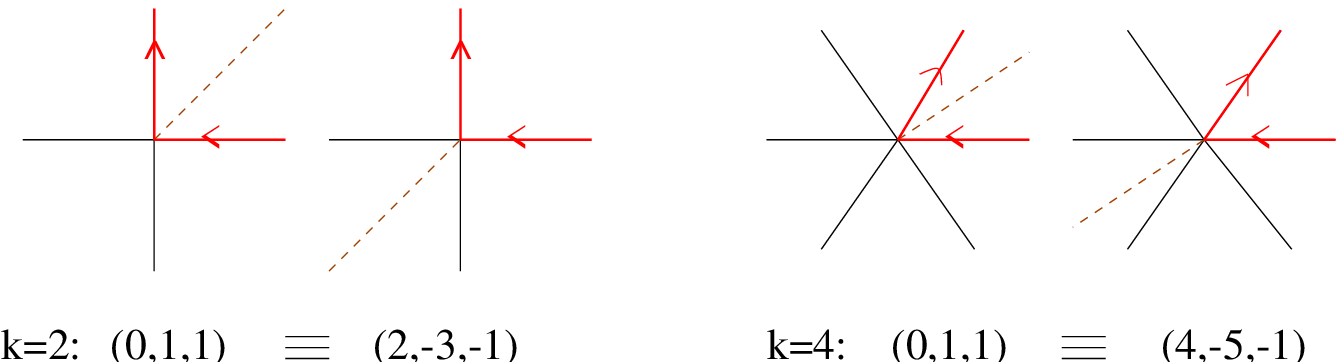}{Field identification}

It is convenient to choose $n_2>n_1 $ (with their differences
less than $k+2$) in which case we have
\be 
n_2-n_1=l+1,~~ n_2+n_1=m,~~  s=1,
\ee
which can be solved for $2n_1$ and $2n_2$ as 
\be
2n_1=m-l-1,\qquad 2n_2=m+l+1.
\label{usef}
\ee
We
will denote the D-brane corresponding to the Ramond boundary state
$|l,m,s\rangle_{RR}$ by $\gamma_{~l,m,s}$.
We will now provide further evidence for this identification.
Along the way we find a simple geometric interpretation
of Verlinde ring for $SU(2)$ level $k$, as well
as certain matrix elements of modular transformations
matrix $S$.

\subsection{Geometric Interpretation of Verlinde Algebra}
We would like to compute the Witten index at the conformal
point for the open string strechted between
two D-branes $\Ggamma_{n_1n_2}$ and $\Ggamma_{r_1r_2}$
and reproduce the index formula (\ref{verl}).
One aspect of the formula is clear.
The ``intersection number'' will not change if we rotate
both branes by integral multiples of $\frac{2\pi}{k+2}$ which implies
that the index will depend on $m_2-m_1$ but not on the other
combination of $m_1$ and $m_2$ (as
$m_1$ and $m_2$ shift by the same amount under the rotation).  
Moreover
the appearance of $(-1)^{\frac{s_2-s_1}{2}}$ in the intersection
is also natural as that correlates with the choice of orientation
on the D-branes.  So without loss of generality we set $s_1=s_2=1$,
i.e.,  as before we choose $n_2>n_1$ and $r_2>r_1$.
Also in checking (\ref{verl}) in computing
the Verlinde algebra coefficients it suffices
to consider the case where $m_2-m_1\geq 0$ which is
the same case as $r_1+r_2\geq n_1+n_2$.
With these set, it is now clear what the conditions are
for obtaining overlap 1, namely we must have
\be
n_1 \leq  r_1< n_2 \leq r_2 < n_1+k+2
\label{inter}
 \ee
and all the other cases vanish.
This is simply the condition that the branes intersect
as shown in \figref{ade-5}.
\onefigure{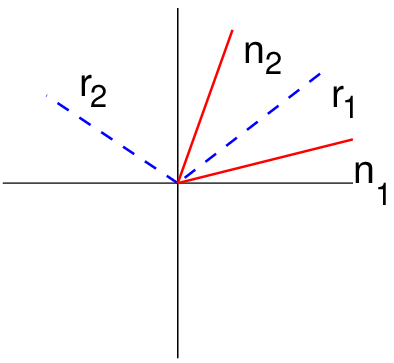}{The intersection number
of the branes is 1 if 
the corresponding asymptotes alternate.  Otherwise
it is zero.}

Note that the condition of getting non-vanishing
results in the case of equality follows from
equation (\ref{equalcase}).  Now we use (\ref{usef})
to rewrite (\ref{inter}) as
\be
m_1-l_1-1\leq m_2-l_2-1 <m_1+l_1+1 \leq m_2+l_2+1 <m_1-l_1+2k+3
\ee
These four conditions can also be written as
\be
|l_2-l_1|\leq m \leq {\rm{min}}[l_1+l_2, 2k-l_1-l_2]
\ee
where $m=m_2-m_1 $ (to show this and write all inequalities in terms
of inequalities
with equal signs we used the fact that $m_2-m_1$ and $l_2+l_1$
are equal mod 2).  This is precisely the condition
for the $SU(2)$ level $k$ algebra and we have thus derived
(\ref{verl}) from a purely LG point of view.

\subsection{Period Integrals and Boundary States}

As was discussed in
the context of LG models
\cite{HV}  and also 
in section 3 of this paper 
there is a natural pairing between the
A-model
boundary states
and B-model chiral fields given by integrating the B-model chiral
fields over the cycles representing the A-type boundary states. 
This kind of pairing was first noticed in \cite{cecotti}
and elaborated further in \cite{CV}.
 For
the $A_{k+1}$ minimal models the chiral primary fields are $X^{l}$ and
therefore the inner product of the boundary state $|l,m,s\rangle_{{}_{\rm RR}}$ and the
state defined by the B-model chiral field is as discussed in section 3,

\begin{eqnarray}
~_{{}_{\rm RR}}\langle l,m,s|X^{l'}\rangle &=&  \int_{\gamma_{l,m,s}}dX~
 X^{l'}
 e^{-W(X)}\,, 
\label{pairing2}\nn
\end{eqnarray}
where the superpotential $W(X)=X^{k+2}$. The image of the cycle
$\gamma_{l,m,s}$ in the W-plane is the positive real axis. Thus from
the discussion of the previous section we can see that we can
parameterize the curve $\gamma_{l,m,1}$ in the following way,
\bea
\gamma_{l,m,1}: X(t)&=&(-t)^{\frac{1}{k+2}}e^{\frac{i \pi (m-l-1)}{k+2}},\,\,\,  t\in [-\infty,0]\,, \\ \nn
&=&t^{\frac{1}{k+2}}e^{\frac{i\pi (m+l+1)}{k+2}},\,\,\,\,\,\,\,t\in[0,\infty]\,.
\eea
Since $m+l+1\equiv 0~(\mbox{mod}~2)$,
the image in the W-plane of $\gamma_{l,m,1}$ is the
positive real axis. We have given the curve for $s=1$ the curve for
$s=-1$ can be obtained from this by reversing the orientation. With
the above parameterization of $\gamma_{l,m,1}$,
\bea\nn
\int_{\gamma_{l,m,1}}dX X^{l\,'}e^{-W(X)}&=&
\fracs{{e^{\frac{i\pi (m-l-1)(l'+1)}{k+2}}}}{{k+2}}\int_{+\infty}^{0}\,\,t^{\frac{l'+1}{k+2}-1}\,e^{-t}\,dt 
+\fracs{{e^{\frac{i\pi (m+l+1)(l'+1)}{k+2}}}}{{k+2}}\int_{0}^{+\infty}\,\,t^{\frac{l'+1}{k+2}-1}\,e^{-t}\,dt \\ \nn 
&=&(-\fracs{{e^{\frac{i\pi (m-l-1)(l'+1)}{k+2}}}}{{k+2}} +
\fracs{{e^{\frac{i\pi (m+l+1)(l'+1)}{k+2}}}}{{k+2}})\int_{0}^{+\infty}\,\,t^{\frac{l'+1}{k+2}-1}\,e^{-t}\,dt \\ 
&=&\fracs{{e^{\frac{i\pi m(l'+1)}{k+2}}}}{{k+2}}(2i\mbox{Sin}(\pi\fracs{{(l+1)(l'+1)}}{{k+2}})\Gamma(\fracs{{l'+1}}{{k+2}})\,.
\label{inte}
\eea
To relate the above integral with the modular S-matrix we need to use the
normalized operator $X_{norm}^{l\,'}$ (\cite{girarce} \cite{CV}) in
the above integral.
 The normalization can be determined by evaluating
the matrix element $\langle \bar{X}^{l\,'}|X^{l\,'}\rangle$.

To determine the matrix element $\langle
\bar{X}^{l\,'}|X^{l\,'}\rangle$ note that 
$\sum_{a_{1},a_{2}}|a_{1}\rangle_{{}_{\rm RR}}\, _{{}_{\rm RR}}\langle
a_{2}|S^{a_{1}a_{2}} =1$ when it is sandwitched by the ground states.
Here $|a_{1}\rangle_{{}_{\rm RR}} $ form a basis
of the Ramond
boundary states and $S^{a_{1}a_{2}}$ is the inverse of the index matrix, 
\be
S^{a_{1}a_{2}}=(S^{\,-1})_{a_{1}a_{2}}\,,\,\,\,S_{a_{1}a_{2}}=
I(a_1,a_2)={}_{{}_{\rm RR}}\langle a_{1}|a_{2}\rangle_{{}_{\rm RR}}\,.
\ee
We choose the basis such that the state $|a\rangle_{{}_{\rm RR}}$ corresponds to the
D-brane ${\cal L}_{a+1}-{\cal L}_{a}$, where $a=0,\cdots,k$. With this
choice of the basis states we see that \footnote{$S_{a_{1}a_{2}}$ is
the intersection matrix not to be confused with the modular
transformation matrix $S_{l,m,s}^{l',m',s'}$ for which the indices
will always be written as subscript and superscript.}
\bea
S_{a_{1}a_{2}}=\delta_{a_{1},a_{2}}-\delta_{a_{1}+1,a_{2}}\,,\,\,\,\,
S^{a_{1}a_{2}}:=(S^{-1})_{a_{1}a_{2}}= \left\{\begin{array}{ll} 1 & \,,\,\,a_{2}\geq a_{1} \\
0 & \,, \,\,\,a_{2} < a_{1} \,. \end{array} \right.,
\label{goodb}
\eea
which follows from eq.\,(\ref{equalcase}) by taking
into account the reversal of the orientation
of the neighboring branes.
We insert this complete set of states in the matrix element 
$\langle \bar{X}^{l\,'}|X^{l\,'}\rangle$,
\be
\langle \bar{X}^{l\,'}|X^{l\,'}\rangle= \sum_{a_{1},a_{2}=0}^{k}
\langle \bar{X}^{l\,'}\,|\,a_{1}\rangle_{{}_{\rm RR}}\,\,S^{a_{1}a_{2}}
\,_{{}_{\rm RR}}\langle a_{2}\,|\,X^{l\,'}\rangle\,.
\label{norm}
\ee
Using eq.\,(\ref{inte}) we see that
\bea \nn
_{{}_{\rm RR}}\langle a_{2}\,|\,X^{l\,'}\rangle &=&  \int_{{\cal L}_{a_{2}+1}}dX\, X^{l\,'}e^{-W(X)}-
 \int_{{\cal L}_{a_{2}}}dX\, X^{l\,'}e^{-W(X)}\\
&=& \fracs{{1}}{{k+2}}\{e^{\frac{2 \pi i (a_{2}+1)(l'+1)}{k+2}}-
e^{\frac{2\pi i a_{2}(l'+1)}{k+2}}\}\Gamma(\fracs{{l'+1}}{{k+2}})\,.
\label{one}
\eea
To calculate $\langle \bar{X}^{l\,'}|a_{1}\rangle_{{}_{\rm RR}}$ we
use the fact that, as discussed in section 3,
\be
\langle \bar{X}^{l\,'}|a_{1}\rangle_{{}_{\rm RR}} = 
\,_{{}_{\rm RR}}\langle {a_{1}}|(-1)^{F_L}|X^{l\,'}\rangle ^{*}\,.
\ee
Using the action 
\bea
(-1)^{F_L}|X^{l\,'}\rangle =e^{i\pi(-\frac{{\widehat c}}{2}-\frac{l'}{k+2})}
|X^{l\,'}\rangle=i\,e^{-\frac{i\pi(l'+1)}{k+2}}|X^{l\,'}\rangle \nn
\eea
where $\widehat{c}=\frac{c}{3}=\frac{k}{k+2}$ we thus obtain
\be
\langle \bar{X}^{l\,'}|a_{1}\rangle_{{}_{\rm RR}} = 
\fracs{{-i\,e^{\frac{i \pi (l'+1)}{k+2}}}}{{k+2}}\{e^{-\frac{2 \pi i (a_{1}+1) (l'+1)}{k+2}}
-e^{-\frac{2\pi i a_{1}(l'+1)}{k+2}}\}\Gamma(\fracs{{l'+1}}{{k+2}})\,. 
\label{two}
\ee
Using eq.\,(\ref{one}) and eq.\,(\ref{two}) in eq.\,(\ref{norm}) we see that
\bea \nn
\langle \bar{X}^{l\,'}| X^{l\,'}\rangle &=&
-4i\fracs{{e^{\frac{i\pi (l'+1)}{k+2}}}}{{(k+2)^2}}\,\mbox{Sin}^{2}(\pi\fracs{{l'+1}}{{k+2}})\,\Gamma(\fracs{{l'+1}}{{k+2}})^{2}
\sum_{a_{1},a_{2}=0}^{k} S^{a_{1}a_{2}} e^{\frac{2 \pi i (a_{2}-a_{1})(l'+1)}{k+2}} \\ \nn
&=&-4i\fracs{{e^{\frac{i\pi (l'+1)}{k+2}}}}{{(k+2)^2}}\,\mbox{Sin}^{2}(\pi\fracs{{l'+1}}{{k+2}})\,\Gamma(\fracs{{l'+1}}{{k+2}})^{2}
\sum_{a_{1}=0}^{k}\sum_{a_{2} =a_{1}}^{k} e^{\frac{2 \pi i (a_{2}-a_{1})(l'+1)}{k+2}} \\ 
&=&  \fracs{{2}}{{k+2}} \,\,\mbox{Sin}(\pi\fracs{{l'+1}}{{k+2}}) \,\,\Gamma(\fracs{{l'+1}}{{k+2}})^{2}\,.
\eea
Thus we see that 
\be
X_{norm}^{l\,'}= \sqrt{\frac{k+2}{2 \mbox{Sin}(\frac{\pi(l'+1)}{k+2})}}\,\,
\frac{X^{l\,'}}{\Gamma(\frac{l'+1}{k+2})}\,.
\ee
Using the normalized operator $X^{l'}_{norm}$ in eq.\,(\ref{inte}) we get 
\be
\int_{\gamma_{l,m,s}}dX X_{norm}^{l'}e^{-W(X)}=
\sqrt{\frac{2}{(k+2)\mbox{Sin}(\pi
\frac{l'+1}{k+2})}}\,\,e^{i\pi s/2}\,
e^{i\pi\frac{m(l'+1)}{k+2}}\,\mbox{Sin}(\pi
\fracs{{(l+1)(l'+1)}}{{k+2}})\,. \nn
\label{pairing}
\ee
We can immediately recognize the r.h.s in eq.\,(\ref{pairing}) as the
coefficient of the Ishibashi state in the expansion of the boundary
state i.e.,
\be
~_{{}_{\rm RR}}\langle l,m,s|X^{l'}\rangle=\sqrt{{2\sqrt{2}}}\,\,\,
 \frac{S^{~l',l'+1,-1}_{l,m,s}}{\sqrt{S^{~l',l'+1,-1}_{0,0,0}}}
 \label{modtr}
\ee
Thus we have found a beautiful realization of the modular
transformation matrix in terms of classical integrals in
the LG theory.\footnote{Computation of boundary entropy
in terms of kinks was carried out in \cite{Fendley}
in a slightly different
context where the modular S-matrix for $SU(1)_k$ appeared in a similar way.
It would be interesting to see whether and how
it is related to the present discussion.}

We can actually check more. Namely we know that the
Ramond states corresponding to chiral fields $|X^{l'}\rangle$
provides a basis where the $R$ charge is diagonal. This  implies
that if we consider a basis for the D-branes, for example the one
given above, $\gamma_{n,n+1}:={\cal L}_{n+1}-{\cal L}_{n}$ 
where $n=0,...,k$ and  compute
the operator $SS^{-t}$ where $S$ is intersection matrix given in
eq.\,(\ref{goodb}), then the corresponding
change of basis to make it diagonal should be given by the matrix
\be
M_{ab}:=~_{{}_{\rm RR}}\langle \gamma_{a,a+1} |X^{b} \rangle \ee where $a,b \in
\{0,\cdots, k\}$. To show this we will calculate
$D:=M^{-1}SS^{-t}M$ and show that it is a diagonal matrix with
eigenvalues equal to the $R$ charge.

From eq.\,(\ref{pairing}) and eq.\,(\ref{modtr})
it follows that the matrix $M$ and its inverse is given by,
\bea
M_{ab}&=&-i \sqrt{\fracs{{2}}{{k+2}}}
\,\,e^{\frac{i \pi (2a+1)(b+1)}{k+2}} \sqrt {\mbox{Sin}(\pi \fracs{{b+1}}{{k+2}})}\,,\\ \nn
(M^{-1})_{ab}&=&\sqrt{\fracs{{2}}{{k+2}}}\sum_{c=0}^{k}S^{c\,b}\,\,e^{-\frac{2\pi i c(a+1)}{k+2}}\,
\sqrt{\mbox{Sin}(\pi \fracs{{a+1}}{{k+2}})}\,.
\eea
Now consider $D_{ab}$,
\bea \nn
D_{ab}&=&\sum_{c,d,g=0}^{k}(M^{-1})_{ac}S_{cd}(S^{-t})_{dg}M_{gb}\,\\ \nn
&=&\fracs{{-2i}}{{k+2}}\sqrt{\mbox{Sin}(\pi \fracs{{a+1}}{{k+2}})\,
\mbox{Sin}(\pi \fracs{{b+1}}{{k+2}})}
\sum_{g,f=0}^{k}e^{-\frac{2\pi i f(a+1)}{k+2}}\,(S^{-t})_{fg}\,
e^{\frac{i\pi (2g+1)(b+1)}{k+2}}\,,\\ \nn
&=&\fracs{{-2i}}{{k+2}}\sqrt{\mbox{Sin}(\pi \fracs{{a+1}}{{k+2}})\,
\mbox{Sin}(\pi \fracs{{b+1}}{{k+2}})}
\sum_{g=0}^{k}\sum_{f \geq g}^{k}e^{-\frac{2\pi i f(a+1)}{k+2}}\,
e^{\frac{i\pi (2g+1)(b+1)}{k+2}}\,,\\ 
&=&\fracs{{-2i}}{{k+2}}\sqrt{\mbox{Sin}(\pi \fracs{{a+1}}{{k+2}})\,
\mbox{Sin}(\pi \fracs{{b+1}}{{k+2}})}\,\,e^{\frac{i\pi(b+1)}{k+2}}\,
\sum_{g=0}^{k}\sum_{f \geq g}^{k}e^{-\frac{2\pi i f(a+1)}{k+2}}\,
e^{\frac{2 \pi i g(b+1)}{k+2}}\,.
\eea
Using the identity
\be
\sum_{e=0}^{k}\sum_{f \geq e}^{k}e^{-\frac{2\pi i f(a+1)}{k+2}}\,
e^{\frac{2 \pi i e(b+1)}{k+2}}= \delta_{a,b} \frac{(k+2) e^{\frac{i \pi(b+1)}{k+2}} }{2\,i\, \mbox{Sin}(\pi\fracs{{b+1}}{{k+2}})}
\ee
we see that 
\be
D_{ab} = -\,e^{\frac{2\pi i (b+1)}{k+2}} \,\,\delta_{a,b}\,,
\ee
which is indeed the spectrum of $\exp(2\pi i R)$ for the ${\cal N}=2$
minimal model. 
This is
morally the analog of the fact that in rational conformal field theory
the modular transformation matrix corresponding to $\tau \rightarrow
-\frac{1}{\tau}$ diagonalizes the fusion algebra $N_{ij}^k$
\cite{verlinde}.  Namely in this
case the intersection matrix $S$ is related to the $N_{ij}^k$ 
coefficients, as already shown, and the $M$ is given
by the overlap of Ishibashi states with chiral fields eq.\,(\ref{modtr})
which is given in terms of the modular transformation matrix
of the rational conformal theory.


\section{Boundary Linear Sigma Models}

In this section, we study $(2,2)$ supersymmetric gauge theories
formulated on a worldsheet with boundary.
We seek for boundary conditions that preserve B-type supersymmetry
and study its relation to non-linear sigma model to which the theory
reduces at low energies.
We also analyze how these boundary conditions are described in the dual
description that was found in \cite{HV}. We include a brief review
of the analysis of \cite{HV}.

\subsection{Supersymmetric Boundary Conditions}

Let us consider a supersymmetric $U(1)$ gauge theory
with chiral multiplets $\Phi_1,\ldots,\Phi_N$ of charge
$Q_1,\ldots,Q_N$.
We formulate the theory on the strip $\Sigma=\R\times I$
where $I$ is a finite interval parametrized by $x^1\in [0,\pi]$
and $\R$ is parametrized by the time coordinate $x^0$. 
The action of the system is given by
\beq
S={1\over 2\pi}
\int\limits_{\Sigma}\Bigl(L_{\it kin}+L_{\it gauge}+L_{{\it FI},\theta}\Bigr)
\dd^2 x.
\label{acti}
\eeq
The terms in the integrand are respectively
the matter kinetic term, gauge kinetic term and
the Fayet-Iliopoulos-Theta term, which are given by
\beqa
&&L_{\it kin}\,=\,-D^{\mu}\bphi D_{\mu}\phi
+{i\over 2}\bpsi_-(\lrD_{\!\! 0}+\lrD_{\!\! 1})\psi_-
+{i\over 2}\bpsi_+(\lrD_{\!\! 0}-\lrD_{\!\! 1})\psi_+
+D|\phi|^2+|F|^2
\nonumber\\
&&~~~~~~~~~\,
-|\sigma|^2|\phi|^2-\bpsi_-\sigma\psi_+-\bpsi_+\bsigma\psi_-
-i\bphi\lambda_-\psi_++i\bphi\lambda_+\psi_-+i\bpsi_+\blambda_-\phi
-i\bpsi_-\blambda_+\phi,
\nonumber\\
&&\\
&&L_{\it gauge}\,=\,
{1\over 2e^2}\,\left[
-\partial^{\mu}\bsigma\partial_{\mu}\sigma
+{i\over 2}\blambda_-(\lrd_{\!\! 0}+\lrd_{\!\! 1})\lambda_-
+{i\over 2}\blambda_+(\lrd_{\!\! 0}-\lrd_{\!\! 1})\lambda_+
+v_{01}^2+D^2
\right],
\nonumber\\
&&\\
&&L_{{\it FI},\theta}\,=\,
-rD+\theta v_{01}.
\eeqa
In the above expressions, the notation
$\bpsi\lrD_{\!\! \mu}\psi=
\bpsi (D_{\mu}\psi)-(D_{\mu}\bpsi)\psi$ is used.
Also, we have written the Lagrangian only in the case
of single matter field of unit charge
($N=1$, $Q_1=1$) to avoid complicated expressions,
but the generalization is obvious.

If there were no boundary of $\Sigma$, the system would be invariant under
$(2,2)$ supersymmetry whose transformation laws are given by
\beqa
&&\delta v_{\pm}=i\bepsilon_{\pm}\lambda_{\pm}
+i\epsilon_{\pm}\blambda_{\pm},
\nonumber\\
&&\delta\sigma=-i\bepsilon_+\lambda_-
-i\epsilon_-\blambda_+,
\nonumber\\
&&\delta D={1\over 2}\Bigl(
-\bepsilon_+(\partial_0-\partial_1)\lambda_+
-\bepsilon_-(\partial_0+\partial_1)\lambda_-
+\epsilon_+(\partial_0-\partial_1)\blambda_+
+\epsilon_-(\partial_0+\partial_1)\blambda_-\Bigr).
\nonumber\\
&&\delta\lambda_+=
i\epsilon_+(D+iv_{01})
+\epsilon_-(\partial_0+\partial_1)\bsigma,
\nonumber\\
&&\delta\lambda_-=
i\epsilon_-(D-iv_{01})
+\epsilon_+(\partial_0-\partial_1)\sigma,
\label{vctSUSY}\\
\mbox{and}&&
\nonumber\\
&&\delta\phi=\epsilon_+\psi_--\epsilon_-\psi_+,
\nonumber\\
&&\delta\psi_+=i\bepsilon_-(D_0+D_1)\phi+\epsilon_+F
-\bepsilon_+\bsigma\phi,
\nonumber\\
&&\delta\psi_-=-i\bepsilon_+(D_0-D_1)\phi+\epsilon_-F
+\bepsilon_-\sigma\phi,
\nonumber\\
&&\delta F=
-i\bepsilon_+(D_0-D_1)\psi_+-i\bepsilon_-(D_0+D_1)\psi_-
+(\bepsilon_+\bsigma\psi_-+\bepsilon_-\sigma\psi_+)
+i(\bepsilon_-\blambda_+-\bepsilon_+\blambda_-)\phi.
\nonumber\\
&&\label{chiSUSY}
\eeqa
In the case where $\Sigma$ has a boundary (where we now consider
the strip $\Sigma=\R\times I$),
the action transforms under (\ref{vctSUSY}) and (\ref{chiSUSY}) as
\beqa
\delta S=
{1\over 4\pi}\int\limits_{\partial\Sigma}
\dd x^0~\Biggl\{\!\!\!\!\!\!\!\!\!\!\!\!\!
&&\,\,\,\,\,
\epsilon_+\Bigl[~T\blambda_+
-(D_0+D_1)\bphi\psi_-+i\bphi\sigma\psi_+
+{i\over 2e^2}\blambda_-(\partial_0+\partial_1)\sigma+i\bpsi_+F~
\Bigr]
\nonumber\\[-0.1cm]
&&+\,\epsilon_-\Bigl[-\overline{T}\blambda_-
-(D_0-D_1)\bphi\psi_++i\bphi\bsigma\psi_-
-{i\over 2e^2}\blambda_+(\partial_0-\partial_1)\bsigma-i\bpsi_-F
\Bigr]
\nonumber\\
&&+\,\bepsilon_+\Bigl[-\overline{T}\lambda_+
+\bpsi_-(D_0+D_1)\phi+i\bpsi_+\bsigma\phi
+{i\over 2e^2}(\partial_0+\partial_1)\bsigma\lambda_-+i\overline{F}\psi_+
\Bigr]
\nonumber\\
&&+\,\bepsilon_-\Bigl[~T\lambda_-
+\bpsi_+(D_0-D_1)\phi+i\bpsi_-\sigma\phi
-{i\over 2e^2}(\partial_0-\partial_1)\sigma\lambda_+-i\overline{F}\psi_-
\Bigr]
~\Biggr\},
\nonumber\\
&&
\label{vari}
\eeqa
where $T$ is defined by
\beq
T=\Biggl(\,r-|\phi|^2-{D\over 2e^2}\,\Biggr)
-i\Biggl(\,\theta+{v_{01}\over 2e^2}\,\Biggr).
\eeq
We look for a boundary condition such that 
B-type supersymmetry generated by
$Q=\bQ_++\e^{i\beta}\bQ_-$ and $Q^{\dag}$ is
unbroken.

Before discussing the detail,
we note that the locality of equation motion for the gauge fields requires
the boundary condition
\beq
{v_{01}\over e^2}=-\theta,~~~\mbox{at}~~\partial\Sigma.
\label{bcv1}
\eeq
Also, the auxiliary fields are solved by
$F=0$ and
\beq
{D\over e^2}=r-|\phi|^2.
\label{solD}
\eeq
If we use these relations we have
$T={1\over 2}(r-|\phi|^2-i\theta)={1\over 2e^2}(D+iv_{01})$
at $\partial\Sigma$.

We also make an important remark.
Since $v_{01}=\partial_0v_1-\partial_1v_0$ is a total
derivative,
the Theta term of the action can na\"\i vely be written as the boundary term
\beq
{\theta\over 2\pi}\int\limits_{\Sigma} v_{01}\dd^2 x
\stackrel{?}{=}-{\theta\over 2\pi}\int\limits_{\partial\Sigma}v_0 \dd x^0.
\label{thea}
\eeq
However, $v_0$ is not gauge invariant whereas $v_{01}$ is.
In particular, when the boundary components are
 compactified on circles,
the right hand side changes by integer multiples of $\theta$
under gauge transformations.
Thus, for a generic $\theta$, (\ref{thea}) is not an allowed thing to do. 
If $\theta$ is an integer multiple of $2\pi$, however,
the right hand side of (\ref{thea}) is gauge invariant
up to $2\pi$ shifts so that
$\exp\Bigl(i{\theta\over 2\pi}\oint v \Bigr)$
is well-defined.
Thus, only for those cases, the manipulation (\ref{thea}) is allowed.
More generally, for a general $\theta$ one can shift
$\theta\to\theta-2\pi n$, with $n$ integer, provided the boundary
term $-n\int_{\partial\Sigma}v_0 \dd x^0$ is added to the action.

\subsection*{\it Pure Maxwell Theory}

We start the study with the simplest case; without
matter.
In this case, the theory has a single twisted chiral (gauge) multiplet
$\Sigma$ with the twisted superpotential
\beq
\widetilde{W}=-t\Sigma,
\label{twistW}
\eeq
where $t$ is the complex combination of the FI and Theta parameters
\beq
t=r-i\theta.
\eeq
B-type boundary condition for twisted chiral multiplet fields
is like A-type boundary condition for chiral multiplet fields.
In particular, the world sheet boundary
must end on a middle dimensional Lagrangian submanifold whose image
in the $\widetilde{W}$-plane is a straight line.
Since (\ref{twistW}) is linear in $\Sigma$, this means that
the worldsheet boundary must end on a straight line in $\Sigma$.
The Lagrangian condition is trivially satisfied.
Thus, if we denote the phase of the FI-Theta parameter $t$ as
\beq
t=|t|\e^{i\gamma},
\eeq
the B-type supersymmetric D-brane is the straight line
in the $\sigma$-plane whose slope is given by $-\gamma$;
\beq
{\rm Im}\bigl(\e^{i\gamma}\sigma\bigr)={\rm constant}~~~\mbox{at}~~
\partial\Sigma.
\eeq
The boundary condition on the component fields is given by
\beq
\begin{array}{l}
\e^{i\gamma}(\partial_0+\partial_1)\sigma
=\e^{-i\gamma}(\partial_0-\partial_1)\bsigma,
\\[0.2cm]
\e^{-i\gamma}\lambda_++\e^{i\gamma}\lambda_-=0,
\\[0.2cm]
\e^{i\gamma}\blambda_++\e^{-i\gamma}\blambda_-=0,
\end{array}~~~\mbox{at}~~\partial\Sigma.
\label{Bbcvct}
\eeq
It is indeed easy to check that the variation (\ref{vari})
vanishes for B-type supersymmetry with
$\epsilon_-=-\epsilon_+$.
For supersymmetry with 
$\epsilon_-=-\e^{i\beta}\epsilon_+$,
we only have to make the replacement $\sigma\to\e^{-i\beta}\sigma$,
$\lambda_{\pm}\to\e^{\pm i\beta/2}\lambda_{\pm}$ in the above expressions.

The zero point energies of $\sigma$ and $\lambda_{\pm}$
cancel against each other and the vacuum energy of the system
comes purely from the gauge field sector.
By the equation of motion (or a Gauss law constraint)
$\partial_1v_{01}=0$,
the field strength $v_{01}$ is a constant
and by the boundary condition (\ref{bcv1}) it is given by
$v_{01}=-e^2\theta$.
The vacuum energy is then given by
\beq
E_0=\pi {e^2|t|^2\over 2}.
\eeq
In particular, the supersymmetry is spontaneously broken
if $t\ne 0$
as can be seen also by the supersymmetry transformation of
$\lambda_{\pm}$ in (\ref{vctSUSY}) (where
$D\pm iv_{01}=e^2(r\mp i\theta)$ by the constraint).
All these are the same as the
elements of the standard story in the bulk theory.

\subsection*{\it The General Case}

Let us now consider the case with matters.
It is known that, under certain conditions,
the bulk theory can be identified as
a non-linear sigma model at low enough energies compared to
$e\sqrt{r}$ (see for example \cite{phases,MP,HV}).
The target space $X$ is a toric manifold defined as
the solution space to $\sum_{i=1}^NQ_i|\phi_i|^2=r$ modded out by the
$U(1)$ gauge transformations.
We look for the boundary conditions corresponding to D-branes
wrapping totally on $X$ (with or without coupling to
gauge fields on $X$).

Theta angle in the gauge theory
is usually identified as the $B$-field.
In non-linear sigma models,
as we have seen in section 3, $B$-field modifies
the boundary condition on the coordinate fields as (\ref{locB}),
from pure Neumann to mixed Dirichlet-Neumann condition.
However, in the gauge theory with action (\ref{acti}),
the condition remains pure Neumann type
$D_1\phi=0$ even if we turn on $\theta$.
Thus, there appears to be a discrepancy between the gauge theory
and the non-linear sigma model when formulated on a worldsheet
with boundary.
This mismatch can be cured by adding the boundary term
\beq
S_{\it boundary}=
{\theta\over 4\pi r}\int\limits_{\partial\Sigma}
\Bigl(\, iD_0\bphi\,\phi-i\bphi D_0\phi\,\Bigr)\,\dd x^0
\label{Sbo}
\eeq
to the action (\ref{acti}).
Then, the boundary condition required from the
locality of equation of motion becomes
\beq
\cos(\gamma)\, D_1\phi-i\sin(\gamma)\, D_0\phi=0,~~~\mbox{at}~~
\partial\Sigma,
\label{bbc}
\eeq
where $t=r-i\theta=|t|\e^{i\gamma}$. This corresponds to
the mixed Dirichlet-Neumann boundary condition of (\ref{locB}).
We note that the addition of (\ref{Sbo}) also alters
the boundary condition (\ref{bcv1}) for the gauge field as
\beq
{v_{01}\over e^2}=-\theta+{|\phi|^2\over r}\theta.
\label{bcv2}
\eeq

The total action
\beq
S_{\rm tot}=S+S_{\it boundary}
\eeq
transforms under
(\ref{vctSUSY}) and (\ref{chiSUSY}) as
\beqa
\delta S_{\rm tot}\!=
{1\over 4\pi}\int\limits_{\partial\Sigma}
\dd x^0~\Biggl\{\!\!\!\!\!\!\!\!\!\!\!\!\!
&&\,\,\,\,\, 
\epsilon_+\Bigl[~\widetilde{T}\blambda_+
-\Bigl((1-{\textstyle{2i\theta\over r}})D_0+D_1\Bigr)\bphi\psi_-
+i\bphi\sigma\psi_+
+{i\over 2e^2}\blambda_-(\partial_0+\partial_1)\sigma\,
\Bigr]
\nonumber\\[-0.1cm]
&&+\,\epsilon_-\Bigl[-\overline{\widetilde{T}}\blambda_-
-\Bigl((1+{\textstyle{2i\theta\over r}})D_0-D_1\Bigr)\bphi\psi_+
+i\bphi\bsigma\psi_-
-{i\over 2e^2}\blambda_+(\partial_0-\partial_1)\bsigma
\Bigr]
\nonumber\\
&&+\,\bepsilon_+\Bigl[-\overline{\widetilde{T}}\lambda_+
+\bpsi_-\Bigl((1+{\textstyle{2i\theta\over r}})D_0+D_1\Bigr)\phi
+i\bpsi_+\bsigma\phi
+{i\over 2e^2}(\partial_0+\partial_1)\bsigma\lambda_-
\Bigr]
\nonumber\\
&&+\,\bepsilon_-\Bigl[~\widetilde{T}\lambda_-
+\bpsi_+\Bigl((1-{\textstyle{2i\theta\over r}})D_0-D_1\Bigr)\phi
+i\bpsi_-\sigma\phi
-{i\over 2e^2}(\partial_0-\partial_1)\sigma\lambda_+
\Bigr]
~\Biggr\}.
\nonumber\\
&&
\label{variation}
\eeqa
Here $\widetilde{T}$ is given by
\beqa
\widetilde{T}&=&\Biggl(\,r-|\phi|^2-{D\over 2e^2}\,\Biggr)
-i\Biggl(\,\theta\Bigl(1-{|\phi|^2\over r}\Bigr)
+{v_{01}\over 2e^2}\,\Biggr)\nonumber\\
&=&
{r-|\phi|^2\over 2r}\Bigl(\,r-i\theta\,\Bigr),
\eeqa
where
we have used (\ref{solD}) and the new boundary condition
(\ref{bcv2}) in the second equality.
Since $\widetilde{T}$ is proportional to $t=r-i\theta$ as
in the pure Maxwell theory, it is obvious that
$\Phi$ independent part of the variation (\ref{variation})
vanishes for $\epsilon_-=-\epsilon_+$
under the same condition (\ref{Bbcvct})
as in the Maxwell theory.
We are now left with the following
terms (for $\epsilon_-=-\epsilon_+$)
\beqa
&&\epsilon_+\Bigl[
-\Bigl((1-{\textstyle{2i\theta\over r}})D_0+D_1\Bigr)\bphi\psi_-
+\Bigl((1+{\textstyle{2i\theta\over r}})D_0-D_1\Bigr)\bphi\psi_+
+i\bphi(\sigma\psi_+-\bsigma\psi_-)\,
\Bigr]\nonumber\\
&&+\bepsilon_+\Bigl[
~\bpsi_-\Bigl((1+{\textstyle{2i\theta\over r}})D_0+D_1\Bigr)\phi
-\bpsi_+\Bigl((1-{\textstyle{2i\theta\over r}})D_0-D_1\Bigr)\phi
+i(\bpsi_+\bsigma-\bpsi_-\sigma)\phi\,\Bigr].
\nonumber
\eeqa
The non-derivative terms vanish if the straight line
of $\sigma$ is of the type:
\beq
{\rm Im}\bigl( \e^{i\gamma}\sigma\bigr)=0~~~\mbox{at}~~\partial\Sigma,
\label{str}
\eeq
and the matter fermions satisfy
the boundary condition
\beq
\begin{array}{l}
\e^{-i\gamma}\psi_+=\e^{i\gamma}\psi_-,\\[0.2cm]
\e^{i\gamma}\bpsi_+=\e^{-i\gamma}\bpsi_-,
\end{array}
~~~~\mbox{at}~~\partial\Sigma.
\label{fbc}
\eeq
It is now straightforward to see that the derivative terms also vanish
under the boundary conditions (\ref{bbc}) and (\ref{fbc}).
It is also easy to see that these boundary conditions (including
(\ref{str})) are
invariant under the B-type supersymmetry.

To summarize, the total action $S_{\rm tot}$
has B-type supersymmetry with
$\epsilon_-=-\epsilon_+$ under the boundary conditions
(\ref{bbc}) and (\ref{fbc}) for the matter fields
and (\ref{Bbcvct}), (\ref{str}), and (\ref{bcv2}) for the gauge multiplet
fields.
These conditions reduce to the ordinary mixed Dirichlet-Neumann boundary
conditions (\ref{locB}) and (\ref{N1B}) of
the non-linear sigma model on $X$.
To recover the phase, $\epsilon_-=-\e^{i\beta}\epsilon_+$,
it is enough to make the replacement
$\psi_{\pm}\to\e^{\pm i\beta/2}\psi_{\pm}$,
$\sigma\to\e^{-i\beta}\sigma$ and
$\lambda_{\pm}\to\e^{\pm i\beta/2}\lambda_{\pm}$.

So far we have been analyzing the boundary condition of
the classical theory.
There are two important quantum effects of the theory
with $\sum_{i=1}^NQ_i\ne 0$;
 the running of the FI parameter
$r$ and the anomaly of the axial $U(1)$ R-symmetry.
 From the running of $r$,
$r_0=\sum_{i=1}^NQ_i\log(\Lambda_{\rm UV}/\Lambda)$,
the phase
$\e^{i\gamma}=t/|t|$
which enters in the boundary condition
changes along the renormalization group flow.
In particular, if $\sum_{i=1}^NQ_i>0$
(which corresponds to an asymptotic free sigma model),
the ``bare phase'' becomes trivial
$\e^{i\gamma_0}\to 1$
in the continuum limit $\Lambda_{\rm UV}/\Lambda\to\infty$.
Also, by the axial anomaly, the axial rotation
can be done not just by the replacement
$\psi_{\pm}\to\e^{\pm i\beta/2}\psi_{\pm}$,
$\sigma\to\e^{-i\beta}\sigma$ and
$\lambda_{\pm}\to\e^{\pm i\beta/2}\lambda_{\pm}$
but {\it together with the shift of the Theta angle}
$\theta\to\theta+\sum_{i=1}^NQ_i\beta$.
These effects should be visible in a quantum effective description.
Here we look at the effective action in terms of $\Sigma$-field
whose scalar component is chosen to have large expectation values. 
This is obtained by integrating out the charged matter fields
and is given (for $Q_i=1$ case)
by
\beq
\widetilde{W}=-N\Sigma(\log\Sigma-1)-t\Sigma.
\label{SlogS}
\eeq
This yields the following effective FI-Theta parameter
\beq
t_{\it eff}=t+N\log\Sigma,
\eeq
where the energy scale is set by the value of $\Sigma$.
This effective theory
is essentially the LG model with the superpotential (\ref{SlogS})\footnote{
Strictly speaking, the theory involves a gauge field.
However, in the absence of light or tachyonic charged matter field,
the effect of the gauge field is simply to create the
vacuum energy $e^2({\rm Im}\,t_{\it eff})^2/2$,
as the standard auxiliary field does.
There is actually a (minor) subtlety; If the theory is formulated on $\R^2$,
the physics is periodic in $\theta$ which is identified as the
constant electric field (divided by $e^2$). This is because of the pair
creation of the electron and positron \cite{coleman} which
run away to opposite infinity in the space.
However, if the theory is formulated on a strip, $\R\times [0,\pi]$,
the electron positron pair, even if they are pair-created,
can never run away to infinity. Thus, the physics is not periodic in
$\theta$.}
which has $N$ non-degenerate
critical points $\Sigma_a=\e^{-t/N+2\pi a i/N}$
($a=0,\ldots,N-1$).
As we have seen, a D-brane preserving the B-type
supercharge $Q=\bQ_++\bQ_-$ is the preimage of the straight line
in the $\widetilde{W}$-plane.
The equation is given by
\beq
{\rm Im}\Bigl(\e^{i\gamma_{\it eff}(\sigma)}\sigma\Bigr)=
{\rm constant},
\eeq 
where $t_{\it eff}-N=\e^{i\gamma_{\it eff}}|t_{\it eff}-N|$.
If we insist the straight line to pass through
a critical value $\widetilde{W}(\Sigma_a)\sim N\e^{-t/N}$,
the constant in the r.h.s.
is of order $\e^{-t/N}$ and can be considered as the
correction to the condition (\ref{str}).
It is in general a non-trivial task to find the explicit
solution to the straight line equation.
However, there is a trivial one if the Theta angle vanishes
$\theta=0$. In this case $\sigma=|\sigma|$
is a solution to the straight line equation with the zero slope
$\e^{i\beta}=1$.
By the axial rotation $\sigma\to \e^{-i\beta}\sigma$,
$\lambda_{\pm}\to \e^{\pm i\beta/2}\lambda_{\pm}$, we obtain the
solution $\sigma=\e^{i\beta}|\sigma|$ with the slope $\beta$.
However, we should note that this axial rotation
shifts the Theta angle from zero to $\theta=N\beta$.
Indeed the image of
$\sigma=\e^{i\beta}|\sigma|$
in the $\widetilde{W}$-plane is a straight line only when this shift
is made.
Thus, we have seen that there is a one parameter family of
explicit solutions
\beq
\begin{array}{l}
\sigma=\e^{i\theta/N}|\sigma|,\\[0.1cm]
\e^{i\theta/2N}\lambda_++\e^{-i\theta/2N}\lambda_-=0,\\[0.1cm]
\e^{-i\theta/2N}\blambda_++\e^{i\theta/2N}\blambda_-=0,
\end{array}
~~~\mbox{at}~~\partial\Sigma,
\label{qcond}
\eeq
parametrized by the worldsheet Theta angle $\theta$.
This preserves the supercharge
\beq
Q=\bQ_++\e^{i\theta/N}\bQ_-
\label{Qcomb}
\eeq
and $Q^{\dag}$.
There are of course
other solutions (especially those with $\beta\ne \theta/N$)
but the quantum correction is non-trivial
and it is not easy to determine them explicitly.
It is easy to extend the above solutions to the general $Q_i$'s:
replace $N$ in these formulae by $\sum_{i=1}^NQ_i$.

There is actually a better quantum effective description of the bulk theory
found in \cite{HV},
using the dual variables $Y_i$ of the charged fields $\Phi_i$.
Later in this section and further in the next section,
we will see how the boundary condition is described
in that theory. This will lead to the map of D-branes under mirror
symmetry.

\subsection*{\it Coupling to Gauge Fields on $X$}

So far, we have been considering a gauge theory
that corresponds to
the non-linear sigma model on $X$ with a $B$-field, but not
including coupling the worldsheet boundary to the
target space gauge fields.
Now it is useful to observe that
the $B$-field obeying a certain quantization
condition can be considered as the curvature of a gauge field $A_I$
on $X$. In such a case, as noted in section 3,
the coupling to $B$ field is equal to
the boundary coupling to the gauge field $A_I$.
The quantized $B$ field
corresponds to the case where the
worldsheet Theta angle becomes an integer multiple of $2\pi$,
$\theta=2\pi n$. We now recall that in such a case
(and only in such a case)
the worldsheet Theta term can be converted into a boundary term
(\ref{thea}). Then, the total boundary term becomes
\beqa
S_{\rm boundary}^{\prime}&=&
{n\over 2 r}\int\limits_{\partial\Sigma}
\Bigl(\, iD_0\bphi\,\phi-i\bphi D_0\phi\,\Bigr)\,\dd x^0
-n\int\limits_{\partial\Sigma}v_0 \,\dd x^0
\nonumber\\
&=&{n\over 2 r}\int\limits_{\partial\Sigma}
\Bigl(\, i\partial_0\bphi\,\phi-i\bphi \partial_0\phi
+2v_0(|\phi|^2-r)\,\Bigr)\,
\dd x^0
\label{Sbop}
\eeqa
In the sigma model limit $e\sqrt{r}\to\infty$,
the constraint $|\phi|^2=r$ is strictly imposed.
Then, the boundary term is given by
\beq
S_{\rm boundary}^{\prime}
=-n\int\limits_{\partial\Sigma}A_I\partial_0\phi^I\dd x^0
\eeq
where
\beq
A_I\dd\phi^I=
{i\over 2}{\displaystyle~\sum \!{}_{i=1}^N
\bphi_i\lrdd \phi_i~\over
\displaystyle \sum \!{}_{i=1}^NQ_i|\phi_i|^2}.
\label{hermi}
\eeq
In this expression,
we have recovered all the $N$ matter fields of charge $Q_1,\ldots,Q_N$
where the constraint is $\sum_{i=1}^NQ_i|\phi_i|^2=r$.

The gauge field $A_I$ in (\ref{hermi})
is nothing but the hermitian connection of the
natural holomorphic line bundle ${\cal O}_X(1)$
on the toric manifold $X$ (where $\phi_i$'s represent
the sections) with respect to
the natural hermitian metric.
To see this, let us make a gauge transformation
$\phi_i\to \e^{iQ_i\lambda}\phi_i$. Then, the gauge field transforms as
\beq
A_I\dd \phi^I\to
A_I\dd \phi^I-\dd\lambda.
\eeq
This is indeed the transformation property of a connection form
of the bundle ${\cal O}_X(1)$.
For example,
let us consider the simplest case $X=\CP^1$ where the gauge theory
has two matters $\Phi_1$, $\Phi_2$ of charge $1$.
In the gauge where $\phi_1=1$ and $\phi_2=z$,
the gauge field (\ref{hermi}) is given by
\beq
A={i\over 2}{\bz\dd z-z\dd\bz\over 1+|z|^2}.
\eeq
This is the gauge field of the line bundle ${\cal O}(1)$ of $\CP^1$.
Indeed, the first Chern class is represented by the curvature
${i\over 2\pi}\cdot i\dd A={i\over 2\pi}\dd z\dd \bz/(1+|z|^2)^2$
which is the positive unit volume form of $\CP^1$.

Thus, we indeed see that
the boundary term (\ref{Sbop}) corresponds to the boundary coupling
to the natural gauge fields of the bundle ${\cal O}_X(-n)$.
Here we have to bear in mind that the boundary condition
should be given by
(\ref{bbc})-(\ref{fbc})
and (\ref{Bbcvct})-(\ref{str})-(\ref{bcv2})
where it is understood that $\gamma=\arg(r-2\pi n i)$.
If we turn on the bulk $\theta$-term anew, the angle is given by
$\gamma=\arg(t-2\pi ni)$ where $t=r-i\theta$.

\subsection*{\it Alternative Formulation}

In the non-linear sigma model,
we have seen that there is an alternative formulation
for coupling to target space gauge fields
where we do not change the boundary condition
but add a fermion bilinear boundary term.
This can also be done in the gauge theory.
The relevant boundary term for the gauge field of the bundle
${\cal O}_X(-n)$ is given by
\beq
S_{\rm boundary}^{\prime\prime}
={n\over 2 r}\int\limits_{\partial\Sigma}
\Bigl(\, iD_0\bphi\,\phi-i\bphi D_0\phi
+(\psi_++\psi_-)(\bpsi_++\bpsi_-)
-(\sigma+\bsigma)|\phi|^2
\,\Bigr)\,\dd x^0.
\eeq
It is straightforward to check that this is by itself
invariant under the B-type
supersymmetry with $\epsilon_-=-\epsilon_+$.
Thus, one can add this to the total action $S_{\rm tot}$
without changing the boundary condition.
We note that, as in non-linear sigma models,
the equations of motion for the worldsheet
fields have boundary contributions in this formulation.

\subsection{A Review of a Derivation of Mirror Symmetry}

We now briefly review the dual description of the gauge theory
found in \cite{HV}.

Let us consider the $U(1)$ gauge theory on $\Sigma=\R^2$
with matters of charge $Q_1,\ldots,Q_N$.
The action is given by (\ref{acti}) (in the case $N=1,Q_1=1$).
In the superfield notation, the Lagrangian is expressed as
\beq
L=\int\dd^4\theta
\left(\,
\sum_{i=1}^N\bPhi_i\,\e^{2Q_iV}\Phi_i
-{1\over 2e^2}\bSigma\Sigma
\,\right)
+{1\over 2}\Biggl(
-\int\dd^2\widetilde{\theta}~t\,\Sigma
\,+\, c.c.\,
\Biggr).
\label{lag}
\eeq

If we dualize the phase of the charged chiral superfield $\Phi_i$,
we obtain a neutral twisted chiral superfield $Y_i$ that 
is periodic with periodicity $2\pi i$,
$Y_i\equiv Y_i+2\pi i$.
The fields $Y_i$ are related to the original charged chiral
superfields $\Phi_i$ by
\beq
Y_i+\overline{Y}_i=2\overline{\Phi_i}\e^{2Q_iV}\Phi_i,
\eeq
or in components,
$Y_i=y_i+\sqrt{2}\theta^+\bchi_{i+}+\sqrt{2}\btheta^-\chi_{i-}+\cdots$,
\beqa
&&y_i=\varrho_i-i\vartheta_i,~~
\left\{
\begin{array}{l}
\varrho_i=|\phi_i|^2,\\
\partial_{\pm}\vartheta_i=
\pm 2\Bigl(-|\phi_i|^2
(\partial_{\pm}\varphi_i+Q_iv_{\pm})
+\bpsi_{i\pm}\psi_{i\pm}\Bigr),
\end{array}\right.
\nonumber\\
&&\chi_{i+}=2\bpsi_{i+}\phi_i,~~\chi_{i-}=-2\bpsi_{i-}\phi_i,
\label{bcchi}\\
&&\bchi_{i+}=2\bphi_i\psi_{i+},~~\bchi_{i-}=-2\bphi_i\psi_{i-},
\nonumber
\eeqa
where $\varphi_i$ is the phase of $\phi_i$,
$\phi_i=|\phi_i|\,\e^{i\varphi_i}$.

The fields $Y_i$ couple to
the gauge field as dynamical Theta angle.
Thus, at the level of dualization we have the twisted superpotential
$\widetilde{W}=\Sigma(\sum_{i=1}^NQ_iY_{i0}-t_0)$
where subscript $0$ stands for the bare parameters and fields.
The FI parameter runs as $r_0=b_1\log(\Lambda_{UV}/\Lambda)$
with
\beq
b_1:=\sum_{i=1}^NQ_i,
\eeq
but one can make the superpotential finite by renormalizing the
bare fields
$\varrho_{i0}$ as $\varrho_{i0}=\varrho_i+\log(\Lambda_{UV}/\mu)$,
where $\mu$ is the renormalization point.
The K\"ahler metric of the $y_i$ variables is given classically
by
\beq
\dd s^2=\sum_{i=1}^N{|\dd y_i|^2\over
2(2r_0/b_1+y_i+\overline{y}_i)}\simeq
{b_1\over 4r_0}\sum_{i=1}^N\,|\dd y_i|^2.
\label{Kmet}
\eeq

This superpotential is corrected by instanton effect
where the instantons are the vortices of the gauge theory.
The correction is of the form $\e^{-Y_i}$ and
the exact twisted superpotential is given by
\beq
\widetilde{W}=\Sigma\left(\sum_{i=1}^NQ_iY_i-t(\mu)\right)
+\sum_{i=1}^N\mu\,\e^{-Y_i}.
\label{Weff}
\eeq
In the case $b_1\ne 0$, the FI parameter is renormalized
and $\Lambda=\mu\e^{-t/b_1}$
is renormalization group invariant,
as can be seen also from (\ref{Weff}) using the shifts of $Y_i$'s.
In the conformal case $b_1=0$,
$t$ is the dimensionless parameter of the theory,
and $\mu$ can be simply absorbed by the shifts of $Y_i$'s.
In what follows, we omit the scale $\mu$.

In the sigma model limit $e\sqrt{r}\to\infty$,
the gauge multiplet fields becomes infinitely heavy and can be 
integrated out.
Then, this yields a constraint
\beq
\sum_{i=1}^NQ_iY_i=t.
\label{constr}
\eeq
Thus, we obtain a theory of $N$ periodic fields $Y_i$ with one constraint
(\ref{constr}) which has a twisted superpotential
\beq
\widetilde{W}=\sum_{i=1}^N\e^{-Y_i}.
\label{dualW}
\eeq
In other words, we obtain a LG model on $(\C^{\times})^{N-1}$.
Since the original gauge theory becomes non-linear sigma model
on the toric manifold $X$ in the limit $e\sqrt{r}\to\infty$,
the above LG model is a
dual description of the non-linear sigma model on $X$.
Since it is described by twisted chiral fields,
it is the mirror of the sigma model on $X$.

It is easy to find the critical point of the superpotential
(\ref{dualW}) under the constraint (\ref{constr}).
There are $b_1$
critical points $p_0,\ldots,p_{b_1-1}$, where at the $a$-th critical point
$\e^{-y_i}(p_a)=Q_i\e^{-t/b_1+2\pi ai/b_1}\prod_{j=1}^NQ_j^{-Q_j}$
with the critical value
\beq
\widetilde{w}_a=b_1\e^{-t/b_1+2\pi ai/b_1}\prod_{j=1}^NQ_j^{-Q_j}.
\eeq
All these are massive vacua at which
the $\Z_{2b_1}$ axial R-symmetry is
spontaneously broken to $\Z_2$.

\subsection{D-branes and Mirror Symmetry: First Example}

We would like to see how the boundary conditions for the linear sigma model
can be described in the quantum effective theory
in terms of the dual variables.
We consider the model with
\beq
b_1=\sum_{i=1}^NQ_i> 0,
\eeq
that corresponds to an asymptotic free non-linear sigma model.
Since the dual theory is a LG model described in terms of twisted chiral
superfields, B-type supersymmetry looks like
A-type supersymmetry for chiral superfields.
In particular, the worldsheet boundary must end on a middle dimensional
Lagrangian submanifold of $(\C^{\times})^{N-1}$
that is mapped to a straight line in the $\widetilde{W}$-plane.

We focus on the family of boundary conditions (\ref{qcond})
parametrized by the worldsheet Theta angle $\theta$.
The boundary conditions on the matter fields are
\beq
\begin{array}{l}
\cos(\gamma_0)\, D_1\phi_i-i\sin(\gamma_0)\, D_0\phi_i=0,\\[0.2cm]
\e^{-i\gamma_0+i\theta/2b_1}\psi_{i+}=\e^{i\gamma_0-i\theta/2b_1}\psi_{i-},
\\[0.2cm]
\e^{i\gamma_0-i\theta/2b_1}\bpsi_{i+}=\e^{-i\gamma_0+i\theta/2b_1}\bpsi_{i-},
\end{array}
~~~~\mbox{at}~~\partial\Sigma.
\label{bcth}
\eeq
We note that the phase $\e^{i\gamma_0}$
defined by $r_0-i\theta=\e^{i\gamma_0}|r_0-i\theta|$
becomes trivial
\beq
\gamma_0\to 0,
\eeq
in the continuum limit $\Lambda_{\rm UV}\to\infty$
where $r_0=b_1\log(\Lambda_{\rm UV}/\Lambda)\to\infty$.
This boundary condition preserves the supercharge
\beq
Q=\bQ_++\e^{i\theta/b_1}\bQ_-
\label{sQ}
\eeq
and its conjugate $Q^{\dag}$.

We recall that there is a boundary term in the action
\beqa
S_{\it boundary}&=&
{\theta\over 4\pi r_0}\int\limits_{\partial\Sigma}
\sum_{i=1}^N\Bigl(\, iD_0\bphi_i\,\phi_i-i\bphi_i D_0\phi_i\,\Bigr)
\,\dd x^0
\nonumber\\
&=&{\theta\over 2\pi r_0}
\int\limits_{\partial\Sigma}
\sum_{i=1}^N\,|\phi_i|^2(\partial_0\varphi_i+Q_iv_0)
\,\dd x^0.
\eeqa
Note that in terms of the renormalized dual fields we have
$|\phi_i|^2=r_0/b_1+\varrho_i$.
Then, in the continuum limit the boundary term can be written as
\beq
S_{\it boundary}=
{\theta\over 2\pi}
\int\limits_{\partial\Sigma}
\left({1\over b_1}\sum_{i=1}^N\partial_0\varphi_i\,+\,v_0\,\right)\dd x^0.
\eeq

Now the relevant part of the action in the dualization is
\beq
S_{\varphi}
={1\over 2\pi}\int\limits_{\Sigma}
\sum_{i=1}^Nr_0^2|\dd\varphi_i+Q_iv|^2
-{i\theta\over 2\pi}\int\limits_{\partial\Sigma}
\Bigl({1\over b_1}\sum_{i=1}^N\dd\varphi_i\,+\,v\,\Bigr)
\label{origi}
\eeq
where we consider Euclidean signature (for simplicity)
and we ignore the fermionic components which are not essential
in this part of the argument.
We consider another system involving one-form fields
${\cal B}_i={\cal B}_{i\mu}\dd x^{\mu}$ with the action given by
\beq
S^{\prime}=\sum_{i=1}^N\left[
{1\over 8\pi r_0^2}\int\limits_{\Sigma}{\cal B}_i\wedge *{\cal B}_i
+{i\over 2\pi}\int\limits_{\Sigma}
{\cal B}_i\wedge (\dd\varphi_i+Q_iv)\right]
-{i\theta\over 2\pi}\int\limits_{\partial\Sigma}
\Bigl({1\over b_1}\sum_{i=1}^N\dd\varphi_i\,+\,v\,\Bigr).
\label{dualiz}
\eeq
We require the boundary condition on the one-form fields ${\cal B}_i$
that they vanish against the tangent vectors of the boundary
\beq
{\cal B}_i|_{\partial\Sigma}=0.
\label{bcB}
\eeq
If we first integrate out the one-form field
${\cal B}_i$, we obtain the constraint
${\cal B}_i=i2r_0^2*(\dd\varphi_i+Q_iv)$
(which is consistent with the boundary condition (\ref{bcth})
in the continuum limit) and we obtain the original action
(\ref{origi}).
Instead, if we first integrate out the variables $\varphi_i$,
we obtain the constraint
\beq
{\cal B}_i=\dd\vartheta_i
\label{Bth}
\eeq
where $\vartheta_i$ are periodic variables of period $2\pi$.
By the boundary condition (\ref{bcB}), we see that
$\vartheta_i$ are constants along the boundary of $\partial\Sigma$.
By the terms $-i(\theta/2\pi b_1)\int_{\partial\Sigma}\dd\varphi_i$
in the action (\ref{dualiz}), we see that the constants are
\beq
\vartheta_i={\theta/b_1}~~~\mbox{at}~~\partial\Sigma,
\label{bcthea}
\eeq
for all $i$.
Now, if we plug the constraint (\ref{Bth}) back into (\ref{dualiz})
we obtain the action
\beqa
S_{\vartheta}
&=&\sum_{i=1}^N\left[
{1\over 8\pi r_0^2}\int\limits_{\Sigma}|\dd\vartheta_i|^2
+{i\over 2\pi}\int\limits_{\Sigma}\dd\vartheta_i\wedge Q_iv\,\right]
-{i\theta\over 2\pi}\int\limits_{\partial\Sigma}v
\nonumber\\
&=&\sum_{i=1}^N\left[
{1\over 8\pi r_0^2}\int\limits_{\Sigma}|\dd\vartheta_i|^2
-{i\over 2\pi}\int\limits_{\Sigma}Q_i\vartheta_i\dd v\,\right]
+{i\over 2\pi}\int\limits_{\partial\Sigma}
\Bigl(\sum_{i=1}^NQ_i\vartheta_i-\theta\,\Bigr)v,
\eeqa
where we have performed partial integrations in the last step.
Note that the boundary term in the right hand side
vanishes if we use the boundary condition (\ref{bcthea}).
Then, the dualization proceeds precisely as in the bulk theory and we will
obtain the superpotential (\ref{Weff}).
In the sigma model limit $e^2\sqrt{r}\to\infty$,
after integrating out the $\Sigma$-field, we obtain the
constraint (\ref{constr}) and the twisted superpotential
(\ref{dualW}).

The boundary condition (\ref{bcthea}) means that $\e^{-y_i}$'s at the
boundary have a fixed
common phase $\theta/b_1$.
Namely, the worldsheet boundary $\partial\Sigma$
is mapped by $(\e^{-y_i})$ to a real $(N-1)$-dimensional cycle
$\cycle_{\theta}$ in the algebraic torus
$(\C^{\times})^{N-1}$ defined by
\beq
(\e^{-y_1},\ldots,\e^{-y_N})
=(\e^{-\varrho_1+i\theta/b_1},\ldots,\e^{-\varrho_N+i\theta/b_1}),
\label{real}
\eeq
where $(\varrho_1,\ldots,\varrho_N)$ are the real coordinates
constrained by $\sum_{i=1}^NQ_i\varrho_i=0$.
By the boundary condition for $\phi_i$
in (\ref{bcth}),
we see that the tangent coordinates $\varrho_i$
obey the Neumann boundary condition
\beq
\partial_1\varrho_i=0~~~\mbox{at}~~\partial\Sigma,
\eeq
in the continuum limit.
The boundary condition on the fermionic components can be read from
(\ref{bcchi}) and (\ref{bcth}) and is given by
\beq
\begin{array}{l}
\e^{-i\theta/2b_1}\chi_{i+}+\e^{i\theta/2b_1}\chi_{i-}=0,\\[0.1cm]
\e^{i\theta/2b_1}\bchi_{i+}+\e^{-i\theta/2b_1}\bchi_{i-}=0,
\end{array}~~~\mbox{at}~~\partial\Sigma.
\label{bcfc}
\eeq
These are the standard boundary condition on the worldsheet
fields corresponding to the D-brane wrapped on $\cycle_{\theta}$.
The phases $\e^{i\theta/2b_1}$ in the condition for the
fermionic components shows that we are performing an
R-rotation.

\begin{figure}[htb]
\begin{center}
\epsfxsize=5.2in\leavevmode\epsfbox{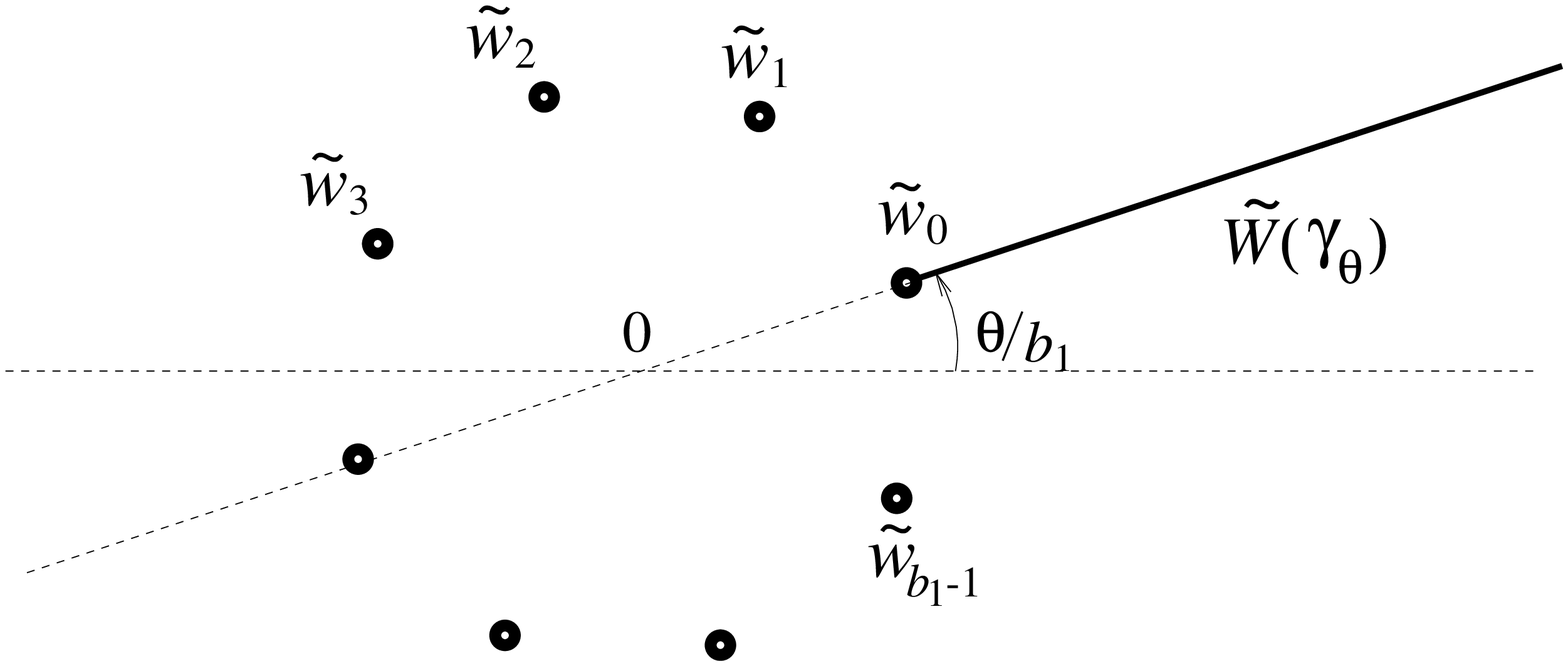}
\end{center}
\caption{The image in the $\widetilde{W}$-plane of
the cycle $\cycle_{\theta}$.}
\label{cyclestr}
\end{figure}

The cycle $\cycle_{\theta}$
is a Lagrangian submanifold of $(\C^{\times})^{N-1}$
with respect to the flat cylinder metric (\ref{Kmet}).
The image of the cycle $\cycle_{\theta}$ in the $\widetilde{W}$
plane is
\beq
\widetilde{W}=\e^{i\theta/b_1}\sum_{i=1}^N|\e^{-y_i}|,
\eeq
and is indeed a straight line (see Figure \ref{cyclestr}).
Moreover,
the cycle passes through the critical point $p_0$
and the straight line in the $\widetilde{W}$-plane emanates outward
from the critical value
$\widetilde{w}_{0}$.
Thus, $\cycle_{\theta}$ is the wavefront trajectory emanating from the
critical point $p_0$.
Since the image in the $\widetilde{W}$-plane
has the slope $\theta/b_1$
and the boundary condition of the fermions
is rotated as (\ref{bcfc}), the boundary condition
indeed preserves the supersymmetry $Q$ and $Q^{\dag}$ in (\ref{sQ}).

When $\theta$ is an integer multiple of $2\pi$, as we have noted before,
in the non-linear sigma model limit the corresponding $B$-field 
is integral and the coupling of the worldsheet to such a $B$-field
can be identified as the
boundary coupling to the gauge field on $X$.
In particular, for $\theta=2\pi n$, it is the gauge field of the bundle
${\cal O}_X(-n)$. 
Thus, the D-brane wrapped on $\cycle_{2\pi n}$ in the LG model
can be considered as the mirror of the D-brane
on $X$ which supports the bundle ${\cal O}_X(-n)$
in the sigma model with trivial $B$-field.
Note that the image of $\cycle_{2\pi n}$ in the $\widetilde{W}$-plane
is the straight line emanating from $\widetilde{w}_n$ in the radial direction
(where the labeling of the critical values is made in the
theory with $\theta=0$).
More generally,
the D-brane which supports the bundle
${\cal O}_X(-n)$ in the sigma model with the $B$-field corresponding to
$\theta\ne 0$ is mirror to
the D-brane wrapped on $\cycle_{\theta+2\pi n}$
whose image in the $\widetilde{W}$-plane is the straight line
emanating from $\widetilde{w}_n$ in the radial direction.

In the next section, we make use of the connection explained in this
section
to determine the relation of the D-branes in the non-linear sigma model
on the toric manifold $X$ (including more general cases
corresponding to the gauge group $U(1)^k$ in the linear sigma model)
and the D-branes of the mirror Landau-Ginzburg model.


\section{D-Branes and Mirror Symmetry: Massive Theories}

In this section we will study how D-branes
transform under mirror symmetry
for sigma models on K\"ahler manifolds with $c_1>0$.  We will
mainly concentrate on the case where the theory
has only massive vacua, and discuss the mirror of D-branes
wrapped on holomorphic cycles on the target
K\"ahler manifold $X$ in terms of Lagrangian submanifolds of the
mirror LG models.  In particular
we concentrate on D-branes which corresponds to exceptional
bundles on $X$ (to be defined below).
It turns out that this
connection explains the observations of Kontsevich 
noting a formal correspondence between
 the properties of Helices and exceptional
bundles on the K\"ahler manifolds and the soliton
numbers of an associated LG model.

In order to do this, we will first review what
exceptional bundles and Helices are.  Afterwards
we discuss how mirror symmetry acts in this context.

\subsection{D-branes, Exceptional bundles and Helices}

The similarities between the structures appearing in the
classification of ${\cal N}=2$ theories \cite{CV} and Helix theory
\cite{rudakov} was observed by Kontsevich \cite{kontsevich}. This 
observation was the starting point of \cite{zaslow} in which the
mysterious correspondence between the soliton numbers of a non-linear
sigma model with $\PP^{N}$ target space and exceptional bundles was
further explored. We will explain this correspondence in this section
as a consequence of mirror symmetry. 

\subsection{Exceptional bundles and mutations}
A vector bundle or a sheaf $E$ on an $N$-dimensional variety
$X$ with $c_1>0$
is called exceptional if
\cite{zaslow,rudakov}
\be
\mbox{Ext}^{0}(E,E)=\c \,,\,\,\,\,\mbox{Ext}^{i}(E,E)=0\,,\,\,\,i\geq 1\,,
\ee
where $\mbox{Ext}^i$ is the sheaf theory
generalization of cohomology
groups $H^i$, i.e., for vector bundles $E$ and $F$,
$\mbox{Ext}^{i}(E,F)=H^{i}(X,E^{*}\otimes F)$ which in tern equals to
the Dolbeault cohomology $H^{0,i}(X,E^*\otimes F)$. An
exceptional collection is a collection of exceptional sheaves 
$\{E_{1},\cdots, E_{n}\}$ such that if $a<b$ then 
\cite{zaslow,rudakov}
\begin{eqnarray}
\mbox{Ext}^{i}(E_{a},E_{b})&=&0\,,\,\,i\not= i_0 \qquad \mbox{for  some}
\ i_0\,,\\ \nn
\mbox{Ext}^{i}(E_{b},E_{a})&=&0\,,\,\,i\geq 0\,.
\end{eqnarray}
Note that the above condition leaves
$\dim\mbox{Ext}^{i_0}(E_a,E_b)$ undetermined
for $a<b$ and that could in principle be any integer.
The alternating sum of dimensions of the groups $\mbox{Ext}^{i}$
defines a bilinear product \cite{zaslow,rudakov},
\be
\chi(E,F)=\sum_{i=0}^{N}(-1)^{i}
\mbox{dim}_{\c}~\mbox{Ext}^{i}(E,F)=
\int_{X}\mbox{ch}(E^{*}\otimes F)\mbox{Td}(X)\,.
\ee
An exceptional sheaf
$E$ has the property that $\chi(E,E)=1$. An important property of an
exceptional collection is that they can be transformed into new
exceptional collections by transformations called mutations.  For an
exceptional collection of sheaves $\{E_{1},\cdots,E_{n}\}$ we can
sometimes
define two transformations, left mutation and right mutation. Given a
neighboring pair of sheaves $(E_{a},E_{a+1})$ in an exceptional
collection, the transformations $L_{E_{a}}$ and $R_{E_{a}}$ are such
that
\begin{eqnarray}
L_{E_{a}}(E_{a},E_{a+1})=
(L_{E_{a}}(E_{a+1}),E_{a})\,,\,\,\, 
R_{a+1}(E_{a},E_{a+1})=
(E_{a+1},R_{E_{a+1}}(E_{a}))\,,
\end{eqnarray}
The transformed sheaf $L_{E_{a}}(E_{a+1})$ is defined through an exact
sequence. The exact sequence used to define the mutated sheaf depends on the 
$\mbox{Ext}^{i}$ groups of the pair $(E_{a},E_{a+1})$ \cite{rudakov,zaslow}, 
 
\noindent
$\bullet$ If $\mbox{Ext}^{0}(E_{a},E_{a+1}) \neq 0$ and
$\mbox{Ext}^{0}(E_{a},E_{a+1})\otimes E_{a} \mapsto E_{a+1}$ is
surjective then $L_{E_{a}}(E_{a+1})$ is defined by the exact sequence,
\be
0 \mapsto L_{E_{a}}(E_{a+1}) \mapsto \mbox{Ext}^{0}
(E_{a},E_{a+1})\otimes E_{a} \mapsto E_{a+1} \mapsto 0 \,,
\ee 

\noindent
$\bullet$  If $\mbox{Ext}^{0}(E_{a},E_{a+1}) \neq 0$ and
$\mbox{Ext}^{0}(E_{a},E_{a+1})\otimes E_{a} \mapsto E_{a+1}$ is
injective then $L_{E_{a}}(E_{a+1})$ is defined by the exact sequence,
\be
0  \mapsto \mbox{Ext}^{0}(E_{a},E_{a+1})
\otimes E_{a} \mapsto E_{a+1} \mapsto L_{E_{a}}(E_{a+1})\mapsto 0 \,,
\ee 

\noindent
$\bullet$ If $\mbox{Ext}^{1}(E_{a},E_{a+1})\neq 0$ then $L_{E_{a}}(E_{a+1})$ is
defined by exact sequence
\be
0 \mapsto E_{a+1} \mapsto L_{E_{a}}(E_{a+1}) 
\mapsto \mbox{Ext}^{1}(E_{a},E_{a+1})\otimes E_{a} \mapsto 0 \,.
\ee

Similarly we can define the right mutated 
sheaf $R_{E_{a+1}}(E_{a})$ by an exact sequence,

\noindent
$\bullet$ If $\mbox{Ext}^{0}(E_{a},E_{a+1}) \neq 0$ and
$\mbox{Ext}^{0}(E_{a},E_{a+1})\otimes E_{a} \mapsto E_{a+1}$ is
surjective then $R_{E_{a+1}}(E_{a})$ is defined by the exact sequence,
\be
0 \mapsto E_{a} \mapsto \mbox{Ext}^{0}(E_{a},E_{a+1})^{*}\otimes E_{a+1} 
\mapsto R_{E_{a+1}}(E_{a}) \mapsto 0 \,,
\ee 

\noindent
$\bullet$ If $\mbox{Ext}^{0}(E_{a},E_{a+1}) \neq 0$ and
$\mbox{Ext}^{0}(E_{a},E_{a+1})\otimes E_{a} \mapsto E_{a+1}$ is
injective then $R_{E_{a+1}}(E_{a})$ is defined by the exact sequence,
\be
0  \mapsto R_{E_{a+1}}(E_{a}) 
\mapsto E_{a} \mapsto \mbox{Ext}^{0}
(E_{a},E_{a+1})^{*}\otimes E_{a+1} \mapsto E_{a+1} \mapsto 0 \,,
\ee 

\noindent
$\bullet$ If $\mbox{Ext}^{1}(E_{a},E_{a+1})\neq 0$ then $L_{E_{a}}(E_{a+1})$ is
defined by exact sequence
\be
0  \mapsto \mbox{Ext}^{1}(E_{a},E_{a+1})^{*}
\otimes E_{a+1} \mapsto R_{E_{a+1}}(E_{a}) \mapsto E_{a} \mapsto 0 \,.
\ee
As far as the Chern characters
are concerned the new sheaves $L_{E_{a}}(E_{a+1})$ and
$R_{E_{a+1}}(E_{a})$ are such that
\bea
\pm \mbox{ch}( L_{E_{a}}(E_{a+1}))&=&
\mbox{ch}(E_{a+1})-\chi(E_{a},E_{a+1})\,\mbox{ch}(E_{a})\,, \nn \\
\pm \mbox{ch}( R_{E_{a+1}}( E_{a}))&=&
\mbox{ch}(E_{a})-\chi(E_{a},E_{a+1})\,\mbox{ch}(E_{a+1})\,.
\label{braid}
\eea
where this follows from the exact sequences
used in the definition of the mutation and $\pm$
depends on which mutation one uses.
The left and the right mutations are inverse of each
other and satisfy the braid group relations \footnote{~$L_{a}\equiv
L_{{\cal E}_{a}}, R_{a}\equiv R_{{\cal E}_{a}}$.}\cite{rudakov}
\begin{eqnarray}
L_{a}L_{b}&=&L_{b}L_{a}\,,\,\,R_{a}R_{b}=R_{b}R_{a}\,,\,\,\,\,\,\mbox{if}\,\,|a-b|>1\,, \nn \\
L_{a}L_{a+1}L_{a}&=&L_{a+1}L_{a}L_{a+1}\,,\,\,R_{a}R_{a+1}R_{a}=R_{a+1}R_{a}R_{a+1}\,. 
\label{braid2}
\end{eqnarray}
These transformations implement the braid group action on the
collection of exceptional sheaves \cite{rudakov}. 

A helix of period $n$, $\{E_{i}~|~i\in \z\}$ is a 
collection of infinitely many
exceptional sheaves such that \cite{zaslow,rudakov}
\begin{eqnarray}
\{E_{i+1},&\cdots&, E_{i+n}\}\,,\,\,\mbox{is an 
exceptional collection for all $i\in \z$},\\
E_{i+n}&=&R_{E_{i+n-1}}\cdots R_{E_{i+2}}R_{E_{i+1}}(E_{i})\,.
\end{eqnarray}
Thus given any exceptional collection $\{E_{1},\cdots, E_{n}\}$ we 
can define a helix by extending the exceptional
collection periodically i.e., 
$E_{i+n}=R_{E_{i+n-1}}\cdots R_{E_{i+1}}(E_{i})$ and 
$ E_{-i+n}=L_{E_{-i+n-1}}\cdots L_{E_{i-1}}(E_{i})$
for $0\leq i\leq n$. The exceptional collection defining a helix is
called the foundation of a helix. Such an exceptional collection generates 
the derived category of $X$ \cite{rudakov}.
For any exceptional collection $\{E_{i}~|~i=1,\cdots,n\}$ which
is the foundation of a helix \cite{rudakov}
\be
R_{ E_{i+n}}\cdots R_{E_{i+1}}(E_{i})=E_{i}\otimes \omega_{X}\,,
\ee
where $\omega_{X}$ is the canonical line bundle of $X$. The collection of
line bundles $\{{\cal O}(0),{\cal O}(1),\cdots,
\linebreak{\cal O}(n)\}$  on $\PP^{n}$ provides an important example of
an exceptional collection which is also the foundation of a helix of
period $n+1$.  Their Chern character is given by
$~\mbox{ch}({\cal O}(k))=e^{k x}$ where
$\int_{\PP^{n}}x^{n}=1$.
In this case \cite{zaslow}
\be
R_{{\cal O}(n)}\cdots R_{{\cal O}(1)}({\cal O}(0))={\cal O}(0)\otimes
\omega_{\PP^{n}}={\cal O}(n+1).
\ee
The bilinear form for the exceptional collection on $\PP^{n}$ is given by
\bea
\chi({\cal O}(a),{\cal O}(b))&=& {n+a-b\choose a-b}\,,\,\,\, a \leq b\\
&=& 0\ \,,\,\,\, ~~~~~~~~~~~~~~~~a >b\,.
\eea

If we consider D-branes corresponding to exceptional
sheaves on the K\"ahler manifold we can couple them
to sigma models.  In this context, as discussed earlier
in this paper (\ref{DolH}), ${\rm Ext}^i(E,F)$ is
interpreted as the ground states in the open string sector
stretched between $E$ and $F$ with 
 fermion number $i$.

The similarities of the objects defined above
and the D-branes we have studied in the context of
LG models is striking 
and as we will discuss below not
accidental:  The D-branes we have constructed in the LG
model turn out to be the mirror of the 
 exceptional bundles.  In particular the property that
$\mbox{dim}\,\mbox{Ext}^i(E,E)=\delta_{i,0}$ is the statement we discussed
before, namely the open string sector of a string stretched between
the same D-brane has only one vacuum with fermion number 0
(the function $W$ gives a Morse function on it with exactly one
critical point corresponding to an absolute minimum). 
Also the system of exceptional collection of sheaves
has a natural parallel in the LG system:  If we consider
the D-branes on the cycles $\gamma_i$ that we constructed,
ordered with decreasing value of ${\rm Im}W$, then the fact that
for $i<j$ the open string stretched between $\gamma_i$
and $\gamma_j$ has no Ramond ground states
and that for $i>j$ there can only be zero
modes in this sector at a fixed fermion number, as discussed
before, is exactly the conditions imposed on a collection
of exceptional sheaves.  Moreover the braiding with left
and right mutations has also a natural parallel:  If we
change the combination of left and right supercharges
that we are preserving the image of the $\gamma_i$ in the
$W$-plane will rotate by the corresponding angle.  Moreover
as we change the angle two neighboring $\gamma_i$ and $\gamma_{i+1}$
might switch order.  In this case the switched $\gamma_i$
define a different basis for $H_n({\bf C}^n,{\rm Re} We^{-i\theta}>0)$
related by Picard-Lefshetz action as discussed before.  This is
exactly the same as the change in the Chern characters
of transmuted exceptional sheaves, up to the
$\pm$ sign, which one can
interpret as the orientation of the corresponding
LG D-brane.  Moreover the left versus
right mutation corresponds to the reversal of the direction of
change of $\theta$. Finally if we consider a massive sigma model
with $N$ isolated vacua, taking any $\gamma_i$ around in the $W$ plane,
as discussed before,
is equivalent to changing the $\theta_i k_i$ B-fields of the sigma model
by $c_1$ of the manifolds, which on the D-brane is realized as a
tensoring with a $U(1)$ connection with curvature given by $c_1$,
i.e. tensoring the D-brane with $\omega_X$.   This is exactly
the condition of having a helix of period $N$. 
Moreover once we discuss why $\gamma_i$ are the mirrors
of the corresponding D-branes it becomes clear
 why, up to braidings, the number of solitons
in Fano varieties (with only massive vacua) are given
by the index of the ${\overline \partial}$ operator
coupled to $E_i^*\otimes E_j$ where $E_i,E_j$ belong
to a collection of exceptional bundles on the Fano variety.

Below we will present many examples of this connection. The
discussions are aimed at giving a sample rather than an exhaustive
search through examples.

\subsection{$\PP^{n}$}

The $\PP^n$ sigma model is realized as
the $U(1)$ gauge theory with $N=n+1$ matters of charge $1$. 
The mirror is the $A_{n}$ affine Toda field theory with the
superpotential
\beq
W=\e^{-Y_1}+\cdots+\e^{-Y_n}+\lambda \,\e^{Y_1+\cdots+Y_n},
\eeq
where $\lambda=\e^{-t}=\e^{-r+i\theta}$.
There are $n+1$ critical points $p_0,\ldots,p_n$
given by
$\e^{-Y_i}(p_k)=\lambda^{1\over n+1}\e^{2\pi k i\over n+1}$
with the critical value
$w_k=(n+1)\lambda^{1\over n+1}\e^{2\pi k i\over n+1}$.

As we saw in
section 6, the mirror of the trivial bundle ${\cal O}(0)$
for the sigma model with $\theta=0$
is the middle dimensional cycle whose
image in the $W$-plane is a straight line starting at the critical value
$w_0$ and extending along the real axis.
We have also seen that the mirror of the bundle
${\cal O}(k)$ for the sigma model with $\theta=0$
is the Lagrangian submanifold whose image in the $W$-plane is a
straight line emanating from $w_{-k}$ in the radial direction.
This has been obtained by the shift $\theta=0\to\theta=-2\pi k$,
i.e. the rotation
$\lambda^{1\over n+1}\to
\lambda^{1\over n+1}\e^{-{2\pi k i\over n+1}}$,
and interpreting the result as the ${\cal O}(k)$ bundle for the
sigma model with $\theta=0$.
In this case, the unbroken supercharge is of the A-type
combination\footnote{Note that we have switched back
to the standard convention of chirality: $Y_i$'s are
{\it chiral} superfield rather than twisted chiral superfield.
That is why the unbroken supercharges are A-type rather than B-type.}
$Q=\bQ_++\e^{-{2\pi ki\over n+1}}Q_-$
and its conjugate $Q^{\dag}$.
At this stage,
one can rotate the Lagrangian cycle (without touching $\theta$)
so that the image in the $W$-plane is parallel to the real axis.
In such a case, the corresponding D-brane preserves the standard
combination $Q=\bQ_++Q_-$ and its conjugate $Q^{\dag}$.

These are depicted in the example of $\PP^5$ in Figure \ref{cpn-c}.
\onefigure{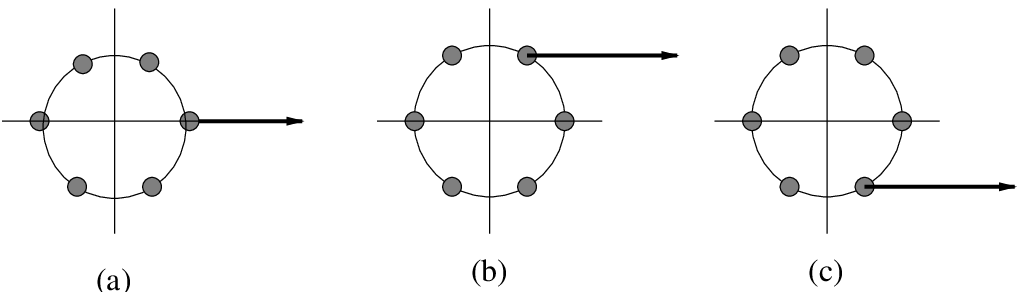}{a) $\{{\cal O}(0),\theta =0\} = {\cal O}(0)$\,, 
b)\,\, $\{ {\cal O}(0), \theta = 2\pi \}$ = ${\cal O}(-1)$,\, c)\,\,
$\{{\cal O}(0), \theta = -2\pi \}$ = ${\cal O}(1)$}
For the bundles ${\cal O}(2)$,
${\cal O}(3)$,... or ${\cal O}(-2)$,
${\cal O}(-3)$,..., it is impossible for rotating the cycle
so that the images in the $W$-plane are parallel to the real axis
without passing through other critical values.
One can avoid this cross-over by bending the branes although
it results in the breaking of the supersymmetry.
Bending in the clockwise direction,
 we obtain the
 collection of bundles $\{{\cal O}(0),\cdots, {\cal O}(n+1)\}$
which are exceptional collections as we have seen above.
By partially changing the direction of bending,
 we can obtain other exceptional collections as shown in
\figref{cpn-d} for the case of $\PP^{5}$.  If we order the lines in terms of
decreasing asymptotic imaginary part then the exceptional collection
in \figref{cpn-d}(b) is $\{{\cal O}(-3),{\cal O}(-2),{\cal
O}(-1),{\cal O}(0),{\cal O}(1), {\cal O}(2)\}$ and the exceptional
collection in
\figref{cpn-d}(c) is $\{{\cal O}(-2),{\cal O}(-1),{\cal O}(0),{\cal O}(1),{\cal
O}(2),{\cal O}(3)\}$.
\onefigure{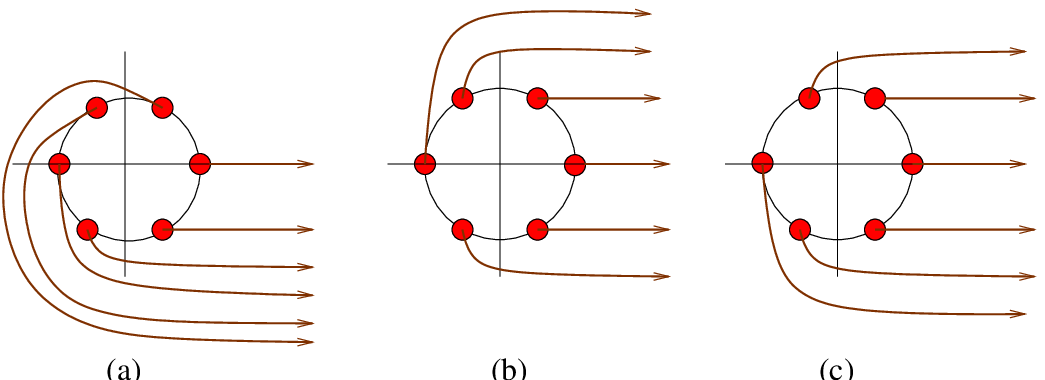}{ 
Three exceptional collections which are related to each other by
mutations.}  

The observation of \cite{kontsevich,zaslow} can now be understood
as a consequence of mirror symmetry. The soliton numbers between
different vacua are given by intersection number of middle dimensional
cycles determined by the superpotential as shown in section 2. Mirror
symmetry relates these cycles and their intersection form to bundles
on $\PP^{n}$ and the bilinear form $\chi({\cal E},{\cal F})$
respectively.  
To find the soliton numbers from this data, as reviewed
in section 2, we need
to choose suitable classes of cycles.  This configuration of
cycles is related
to the D-branes we have by some Picard-Lefshetz action,
which is the mirror realization of Left/Right
mutations discussed in the case of exceptional bundles. 
Let us first discuss LG analog of mutation and then 
return to the computation of soliton numbers
using the $\chi({\cal E},{\cal F})$.

Let us denote the D-brane corresponding to the $-i$-th critical
point by $C_i$.
Note that from the mirror symmetry map we have
for $i>j$, $C_j\circ C_i=\chi({\cal O}(j),{\cal O}(i))=
(n+i-j)!/n!(i-j)!$.  Consider as an example
 the case of $\PP^{2}$ and the
bundle ${\cal O}(2)$ shown in \figref{cpn-e}. Making the middle dimensional
$C_{2}$ cycle pass through the the critical value $w_{2}$ and using
the Picard Lefshetz formula 
\be
C_{2}'=C_{2}-(C_{1}\circ C_{2})C_{1}=C_{2}-3\,C_{1}\,.
\ee 
which is exactly the same as how the left mutation
acts on Chern character upon left mutation of the ${\cal O}(2)$
bundle over ${\cal O}(1)$. We thus identify
$C_{2}'$ as the mirror of the ${\bf L_1}{\cal O}(2)$ (with
the opposite orientation).
 Moreover in order
for charges not to change, 
we see that we create three new D-branes $+3C_1$. 
If we again make the cycle
$C_{2}'$ pass through $w_{0}$ we see using the Picard Lefshetz formula
that,
\be
C_{2}''=C_{2}'-(C_{0}\circ C_{2}')C_{0}= C_{2}-3\,C_{1}+3\,C_{0}\,.
\ee  
\onefigure{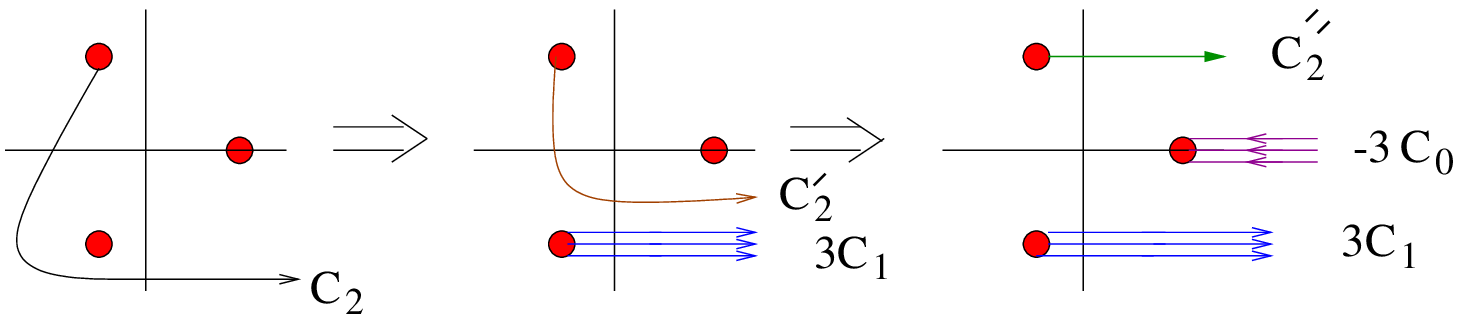}{The Picard-Lefshetz Monodromy, leading
to brane creation, can be interpreted in terms of
left mutations on the mirror side}
Thus we see that we have created $-3C_0$ branes, for charge
conservation.  Again we see that we can identify
$C_{2}''$ with the mirror of ${\bf L_0}{\bf L_1}({\cal O}(2))$.
In other words the above
process viewed in terms of the mutation of ${\cal O}(2)$ in the exceptional
collection $\{{\cal O}(0),{\cal O}(1),{\cal O}(2)\}$, is given as
\bea
\{{\cal O}(0),{\cal O}(1), 
{\bf {\cal O}(2)}\} &\mapsto&  
\{{\cal O}(0),{\bf L_{1}
{\cal O}(2)},{\cal O}(1)\}\mapsto 
\{{\bf L_{0}L_{1}{\cal O}(2)},{\cal O}(0),{\cal O}(1)\}  \nn  \\
\mbox{ch}(L_{1}{\cal O}(2))&=
&\mbox{ch}({\cal O}(2))-3\, \mbox{ch}({\cal O}(1))\,,\,\, \\ \nn 
\mbox{ch}(L_{0}L_{1}{\cal O}(2))&=
&\mbox{ch}(L_{0}{\cal O}(2))-3\, \mbox{ch}(L_{0}{\cal O}(1)) \\ \nn
&=& \mbox{ch}({\cal O}(2))-3\, 
\mbox{ch}({\cal O}(1))+3 
\mbox{ch}({\cal O}(0))=\mbox{ch}({\cal O}(-1))  \\ \nn
\eea
The fact that mutating ${\cal O}(2)$ through all the other critical
points gives ${\cal O}(-1)={\cal O}(2)\otimes {\cal O}(-3)$
and that ${\cal O}(-3)$ is the inverse of $c_1$ of the canonical
bundles, is related to the axial anomaly of the $\PP^{2}$
sigma model.

The left mutated bundle $L_{1}({\cal O}(2))$ is shown in the figure
below.
\onefigure{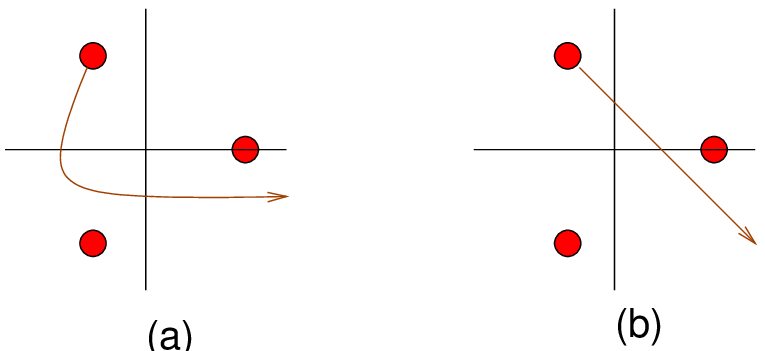}{ The Branes depicted
here are mirrors of $L_{1}({\cal O}(2))$}
\figref{mutation-2}(a) and \figref{mutation-2}(b) represent two
 different representatives of the homology class mirror to
 $L_{1}({\cal O}(2))$. In the case of
\figref{mutation-2}(a) the representative is not a supersymmetric cycle 
since its image in the $W$-plane is not a straight line. The
representative shown in \figref{mutation-2}(b), however, is
supersymmetric and preserves A-model supercharge
$\bar{Q}_{+}+e^{i\alpha}Q_{-}$, where $\alpha$ is the angle that the
straight line makes with the real axis.  This comment also
applies to the branes depicted in figure 19, and other branes
that will be discussed in this section.

As another interesting example of mutation consider the right
mutation of the pair $\{{\cal O}(0),{\cal O}(1)\}$ on $\PP^{n}$ as
shown in \figref{tangent}.
\onefigure{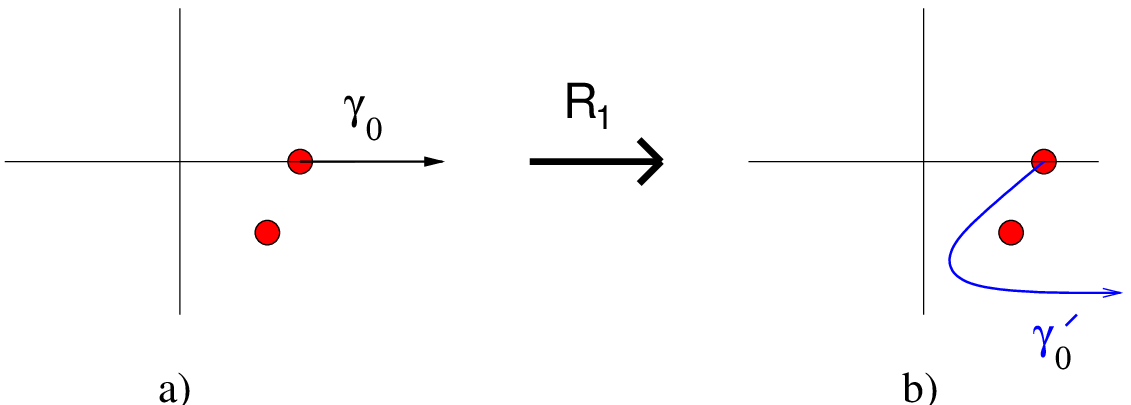}{The mirror of the tangent bundle
is the D-brane shown in (b).}
Since $\mbox{Ext}^{0}({\cal O}(0),{\cal O}(1)) =H^{0}({\cal O}(0),{\cal
O}(1)) \neq 0$, we can use the Euler sequence \cite{zaslow}
\be
0 \mapsto {\cal O}(0) 
\mapsto H^{0}({\cal O}(0),{\cal O}(1))^{*}\otimes 
{\cal O}(1) \mapsto T \mapsto 0
\ee 
In fact $R_{{\cal O}(1)}({\cal O}(0))$ is $T$, the tangent bundle, and
its mirror is identified in the LG theory with the brane $\gamma_{0}\,'$
, shown in \figref{tangent}(b), which is obtained by Picard-Lefshetz
monodromy action on $\gamma_{0}$.\footnote{ It is natural to conjecture
that the mirror of all bundles on $\PP^{n}$ is given on the LG mirror
by the D-branes corresponding to exceptional bundles with
multiplicities given by the decomposition of its Chern character.}

 Now we come back to the question of counting the soliton numbers.
The soliton number between $w_0$ and $w_{-k}$ can be
computed by first left mutating the ${\cal O}(k)$ brane
through ${\cal O}(k-1),...,{\cal O}(1)$ and then taking 
its inner product with ${\cal O}(0)$.
Since $\mbox{ch}(L_{1}\cdots
L_{k-1}{\cal O}(k))=\sum_{i=0}^{k-1}(-1)^{i}{n+1\choose i}
\mbox{ch}({\cal O}(k-i))$, hence
 soliton numbers between two different vacua
$w_{0}$ and $w_{-k}$ is equal to
\bea 
\mu_{i,i-k}=\mu_{0,-k}&=&\chi({\cal O}(0),
L_{1}\cdots L_{k-1}{\cal O}(k))\\ \nn
&=& \sum_{i=0}^{k-1} (-1)^i{n+1 \choose i}{n+k-i \choose k-i}=
       (-1)^{k-1}{n+1 \choose k}
\eea
in agreement with what we had obtained before for the soliton
numbers.

\subsection{Toric del Pezzo Surfaces}
 
We next consider the non-linear sigma model
with two dimensional toric Fano varieties as the target space. We
will show that the supersymmetric cycles of the mirror LG theory,
which are the preimages of the straight lines in the $W$-plane defined
by the superpotential, are related to an exceptional collection of
bundles on the target space.

{}From the classification toric Fano varieties it is known that there
are five toric Fano surfaces. The toric diagram of these surfaces is
captured by a dual lattice
shown in \figref{toric} (see \cite{kmv} and \cite{toricref}
for a detailed discussion)
which is obtained naturally from
the mirror symmetry description derived in \cite{HV}.
 The first four diagram are that of $\PP^{2}$
and its three blow ups respectively. \figref{toric}(e) 
is the toric diagram of $F_{0}=\PP^{1}\times
\PP^{1}$.
\onefigure{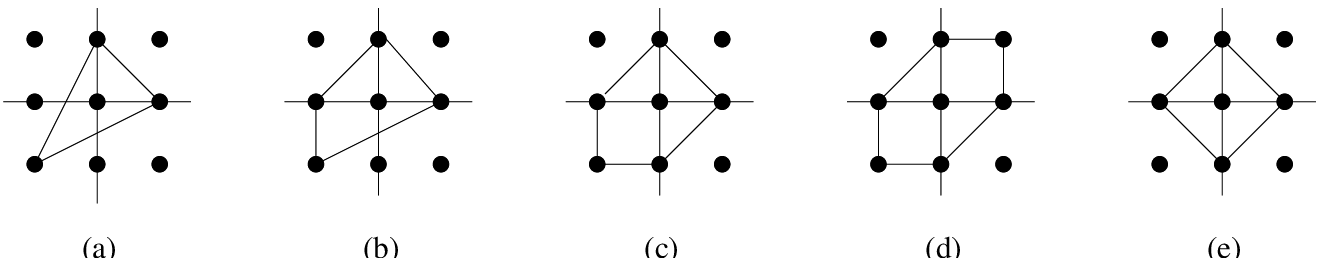}{Toric diagram for $\PP^{2}$, its three blow ups and
$\PP^{1}\times \PP^{1}$.}

The superpotential of the mirror LG theory can be written from the
toric data given in the figure. Let
$\{v^{(a)}=(v^{(a)}_{1},v^{(a)}_{2})~|~a=1,\cdots,N\}$ be the set of
vertices. Then \cite{HV}
\be
W(X)=\sum_{a=1}^{N}C_{a}X_{1}^{v^{(a)}_{1}}\,X_{2}^{v^{(a)}_{2}}\,,
\ee
where $C_{a}$ are complex numbers. Only a subset of $C_{a}$ are actually 
physical since some of them can be absorbed by rescaling $X_{i}$.

\subsubsection{$\PP^{2}$}
{}From the discussion in the previous section we know that the lines
shown in the \figref{b0-b}(a) correspond to the exceptional collection
$\{{\cal O}(-1),{\cal O}(0),{\cal O}(1)\}$. The exceptional collection
shown in \figref{b0-b}(b) is $\{{\cal O}(-1),V, {\cal O}(0)\}$ where
$V=L_{0}{\cal O}(1)$ is such that $\mbox{ch}(V)=(-2,1,\frac{1}{2})$.
Here we are using the notation such that $\mbox{ch}(V)=
(c_{0}(V), c_{1}(V), \int_{X}\mbox{ch}_{2}(V))$.
What we actually mean by the negative number for $c_0$ is
that if we reverse the orientation of the D-brane we obtain
the corresponding mirror of the bundle.  In other words the
Chern character is multiplied by a minus sign, when comparing
with the charges of the LG D-brane.
  We shall be somewhat implicit
about this in this section, 
but it should be clear from the context what we mean--
namely $c_0$ of the bundle should always be positive.  
\onefigure{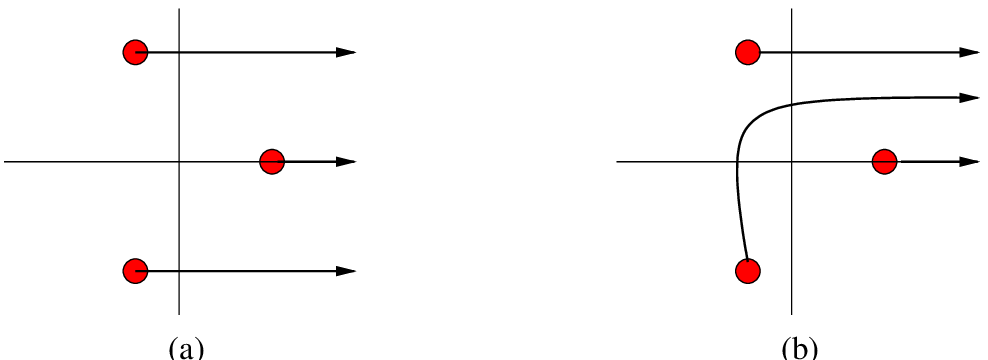}{Two exceptional collections related by a 
left mutation.}
The soliton counting matrix can be determined from the exceptional
collection shown in \figref{b0-b}(b) and is given by
$\chi(E_{i},E_{j})$ where $E_{i}\in \{{\cal O}(-1),V,{\cal O}(0)\}$,
\be
S_{ij}=\chi(E_{i},E_{j})=\pmatrix{1 & -3 & 3\cr 0 &1 &-3\cr 0 & 0 &1}\,.
\ee

\subsubsection{${\cal B}_{1}$}

The superpotential is given by
\be
W(X_{1},X_{2})=X_{1}+X_{2}+\frac{e^{-t}}{X_{1}X_{2}}+\frac{e^{-t_{E_{1}}}}{X_{1}}\,,
\ee
where $t_{E_{1}}=A(E)-i\,\theta(E)$ is the complexified K\"ahler parameter
of the exceptional cycle, $E_{1}$. After rescaling the coordinates we
can express the superpotential in the following form which will be
more useful for the later discussion,
\be
W(X_{1},X_{2})=e^{-\frac{t}{3}}(X_{1}+X_{2}+\frac{1}{X_{1}X_{2}}+\frac{e^{-t_{E_{1}}+\frac{2}{3}t}}{X_{1}})\,.
\label{b1potential}
\ee
We define $\mu_{1}=e^{-t_{E_{1}}+\frac{2}{3}t}$, then the critical
values of the superpotential given above are
\begin{eqnarray}
w=e^{-\frac{t}{3}}\,(3\,y_{*}+2\,\mu_{1}\,y_{*}^{2})\,,
\,\,\,\,\mbox{where}\,\,\,\,\mu_{1}y_{*}^{4}+y_{*}^{3}-1=0\,.
\end{eqnarray}
For $|\mu_{1}|<<1$ we see that 
\begin{eqnarray}
y_{*}&\approx& \{e^{2\pi i k/3}+ \, O(\mu_{1}),
-\fracs{{1}}{{\mu_{1}}}+\,O(\mu_{1})~|~k=0,1,2\}\,,\\
w &\approx& e^{-\frac{t}{3}}\{3\,e^{2\pi i
k/3}+\,O(\mu_{1})\,,\,-\fracs{{1}}{{\mu_{1}}}+\,O(\mu_{1})~|~k=0,1,2\}.
\end{eqnarray}
We will denote the first three critical values as
$\{w_{k}~|~k=0,1,2\}$ and the fourth one as $\widehat{w}_{1}$. Thus we
see that for $|\mu_{1}|$ very small the transformation $\mu \mapsto
\mu e^{i\theta}$ does not change the three symmetrically located
critical values $w_{k}$ much but the critical value $\widehat{w}_{1}$
undergoes a clockwise rotation by an angle $\theta$ as shown
in \figref{b1-d}.
\onefigure{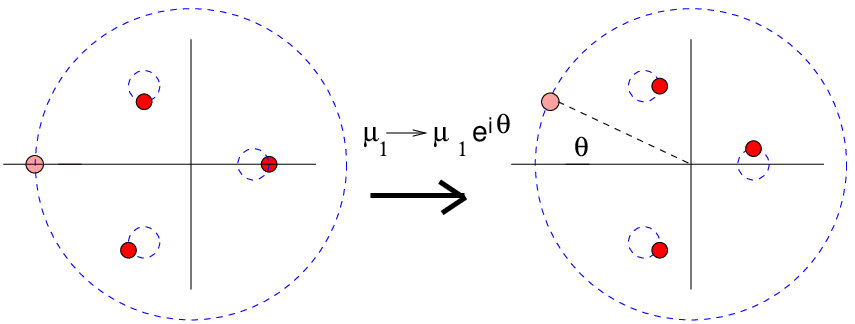}{The effect of transformation 
$\mu_{1}\mapsto \mu_{1}e^{i\theta}$ on the critical values.}  
Thus to determine the bundles associated with the semi-infinite lines
we first consider $\mu_{1}\mapsto 0$. The critical value
$\widehat{w}_{1}$ goes to $-\infty$ and we are left with the case of
$\PP^{2}$ for which we know the correspondence. Thus the three bundles
$V_{1},V_{2}$ and $V_{3}$ shown in \figref{b1-e} correspond to the
pull back of ${\cal O}(-1), {\cal O}(0)$ and ${\cal O}(1)$
from $\PP^{2}$ to ${\cal B}_{1}$ respectively.
\onefigure{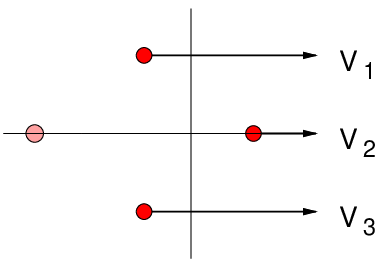}{The pull back from $\PP^{2}$ to ${\cal B}_{1}$ of 
$\{{\cal O}(-1),{\cal O}(0),{\cal O}(1)\}$.}
To determine the bundle associated with the critical value
$\widehat{w}_{1}$ we take $\mu_{1}$ to be very small and positive so
that $\widehat{w}_{1}$ is far from other critical values. In this case
the critical value is on the negative real axis as shown in the
\figref{b1-e}. Consider the case, as shown in \figref{b1-c}, when
there is the D-brane mirror to ${\cal O}(0)$, represented by the
straight line starting from $w_{0}$, present. As mentioned before under
the transformation $\mu_{1}\mapsto \mu_{1}e^{2\pi i}$ the critical
value $\widehat{w}_{1}$ makes a clockwise rotation around the
origin,
\figref{b1-c}. As it passes through the D-brane associated with the
bundle ${\cal O}(0)$, due to brane creation effect discussed earlier,
it acquires a D-brane charge consistent with the homology class of the
cycle associated with this critical point.
\onefigure{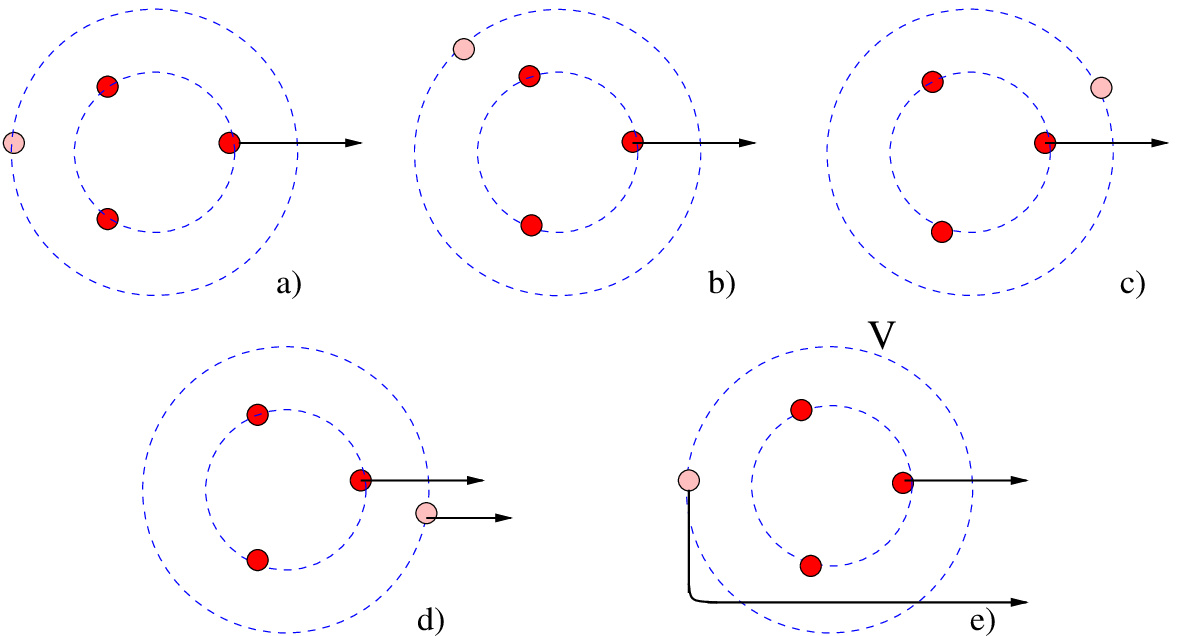}{ Brane creation as the critical value 
$\widehat{w}_{1}$ undergoes a full rotation for
$\mu_{1}\mapsto \mu_{1}e^{2\pi i}$.}
To determine the bundle corresponding to this new D-brane we use
charge conservation. Recall that $\mu_{1}$ was defined in terms of
$t_{E_{1}}$ and $t$, the complexified K\"ahler parameters of ${\cal
B}_{1}$. We have kept $t$ fixed in above discussion therefore since
the imaginary part of $t_{E_{1}}$ is minus the B-field integrated over
the exceptional curve $E_{1}$ the transformation $\mu_{1}\mapsto
\mu_{1}e^{2\pi i}$ corresponds to turning on the B-field through 
$E_{1}$. Thus we interpret \figref{b1-c}(e) as the ${\cal O}(0)$
bundle in the B-field background. This implies that if we denote the
cohomology class dual to the exceptional curve $E_{1}$ as $[E_{1}]$,
the first Chern class of this bundle (${\cal O}(0)$ in the B-field
background) is $c_{1}({\cal O}(0))-[E_{1}]$. Denoting by $\widehat{V}_{1}$
the bundle mirror to the line starting at $\widehat{w}_{1}$ we get,
\begin{eqnarray} 
\mbox{ch}({\cal O}(0)\oplus \widehat{V}_{1})&=&e^{c_{1}({\cal O}(0))-[E_{1}]} \,,\\  \nn
\mbox{ch}(\widehat{V}_{1})&=&e^{c_{1}({\cal O}(0)-[E_{1}]}-\mbox{ch}({\cal O}(0))\\ \nn
&=&(1,[E]_{1},-\shalf)-(1,0,0)=(0,-[E_{1}],-\shalf)\,.
\end{eqnarray}
Thus $\widehat{V}_{1}$ is actually a sheaf with support on the
exceptional divisor $E_{1}$.  The collection of sheaves
$\{V_{1},V_{2},V_{3},\widehat{V}_{1}\}$ is an exceptional collection
since $\chi(\widehat{V}_{1},\widehat{V}_{1})=1$ and
$\chi(\widehat{V}_{1},V_{i})=0$ for $i=1,2,3$.
\onefigure{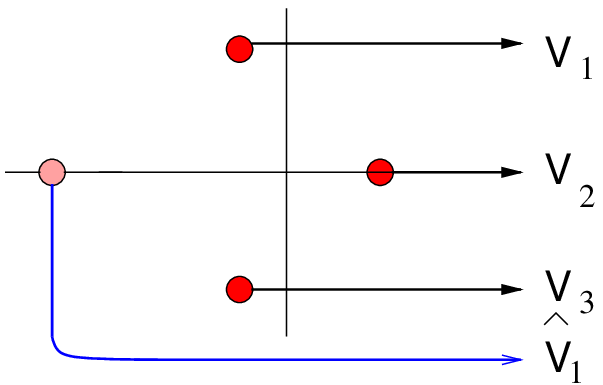}{The mirror of an exceptional collection of bundles on ${\cal B}_{1}$.}

It is interesting to consider the effect of transformation
$e^{-t}\mapsto e^{-t+2\pi i}$. Since
$\mu_{1}=e^{-t_{E_{1}}+\frac{2}{3}t}$ we see that under the above
transformation $\mu_{1}\mapsto \mu_{1}e^{\frac{4\pi i}{3}}$. From the
previous discussion it would seem that the critical point
$\widehat{w}_{1}$ undergoes a rotation by an angle
$\frac{4\pi}{3}$. This, however, is not the case.  From
eq.\,(\ref{b1potential}) it is clear that because of the overall factor
$e^{-\frac{t}{3}}$ a phase transformation $e^{-t}-\mapsto e^{-t+2\pi
i}$ rotates all the critical values by an angle $\frac{2\pi}{3}$.
Thus the critical value $\widehat{w}_{1}$ gets rotated by
$\frac{4\pi}{3}+\frac{2\pi}{3}=2\pi$,
\be
e^{-t}\mapsto e^{-t+2\pi i}:\,\,w_{1}\mapsto w_{2}\mapsto w_{3}\mapsto w_{1}\,,\,\,\widehat{w}_{1} \mapsto \widehat{w}_{1}\,.
\ee
The effect on the bundles, however, is more non-trivial and is shown in
\figref{b1-f}{}.\onefigure{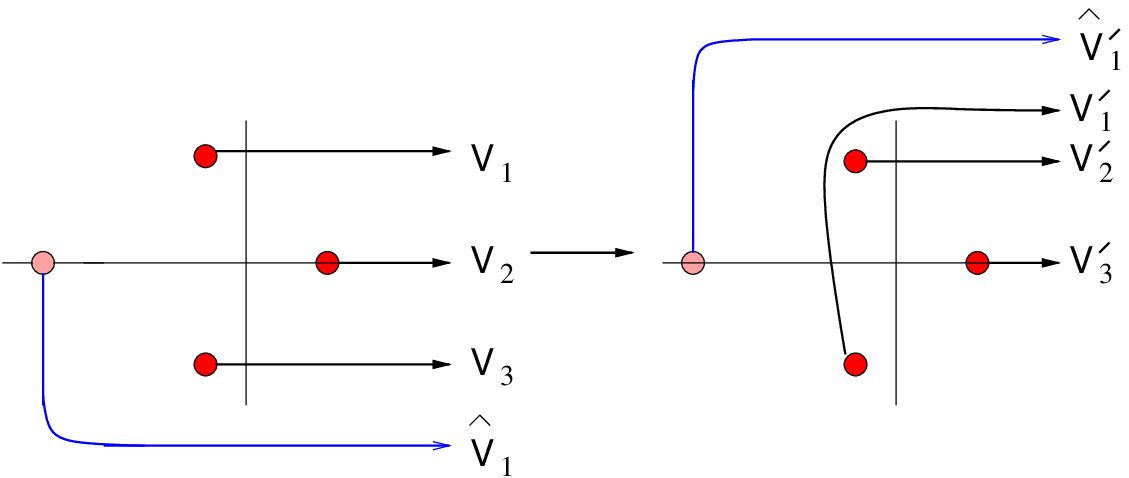}{The effect of the transformation 
$e^{-t}\mapsto e^{-t+2\pi i}$ on the exceptional collection.}
$V_{1}',V_{2}',V_{3}'$ and $\widehat{V}_{1}'$ are $V_{1},V_{2},V_{3}$
and $\widehat{V}_{1}$ bundles in the background where the B-field
through the cycle $l$ has been turned on, where $l$ along with $E_{1}$
forms a basis of $H_{2}({\cal B}_{1},\z)$ such that the
self-intersection number of $l$ is plus one, $l^{2}=1$. Denoting the
cohomology class dual to $l$ by $[l]$ we get
\begin{eqnarray} 
\mbox{ch}(V_{k}')=e^{c_{1}(V_{k})-[l]}=e^{(k-3)[l]}=(1,(k-3)[l],\fracs{{1}}{{2}}(k-3)^{2})\,
\end{eqnarray}
Thus we see that $V_{k}'$ is the pull back of ${\cal O}(k-3)$ bundle
from $\PP^{2}$ to ${\cal B}_{1}$ and can be written as
$\{V_{1}',V_{2}',V_{3}'\}=\{V_{1}\otimes V_{1} ,V_{2}\otimes V_{1},
V_{3}\otimes V_{1}\}$.  The bundle $\widehat{V}_{1}'$ is easy to
determine since under the transformation $t_{E_{1}}\mapsto
t_{E_{1}}+2\pi i$ the critical value $\widehat{w}_{1}$ rotates in the
counter clockwise direction therefore  argument similar to the one
used to determine $\widehat{V}_{1}$ shows that $\widehat{V}_{1}'$ is such that
\be
\mbox{ch}(V_{2}\oplus \widehat{V}_{1}')=e^{c_{1}(V_{2})+[E_{1}]}\,\, 
\Longrightarrow \,\,\mbox{ch}(\widehat{V}_{1}')=(0,[E_{1}],-\fracs{{1}}{{2}})\,.
\ee

Using this exceptional collection we can calculate the the number of
solitons between various vacua. The soliton number between two vacua
is given by $\chi(E,F)$ where $E$ and $F$ are the bundle corresponding
to the semi-infinite straight lines starting from the two vacua we are
studying. The semi-infinite lines must be such that they together
do not enclose another critical value. Otherwise the intersection will
get contribution from the critical value enclosed by the lines. Thus
we first transform to accomplish the exceptional collection given
above into one for which no two lines enclose another critical
value. This exceptional collection is shown in \figref{b1-g} and is
obtained from $\{V_{1},V_{2},V_{3},\widehat{V}_{1},\}$ by successive left 
mutations shown in \figref{b1-g},
\begin{eqnarray}
\{V_{1},V_{2},V_{3},\widehat{V}_{1}\}&\mapsto& \{V_{1},V_{2},L_{V_{3}}(\widehat{V}_{1}),
V_{3}\} \\ \nn
 &\mapsto& \{V_{1},L_{V_{2}}L_{V_{3}}(\widehat{V}_{1}),V_{2},V_{3}\} \\ \nn
&\mapsto &
\{V_{1},L_{V_{2}}L_{V_{3}}(\widehat{V}_{1}),L_{V_{2}}(V_{3}),V_{2}\} =: \{E_{1},E_{2},E_{3},E_{4}\}\,.
\end{eqnarray}
As far as the Chern characters of $E_{i}$ are concerned we have,
\begin{eqnarray} \nn
\mbox{ch}(E_{1})=(1,-1,\fracs{{1}}{{2}})\,,\,\,
\mbox{ch}(E_{2})=(1,-l+[E_{1}],0)\,,\,\,
\mbox{ch}(E_{3})=(2,-l ,-\fracs{{1}}{{2}})\,,\,\,\mbox{ch}(E_{4})=(1,0,0)\,.
\end{eqnarray}
\onefigure{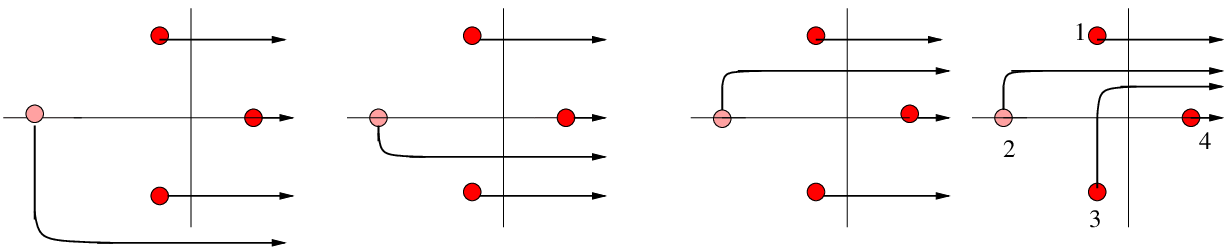}{Left mutations of the exceptional collection to obtain
an exceptional collection for soliton counting.}
The soliton number between the vacua are now given by $\chi(E_{i},E_{j})$,
\be
S_{ij}=\chi(E_{i},E_{j})=\pmatrix{1 & -1 & -3 & +3 \cr 0 & ~~1 & +1
&-2 \cr 0 & ~~0& ~~1 &-3\cr 0 & ~~0 & ~~0 &~~1}
\ee

In other words there is one soliton
 between vacua $2,1$ and $2,3$, three solitons
between $1,3$ and $1,4$ and $3,4$ and two solitons
between vacua $2,4$.  Of course, as we change the K\"ahler
parameters of the manifold the position of the vacua
change and the number of solitons change, as reviewed
in section 2.

\subsubsection{${\cal B}_{2}$}

The superpotential of the LG theory mirror to non-linear sigma model
with ${\cal B}_{2}$ is given by
\be
W(X_{1},X_{2})=X_{1}+X_{2}+\frac{e^{-t}}{X_{1}X_{2}}+\frac{e^{-t_{E_{1}}}}{X_{1}}+\frac{e^{-t_{E_{2}}}}{X_{2}}\,.
\ee
$t_{E_{1}}$ and $t_{E_{2}}$ are the complexified K\"ahler parameters
of the exceptional curves $E_{1}$ and $E_{2}$ respectively.  As in the
case of ${\cal B}_{1}$ we rescale the coordinates $X_{1}$ and $X_{2}$
to obtain the following form for the superpotential which will be
useful for later discussion,
\be
W(X_{1},X_{2})=e^{-\frac{t}{3}}(X_{1}+X_{2}+\frac{1}{X_{1}X_{2}}+\frac{e^{-t_{E_{1}}+\frac{2}{3}t}}{X_{1}}+\frac{e^{-t_{E_{2}}+\frac{2}{3}t}}{X_{2}})\,.
\ee
We define $\mu_{i}=e^{-t_{E_{i}}+\frac{2}{3}t}$ for $i=1,2$.  The critical values of the superpotential are given by
\be
w=\fracs{{1}}{{y_{*}^{2}-\mu_{2}}}+2y_{*}+\mu_{1}(y_{*}^{2}-\mu_{2})\,,\,\mbox{where}\,\,\,\,(y_{*}^{2}-\mu_{2})^{2}(\mu_{1}y_{*}+1)-y_{*}=0\,.
\ee
For $|\mu_{1}|,\mu_{2}|<<1$ we can see that leading order terms for
the critical points and critical values are,
\begin{eqnarray}\nn
y_{*}&=&\{e^{\frac{2\pi i k}{3}}+O(\mu_{1},\mu_{2})\,,\,-\fracs{{1}}{{\mu_{1}}}+O(\mu_{1},\mu_{2})\,,\,\mu_{2}+O(\mu_{1},\mu_{2}^{2})~|~k=0,1,2\}\,,\\ \nn
w&=&\{3\,e^{\frac{2\pi i k}{3}}+O(\mu_{1},\mu_{2})\,,\,-\fracs{{1}}{{\mu_{1}}}+O(\mu_{1},\mu_{2})\,,\,-\fracs{{1}}{{\mu_{2}}}+O(\mu_{1},\mu_{2})~|~k=0,1,2\}\,.
\end{eqnarray}
We will denote the above critical values by
$w_{0},w_{1},w_{2},\widehat{w}_{1}$ and $\widehat{w}_{2}$
respectively.

To determine the bundle corresponding to $\widehat{w}_{1}$
 we use the same argument as
for the case of ${\cal B}_{1}$. As $\mu_{1},\mu_{2}\mapsto 0$ we
recover the $\PP^{2}$ configuration and therefore the the three
bundles corresponding to the semi-infinite lines starting at
$w_{1},w_{2}$ and $w_{3}$ are the pull backs of the ${\cal
O}(-1),{\cal O}(0)$ and ${\cal O}(1)$ bundles from $\PP^{2}$ to ${\cal
B}_{2}$ respectively. We will continue to denote these bundle as
$V_{1},V_{2}$ and $V_{3}$ respectively as before even though they are
different bundles than the ones considered in the last section. However, the
Chern classes of these bundles are the same as before. Since in the
limit $\mu_{2}\mapsto 0$ we recover the ${\cal B}_{1}$ configuration
therefore the bundle corresponding to the line starting at
$\widehat{w}_{1}$ is the sheaf with support on the exceptional curve
$E_{1}$. We will denote it, as before, by $\widehat{V}_{1}$.

As shown in \figref{b2-c} the critical value $\widehat{w}_{2}$ rotates
in a clockwise direction as $\mu_{2}\mapsto \mu_{2}\,e^{2\pi i}$. In
the presence of D-brane corresponding to $V_{2}$ such a rotation of
$\widehat{w}_{2}$ creates a D-brane whose image in the W-plane is the
line starting at $\widehat{w}_{2}$ as shown in \figref{b2-c}. We will
denote the corresponding bundle by $\widehat{V}_{2}$. The
transformation $\mu_{2}\mapsto
\mu_{2}\,e^{2\pi i}$ corresponds to turning on the 
B-field through the exceptional curve $E_{2}$.  Denoting the
cohomology class dual to $E_{2}$ by $[E_{2}]$, charge conservation 
implies that
\begin{eqnarray} \nn
\mbox{ch}(V_{2}\oplus \widehat{V}_{2})=e^{c_{1}(V_{2})-[E_{2}]}\,, \\ \,
\mbox{ch}(\widehat{V}_{2})=(0,-[E_{2}],-\fracs{{1}}{{2}})\,.
\end{eqnarray}
\onefigure{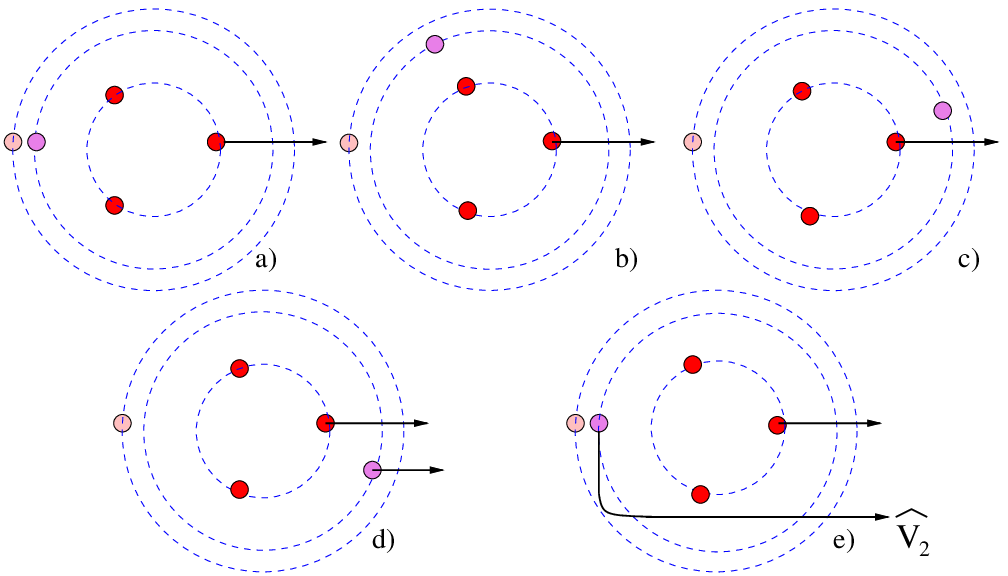}{Brane creation as the critical value $\widehat{w}_{2}$
undergoes a rotation for  
$\mu_{2}\mapsto \mu_{2}e^{2\pi i}$.}

The collection of bundles
$\{V_{1},V_{2},V_{3},\widehat{V}_{1},\widehat{V}_{2},\}$ is an
exceptional collection. As explained before to calculate the soliton
numbers we have to mutate this exceptional collection into an
exceptional collections for which the corresponding semi-infinite
lines are such that any two of them do not enclose a critical point. To
obtain such an exceptional collection we consider the right mutations
shown in \figref{b2-e}.
\onefigure{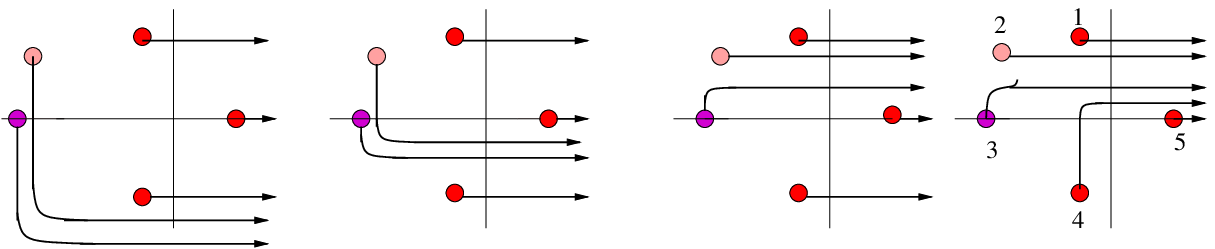}{Left mutations of the exceptional collection 
on ${\cal B}_{2}$ to obtain an exceptional 
collection for soliton counting.}
\begin{eqnarray} 
\{V_{1},V_{2},V_{3},\widehat{V}_{1},\widehat{V}_{2}\}&\mapsto& 
\{V_{1},V_{2},L_{V_{3}}(\widehat{V}_{1}),L_{V_{3}}(\widehat{V}_{2}),V_{3}\}
\\ \nn
&\mapsto& \{V_{1},L_{V_{2}}L_{V_{3}}(\widehat{V}_{1}),L_{V_{2}}L_{V_{3}}(\widehat{V}_{2}),V_{2},V_{3}\}\\ \nn
&\mapsto& \{V_{1},L_{V_{2}}L_{V_{3}}(\widehat{V}_{1}),L_{V_{2}}L_{V_{2}}(\widehat{V}_{2}),L_{V_{2}}(V_{3}),V_{2}\}=: \{E_{1},E_{2},E_{3},E_{4},E_{5}\}\nn\,.
\end{eqnarray}
The soliton numbers between different vacua are now given by
$\chi(E_{i},E_{j})$ where the vacua are labeled in the counter clockwise
direction as shown in \figref{b2-e},
\be
S_{ij}=\chi(E_{i},E_{j})=\pmatrix{1 & -1 & -1 &  -3  & +3 
\cr 0 & ~~1 & ~~0 & ~~1 & -2 \cr 0 & ~~0 & ~~1 & ~~1 & -2
\cr 0 &~~0 & ~~0 & ~~1 & -3\cr 0 &~~0 &~~0 &~~0& ~~1}\,.
\ee
We can read off the number of solitons from the above matrix.
Note that the matrix $S$ will change by a braid transformation
under the change of parameters. Thus the matrix
$S$ given above is the soliton counting matrix as long as the
convexity of the critical values shown in \figref{b2-e} is maintained.

\subsubsection{${\cal B}_{3}$}
Now we will consider the case of $\PP^{2}$ blown up at three points, ${\cal
B}_{3}$. The superpotential of the mirror LG theory is
\be
W_{{\cal B}_{3}}(X_{1},X_{2})=X_{1}+X_{2}+\frac{e^{-t}}{X_{1}X_{2}}+\frac{e^{-t_{E_{1}}}}{X_{1}}+\frac{e^{-t_{E_{2}}}}{X_{2}}+e^{-t_{E_{3}}}X_{1}X_{2}\,.
\ee
$t_{E_{i}}$ are the complexified K\"ahler parameters of the
exceptional curves $E_{i}$. After rescaling the coordinates we can write the above 
superpotential as,
\be
W_{{\cal
B}_{3}}(X_{1},X_{2})=e^{-\frac{t}{3}}(X_{1}+X_{2}+\frac{1}{X_{1}X_{2}}+
\frac{e^{-t_{E_{1}}+\frac{2}{3}t}}{X_{1}}+
\frac{e^{-t_{E_{2}}+\frac{2}{3}t}}{X_{2}}+e^{-t_{E_{3}}-\frac{1}{3}t}X_{1}X_{2})\,,
\ee
We denote by $\mu_{1},\mu_{2}$ and $\mu_{3}$ the three parameters
$e^{-t_{E_{1}}+\frac{2}{3}t}, e^{-t_{E_{2}}+\frac{2}{3}t}$ and
$e^{-t_{E_{3}}-\frac{1}{3}t}$ respectively.  The positions of the
critical values are determined by the $\mu_{i}$ and $e^{-\frac{t}{3}}$
determines the overall scale.  To determine the critical values let
$\mu_{1}=\mu_{2}=\mu_{3}=\mu$.  Then the critical points and the
critical values are
\begin{eqnarray} \nn
(X_{1},X_{2})&=&\{(e^{\frac{2\pi i k}{3}},e^{\frac{2\pi i k}{3}})\,,\,(\mu^{2},-\fracs{{1}}{{\mu}})~|~k=0,1,2\}\,, \\
W_{{\cal B}_{3}}(X_{1},X_{2})&=&\{3\,e^{-\frac{t}{3}}(e^{\frac{2\pi i k}{3}}+\mu e^{-\frac{2\pi i k}{3}}),-e^{-\frac{t}{3}}(\fracs{{1}}{{\mu}}+\mu^{2}) ~|~k=0,1,2\}.
\end{eqnarray}
The critical value $w_{4}$ is degenerate and the three critical values
at this point can be separated by taking $\mu_{i}\neq \mu_{j}$.  For
$\mu$ close to zero we see that as $\mu\mapsto \mu e^{2\pi i}$ the
critical values $w_{1},w_{2},w_{3}$ move in a closed path close to the
original critical value. On the other hand the critical value $w_{4}$
moves around the origin in the clockwise direction.  The solution
given above for the critical values in terms of $\mu$ is an exact
solution. For $\mu_{i}\neq \mu_{j}$ we can construct approximate
solutions that specify the behavior of critical values for
$|\mu_{i}|$ small. This was done for the case of ${\cal B}_{1}$ and
${\cal B}_{2}$ and the result here is similar to that case. We denote
the three degenerate critical values by $\widehat{w}_{i}$ such that
$\mu_{i} \mapsto 0$ implies $\widehat{w}_{i}\mapsto -\infty$. Under
the transformation $\mu_{3}\mapsto \mu_{3}\,e^{2\pi i}$,
$\widehat{w}_{3}$ undergoes a clockwise rotation around the origin as
shown in \figref{b3-c}. Since taking $\mu_{1},\mu_{2}\mapsto 0$ has
no effect on the three symmetrically located critical values therefore
we can identify the corresponding bundles as the pull backs of ${\cal
O}(-1),{\cal O}(0)$ and ${\cal O}(1)$ from $\PP^{2}$ to ${\cal
B}_{3}$, we will denote these bundles by $V_{1},V_{2}$ and $V_{3}$
respectively. Similarly from \figref{b3-c} and charge conservation we see that the bundle associated with
$\widehat{w}_{i}$ denoted by $\widehat{V}_{i}$ is such that,
\be
\mbox{ch}(\widehat{V}_{3})=(0,-E_{3},-\frac{1}{2})\,.
\ee
\onefigure{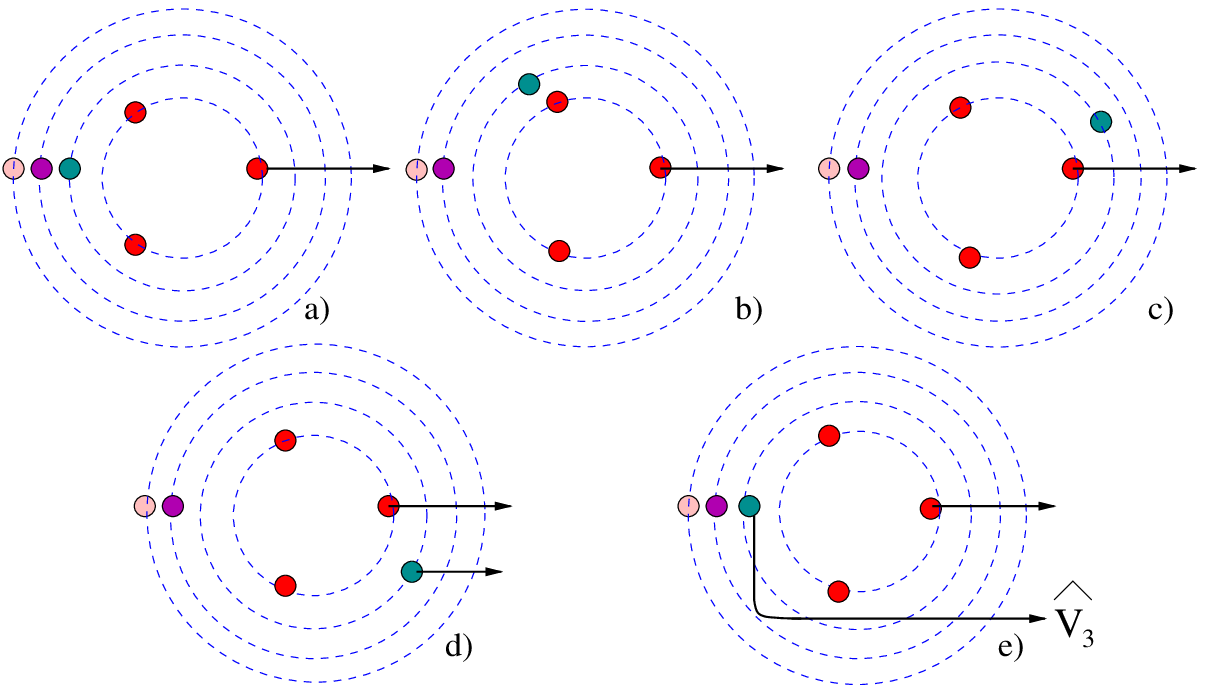}{Brane creation as the critical value $\widehat{w}_{3}$
undergoes a rotation for $\mu_{3}\mapsto \mu_{3}e^{2\pi i}$.}
It is easy to check that
$\{V_{1},V_{2},V_{3},\widehat{V}_{1},\widehat{V}_{2},\widehat{V}_{3}\}$
is an exceptional collection. The semi-infinite lines corresponding to
this collection are shown in \figref{b3-d}{}.
\onefigure{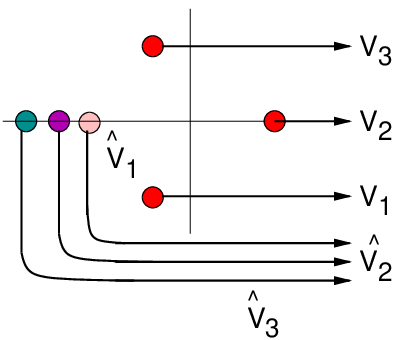}{The mirror of the exceptional collection 
$\{V_{1},V_{2},V_{3},\widehat{V}_{1},\widehat{V}_{2},\widehat{V}_{3}\}$.}

To calculate the number of solitons between the vacua we need to
transform this collection into the one shown in \figref{b3-e}.
\onefigure{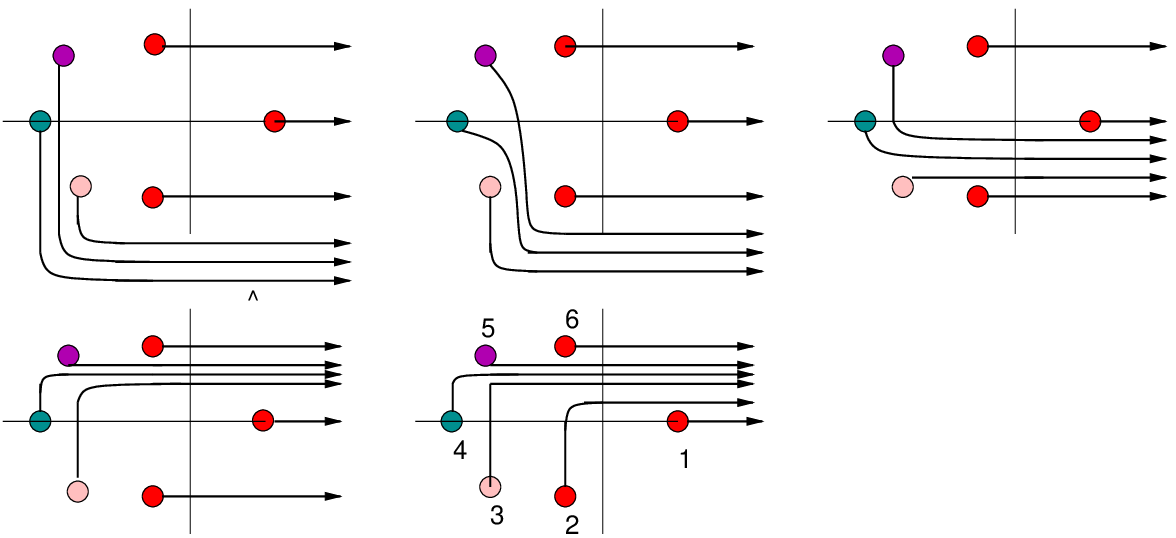}{Left mutations of the exceptional collection 
(shown in \figref{b3-d}) to obtain an
exceptional collection for soliton counting.} We see that the exceptional 
collection shown
in \figref{b3-d} is obtained from
$\{V_{1},V_{2},V_{3},\widehat{V}_{1},\widehat{V}_{2},\widehat{V}_{3}\}$
by successive left mutations. Note that since
$\chi(\widehat{V}_{i},\widehat{V}_{j})=\delta_{ij}$ therefore we can
move the corresponding branes through each other without generating
any new branes,
\begin{eqnarray} 
\{V_{1},V_{2},V_{3},\widehat{V}_{1},\widehat{V}_{2},\widehat{V}_{3}\} 
&\mapsto &
\{V_{1},V_{2},V_{3},\widehat{V}_{2}, \widehat{V}_{3}, \widehat{V}_{1}\} 
\\ \nn
&\mapsto&\{V_{1}, V_{2},L_{V_{3}}(\widehat{V}_{2}),
L_{V_{3}}(\widehat{V}_{3}),L_{V_{3}}(\widehat{V}_{1}),V_{3}\}  \\ \nn
&\mapsto& \{V_{1},L_{V_{2}}L_{V_{3}}(\widehat{V}_{2}),L_{V_{2}}L_{V_{3}}(\widehat{V}_{3}),L_{V_{2}}L_{V_{3}}(\widehat{V}_{1}),V_{2},V_{3}\} \\ \nn
&\mapsto& \{V_{2},L_{V_{2}}L_{V_{3}}(\widehat{V}_{2}),
L_{V_{2}}L_{V_{3}}(\widehat{V}_{3}),L_{V_{2}}L_{V_{3}}(\widehat{V}_{1}),L_{V_{2}}(V_{3}),V_{2}\}  \\ 
&=:&\{E_{1},E_{2},E_{3},E_{4},E_{5},E_{6}\}\,. \nn
\end{eqnarray}
The soliton counting matrix is given by
\be
S_{ij}=\chi(E_{i},E_{j})=\pmatrix{1 & -1 & -1 & -1 &-3 & +3 \cr 0 & ~~1
& ~~0 &~~0 & +1 & -2 \cr 0 & ~~0 & ~~1 & ~~0 & ~~1 & -2 \cr 0 & ~~0 &~~0 & ~~1 & ~~1 &
-2\cr 0 & ~~0 & ~~0 &~~0 & ~~1 & -3 \cr 0 & ~~0 &~~0 &~~0 &~~0& ~~1}
\ee
Note that this matrix gives the number of solitons only for those
values of the parameters for which we can have the convex
configuration shown in \figref{b3-e}. 

\subsection{$F_{0}=\PP^{1}\times \PP^{1}$}

Since $F_{0}$ is a product manifold the superpotential of the mirror
LG theory is just two copies of the superpotential of LG theory mirror
to $\PP^{1}$,
\bea \nn
W_{F_{0}}(X_{1},X_{2})=X_{1}+X_{2}+\frac{e^{-t_{1}}}{X_{1}}+\frac{e^{-t_{2}}}{X_{2}}\,.
\eea
$t_{1}$ and $t_{2}$ are complexified K\"ahler parameters of the two $\PP^{1}$'s.
After rescaling the variables we can write the above superpotential as
\be
W_{F_{0}}(X_{1},X_{2})=e^{-\frac{t_{1}}{2}}(X_{1}+\frac{1}{X_{1}})+
e^{-\frac{t_{2}}{2}}(X_{2}+\frac{1}{X_{2}})\,.
\ee
The critical points and the corresponding critical values of above
superpotential are,
\bea \nn
(X_{1},X_{2})&=&\{(1,1)\,,\,(1,-1)\,,\,(-1,1)\,,\,(-1,-1)\}\,, \\ \nn
W(X_{1},X_{2})&=&w_{i}=\{ 2e^{-\frac{t_{1}}{2}}+2e^{-\frac{t_{2}}{2}}\,,\,
2e^{-\frac{t_{1}}{2}}-2e^{-\frac{t_{2}}{2}}\,,\,
-2e^{-\frac{t_{1}}{2}}+2e^{-\frac{t_{2}}{2}}\,,\,
-2e^{-\frac{t_{1}}{2}}-2e^{-\frac{t_{2}}{2}}\}\,.
\eea
Without loss of generality we assume that $|e^{-\frac{t_{1}}{2}}|\geq
|e^{-\frac{t_{2}}{2}}|$. To determine the bundles associated with the
cycles ${\cal C}_{i}$ (which are the preimages of the semi-infinite
lines in the W-plane) we consider the configuration with
$e^{-\frac{t_{1}}{2}}$ and $e^{-\frac{t_{2}}{2}}$ real. The critical
values in the W-plane in this case lie on the real axis and are
 non-degenerate as long as $e^{-\frac{t_{1}}{2}} \neq
e^{-\frac{t_{2}}{2}}$ as shown in \figref{p1p1-c}.
\onefigure{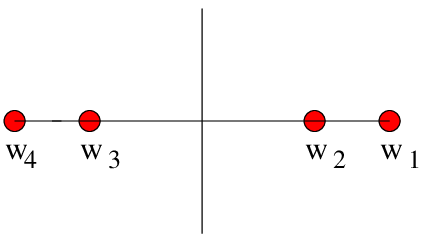}{Critical values when $e^{-\frac{t_{1}}{2}}$ and 
$e^{-\frac{t_{2}}{2}}$ are real.}
We know that the cycle ${\cal C}_{1}$ (whose image in the W-plane is the
semi-infinite line starting at $w_{1}$) is mirror to the trivial
bundle ${\cal O}(0,0)$ \footnote{A bundle ${\cal O}(a,b)$ on $F_{0}$
is a rank one bundle with first Chern class $c_{1}=a\,l_{1}+b\,l_{2}$,
where $l_{1}$ and $l_{2}$ are the generators of $H_{2}(F_{0})$ such
that $l_{1}\circ l_{1}=l_{2}\circ l_{2}=0$ and $l_{1}\circ
l_{2}=1$.}. Note that the transformation $e^{-t_{i}}\mapsto e^{-t_{i}}
e^{2\pi i}$ has the following effect on the critical values,
\bea \nn
e^{-t_{1}}\mapsto e^{-t_{1}} e^{2\pi i}\,\,\,\, \, \Longrightarrow \,\,\, \,\,\, w_{1} \leftrightarrow w_{3}\,,\,\, w_{2} \leftrightarrow w_{4}\,,\\
e^{-t_{2}}\mapsto e^{-t_{2}} e^{2\pi i} \,\,\,\,\, \Longrightarrow  \,\,\,\,\,\, w_{1} \leftrightarrow w_{2}\,,\,\, w_{3} \leftrightarrow w_{4}\,.
\label{trans}
\eea
To determine the bundle mirror to the cycle ${\cal C}_{2}$ (whose
image in the W-plane is the semi-infinite line starting at $w_{2}$ and
going to infinity along $e^{i\epsilon}$ ($\epsilon <<1$) we perform
the second transformation given above. The effect of this
transformation on the image of cycle ${\cal C}_{1}$ in the W-plane is
shown in \figref{p1p1-d}.
\onefigure{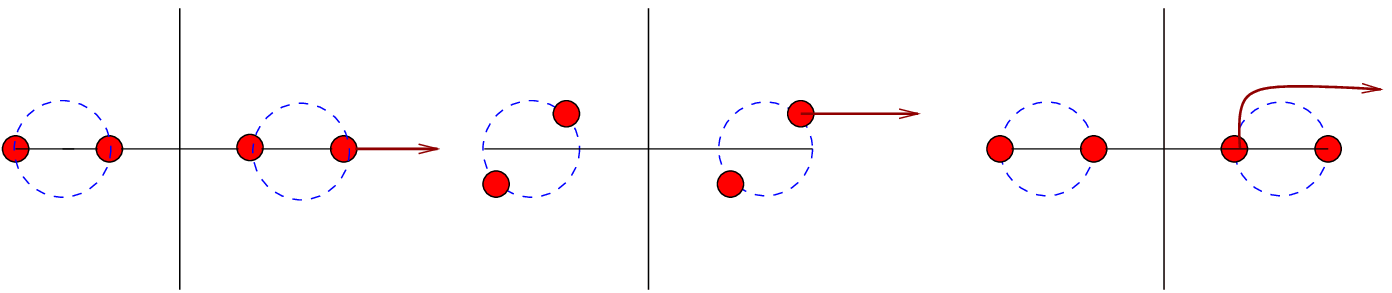}{The effect of transformation $e^{-t_{2}}\mapsto e^{-t_{2}}e^{2\pi i}$
on the critical values.}  Thus the bundle,$V_{2}$, mirror to ${\cal
C}_{2}$ is such that,
\be
\mbox{ch}(V_{2})=\mbox{ch}({\cal O}(0,0))+\mbox{B-field}=e^{c_{1}({\cal O}(0,0))-l_{2}}=1-l_{2}=(1,-l_{2},0)\,.
\ee
Thus we can identify $V_{2}$ with ${\cal O}(0,-1)$.
The effect of the first transformation of eq.\,(\ref{trans}) on the cycle
${\cal C}_{1}$ is shown in \figref{p1p1-e}.
\onefigure{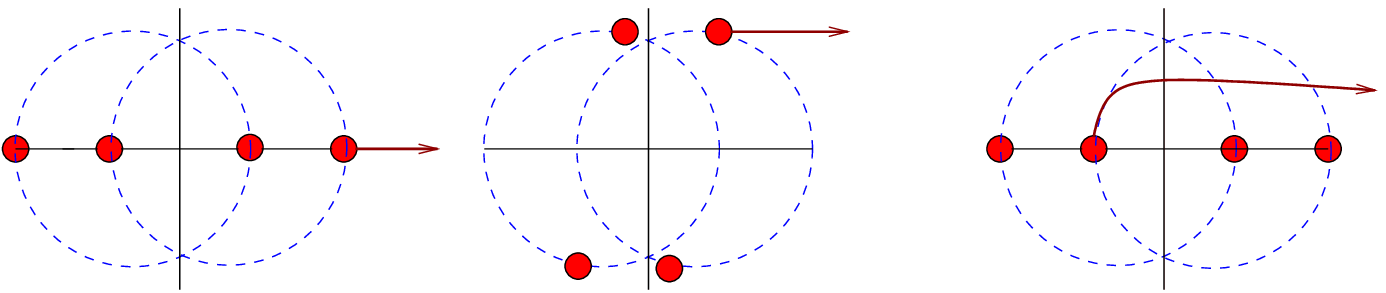}{The effect of transformation $e^{-t_{1}}\mapsto e^{-t_{1}}e^{2\pi i}$ on
the critical values.}  Thus by charge conservation we see that the
bundle,$V_{3}$, mirror to ${\cal C}_{3}$ is such that
\be
\mbox{ch}(V_{3})=\mbox{ch}({\cal O}(0,0))+\mbox{B-field}=e^{c_{1}({\cal O}(0,0))-l_{1}}=1-l_{1}=(1,-l_{1},0)\,.
\ee
Thus we can identify $V_{3}$ with ${\cal O}(-1,0)$. Now consider the
transformation 
\bea
(e^{-t_{1}},e^{-t_{2}})\mapsto (e^{i\theta} e^{-t_{1}},e^{i\theta}e^{-t_{2}})\,,\,\,\,\theta\in[0,2\pi]\,.
\label{both}
\eea
 The effect of this transformation on the critical values and the cycle
 ${\cal C}_{1}$ is shown in \figref{p1p1-f}.
\onefigure{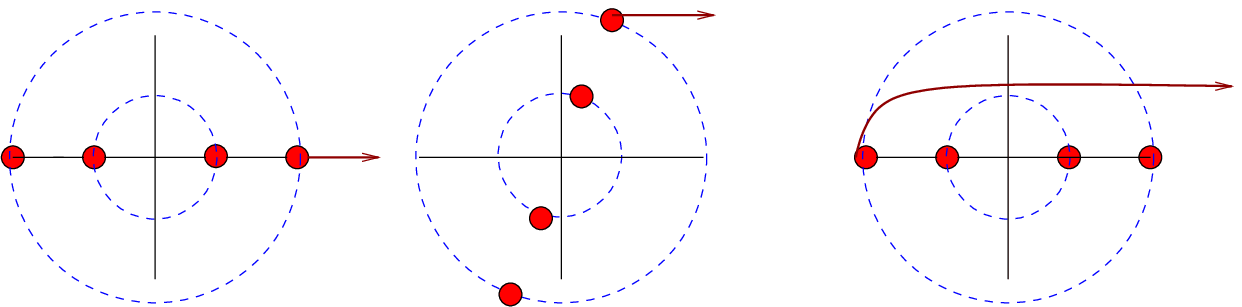}{The effect of the combined transformation
$(e^{-t_{1}},e^{-t_{2}})\mapsto e^{2\pi i}(e^{-t_{1}},e^{-t_{2}})$.}
Thus charge conservation implies that the bundle $V_{4}$ mirror to
the cycle ${\cal C}_{4}$ is such that
\bea
\mbox{ch}(V_{4})=\mbox{ch}({\cal O}(0,0))+\mbox{B-field}=e^{c_{1}({\cal O}(0,0))-l_{1}-l_{2}}=(1,-l_{1}-l_{2},1)\,.
\eea
Thus we can identify $V_{4}$ with ${\cal O}(-1,-1)$.

The set of bundles $\{{\cal O}(-1,-1),{\cal O}(-1,0),{\cal O}(0,-1),{\cal
O}(0,0)\}$ is an exceptional collection and using it we can calculate
the soliton counting matrix. But first since we do not want any three
vacua to be collinear we deform the configuration using eq.\,(\ref{both}) for
$|\theta|<<1$. And also we need to transform this collection of
exceptional bundles into the one shown in \figref{p1p1-g} 
by left or right mutations.
\onefigure{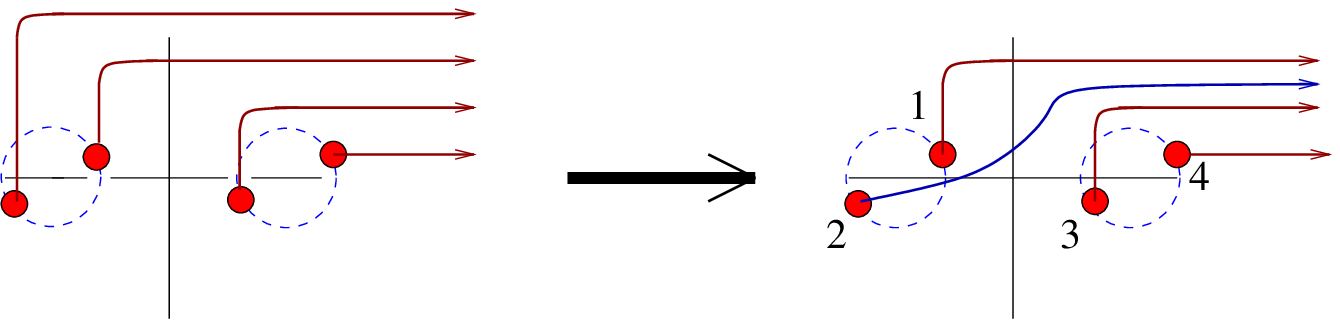}{A right mutation of the exceptional collection
to obtain another exceptional collection for soliton counting.}
\bea \nn
\{ {\cal O}(-1,-1)\,,\,{\cal O}(-1,0)\,,\,{\cal O}(0,-1)\,,\,{\cal O}(0,0)\} 
\mapsto ~~~~~~~~~~~~~~~~~~~~~~~~~~~~~~~~~~~~~~~~~~~~\\ \nn
 \{{\cal O}(-1,0)\,, R_{{\cal O}(-1,0)}({\cal O}(-1,-1))\,,\,
{\cal O}(0,-1)\,,\,{\cal O}(0,0)\}=:\{E_{1}\,,\,E_{2}\,,\,E_{3}\,,\,E_{4}\}\,.
\eea
The soliton counting matrix $S_{ij}$ is then given by
\bea
S_{ij}=\chi(E_{i},E_{j})=\pmatrix{1 & -2 & 0 & 2\cr 0 & ~~1 &2 &0 \cr 0 &~~0& 1 & 2\cr 0 &~~0 &0 &1}\,.
\eea
\subsection{Higher dimensional toric Fano varieties}

As an example of higher dimensional toric Fano varieties we consider
the blow ups of projective spaces. Blow up of $\PP^{n-1}$ upto $n$
points is a toric Fano variety \cite{sato}.  Each blow up corresponds
to replacing a point by $\PP^{n-2}$ and thus each blow
up introduces $n-2$ new
cohomology elements.
 We consider the case of
maximal blow ups since others can be obtained from this one as we saw
for the case of two dimensional del Pezzo surfaces. 

The linear sigma model is a $U(1)^{n}$ gauge theory. The $2n$ chiral
superfields have the following charge assignment under the $U(1)^{n}$
gauge group,
\bea
Q_{1}&=&(1,1,1\cdots1,1; ~~0,~~0,0,\cdots0,~~0)\,,\\ \nn
Q_{2}&=&(0,1,1\cdots1,1;-1,~~0,0,\cdots0,~~0),\,\\ \nn
Q_{3}&=&(1,0,1\cdots1,1;~~0,-1,0,\cdots0,~~0), \\ \nn
\vdots \\ \nn
Q_{n}&=&\underbrace{(1,1,1\cdots1,0}_{n};\underbrace{~~0,~~0,0, \cdots,0,-1}_{n})\,. \nn
\eea The LG superpotential is then given by
\bea 
W(X)=\sum_{i=1}^{n-1}X_{i}+\frac{e^{-t}}{X_{1}\cdots X_{n-1}}+\sum_{i=1}^{n-1}
\frac{e^{-t_{i}}}{X_{i}}+e^{-t_{n-1}}X_{1}\cdots X_{n-1}\,. \nn
\eea
After rescaling $X_{i}$ we can write the above superpotential as
\be
W(X)=e^{-\frac{t}{n}}(\sum_{i=1}^{n-1}X_{i}+\frac{1}{X_{1}\cdots X_{n-1}}
+\sum_{i=1}^{n-1}\frac{e^{-t_{i}+\frac{2}{n}t}}{X_{i}}+e^{-t_{n-1}+\frac{n-2}{n}t}X_{1}\cdots X_{n-1})\,.
\ee
Let $\mu_{i}=e^{-t_{i}+\frac{2}{n}t}$ for $i=1,\cdots,n-2$ and
$\mu_{n-1}=e^{-t_{n-1}+\frac{n-2}{n}t}$. Consider the case when
$\{\mu_{i}=\mu~|~i=1,\cdots,n-1\}$, in this case there are $2(n-1)$
critical points given by
\be
X_{i}=f\,,\,\,(\mu\,f^{n-2}+1)(f^{n}-1)=0\,.
\ee
$n$ of these critical points are also the critical point of the
$\PP^{n-1}$ superpotential.  The new critical points and critical
values are
\bea \nn
X^{(k)}_{i}&=& \mu^{-\frac{1}{n-2}}e^{ \frac{i\pi (2k+1)}{n-2}} \,,\,\, i=1,\cdots,n-1,\,\,k=1,\cdots,n-2,\\ 
\widehat{w}_{k}&=&W(X^{(k)})=n( \mu^{-\frac{1}{n-2}}e^{ \frac{i\pi (2k+1)}{n-2}}+\mu^{n-1}e^{-\frac{i\pi (2k+1)}{n-2}})
\eea
For $\mu_{i}\neq \mu_{j}$ each of the above new critical points splits
up into $n-1$ critical points. Thus the critical value
$\widehat{w}_{k}$ is degenerate with multiplicity $n-1$. If
$\mu_{i}\mapsto 0$ then $n-2$ critical values go to infinity and the
multiplicity of $\widehat{w}_{k}$ reduces to $n-2$. To determine the
bundles corresponding to the lines ending on these new critical values
we only need to consider the case when $\mu_{1}\neq 0$ and $\mu_{i}=0$
for $i=2,\cdots,n-1$.  This is the case of $\PP^{n-1}$ blown up at one
point. In this case there are $2n-2$ non-degenerate critical points
given by the following equation
\be \nn
X_{i}=f\,,\,\, \mu_{1}f^{2n-2}+f^n-1=0\,,
\ee
For $\mu_{1}$ very small we can write the leading terms in the solution as
\bea \nn
X^{(k)}_{i}&=&f_{k}=e^{\frac{2\pi i k}{n}}+ O(\mu_{1})\,, \, X^{(k')}_{i}=f_{k'}=\mu_{1}^{-\frac{1}{n-2}}e^{\frac{i \pi (2k'-1)}{n-2}} + O(\mu_{1})\,,\\ \nn
w_{k}&=&W(X^{(k)}_{i})=n\,e^{\frac{2\pi i k}{n} }+ O(\mu_{1})\,, \,
\widehat{w}_{k'}=W(X^{(k')}_{i})=\mu_{1}^{-\frac{1}{n-2}}\,e^{\frac{i\pi (2k'-1)}{n-2}} + O(\mu_{1})\,,
\eea
where $k\in \{0,\cdots,n-1\}$ and $k'\in \{1,\cdots ,n-2\}$.  Thus we
see that as $\mu_{1}\mapsto \mu_{1}e^{2\pi i}$ the critical values
$\widehat{w}_{k'}$ are rotated by $e^{-\frac{2\pi i}{n-2}}$.  Hence
$w'_{k'}$ is mapped to $w'_{k'-1}$ by this transformation as shown in \figref{cpn-f} for the case of $\PP^{5}$.
\onefigure{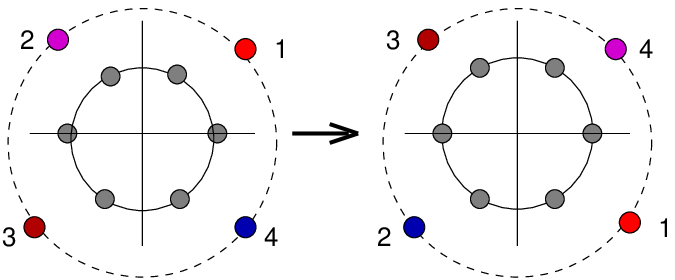}{The effect of the transformation 
$\mu_{1}\mapsto \mu_{1}e^{2\pi i}$ on the critical 
values $\widehat{w}_{i}$ for $\PP^{5}$.}

Consider the case of \figref{cpn-g} where we have the D-brane
corresponding to the bundle ${\cal O}(0)$ ending on the critical value
$w_{0}$ on the positive real axis. After the transformation
$\mu_{1}\mapsto \mu_{1}e^{2\pi i}$ we create another D-brane whose
image in the W-plane is the semi-infinite line starting at $w'_{n-2}$.
\onefigure{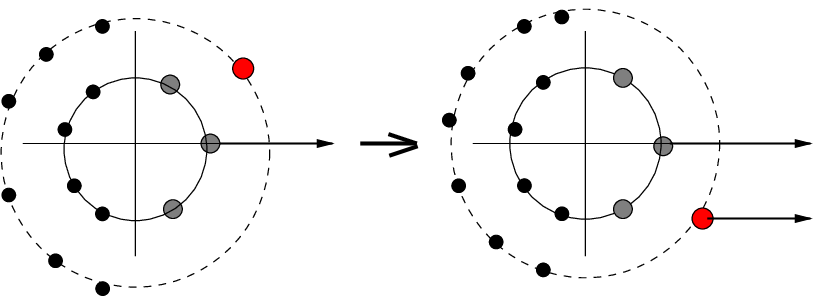}{Brane creation as $\widehat{w}_{1}$ passes through
the mirror of the trivial line bundle.}
Denote by $B$ the cohomology class dual to the exceptional divisor and by $V_{1}$ the new bundle created  
then by charge conservation it follows that,
\bea
\mbox{ch}({\cal O}(0)\oplus V)=e^{c_{1}({\cal O}(0))-B} \Longrightarrow \mbox{ch}(V)=e^{-B}-1\,,\\
c_{0}(V)=0\,,\,c_{1}(V)=-B\,,\,c_{i}(V)=0\,,\,i=2,\cdots,n\,.
\eea
Thus $V$ is the $-[{\cal O}(0)]$ bundle on $\PP^{n-2}$.  A transformation
$\mu_{1}\mapsto \mu_{1}\,e^{2\pi i m}$ maps it to
$\widehat{w}_{n-2-m}$,
thus the bundle which corresponds to the semi-infinite line starting
at $\widehat{w}_{k'}$ is the line bundle ${\cal O}(n-2-k')$ on
$\PP^{n-2}$, a sheaf on $\PP^{n-1}$ with support on the exceptional divisor. The case of $\PP^{5}$ is shown in \figref{cpn-h}.
\onefigure{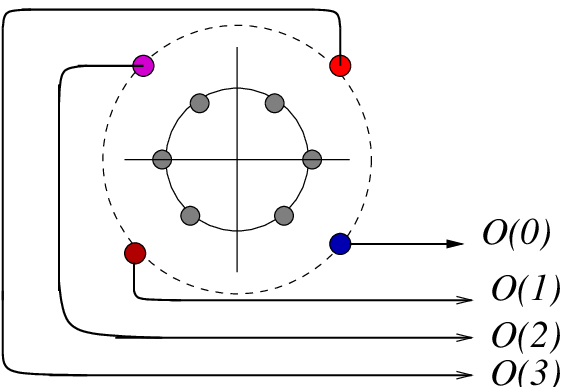}{Bundles which are mirror of the lines starting at $\widehat{w}_{i}$.}

\section{D-Brane in String Theory and Mirror Symmetry}

It is natural to ask how the map between
D-branes corresponding to sheaves and the Lagrangian
submanifolds in the Landau-Ginzburg model
 works in the case of conformal theories.
In the context of Gepner models, the structure of Cardy
states and their sigma model interpretation have
been studied \cite{gepcar}.
 our constructions of Cardy
states in terms of Lagrangian submanifolds
of LG models
lead to a deeper geometric insight in this regard.
Since we have considered the case of minimal models
in detail, and Gepner model is an orbifold of
their tensor products, it is straightforward to
identify the relevant D-branes in the orbifold LG
model.

There is however, another case of the conformal theory we can consider
namely the non-compact CY manifolds.
 These are the cases of most interest
in the context of geometric engineering of QFT's.
As an example of this class consider 
the total space of the canonical line bundle over
a compact Fano variety.  
This space is a non-compact
Calabi-Yau manifold.
We find relations between the  LG theories
mirror to the superconformal
sigma
model on the non-compact Calabi-Yau and the sigma model on the
Fano variety.

For simplicity let us assume
that the Fano variety is given by the weighted projective
space with weights ($q_i>0$).  Then the total space of the canonical bundle
is
captured by a linear sigma model with a single $U(1)$
gauge theory with matter fields with charges
$(-\sum_i q_i,q_1,q_2,...,q_n)$.  The mirror of this
is an LG theory \cite{HV} with of $n+1$
variables with superpotential,
\be 
W=\sum_{i=0}^n x_i \qquad {\rm subject}\quad {\rm to} \ \ \prod_{i=1}^{n} 
x_i^{q_i}
=e^{-t}x_0^{\sum q_i}\,.
\ee
Recall that the correct field variables are $Y_i$ where $x_i=e^{-Y_i}$.
For simplicity let us assume one of the charges say $q_n=1$ (the more general
case can also be done with the additional complication
of introducing orbifold groups).  Then we can also write the
above superpotential as
\be
W=x_0[1+\sum_{i=1}^{n-1} {\widehat x}_i+{e^{-t}\over 
\prod_{i=1}^{n-1} {\widehat x}_i^{\,q_i}}]\,.
\ee
Where ${\hat x_i}=e^{-{\widehat Y}_i} $ and ${\widehat Y}_i=Y_i-Y_0$.
This change of fields is linear and introduces no Jacobians in the
field measure.   Now, as far as
periods and BPS states which are sensitive
only to period integrals
 $\int \prod d\phi_j e^{-W}$  are concerned this LG theory is equivalent
to the LG theory given by
\be
W=x_0[1+\sum_{i=1}^{n-1} {\widehat x}_i+{e^{-t}\over 
\prod_{i=1}^{n-1} {\widehat x}_i^{\,q_i}}-uv]\,,
\label{compact}
\ee
where now $x_0$ is the right field variable (i.e. $x_0\in {\bf C}$ rather
than ${\bf C}^*$), and $u,v$ are chiral fields also taking value
in ${\bf C}$.  To see this note that in the BPS computations
integrating over the $u,v$ fields leads to a $\frac{1}{x_0}$ in the measure
which combined with $dx_0$ converts it back to the measure appropriate
for $x_0$ taking values in ${\bf C}^*$, and leading to the previous
LG periods.  Having established their equivalence (at least in the
weak sense discussed in \cite{HV}), in the period integrals we
can integrate out $x_0$ in the new version of the LG theory, and
obtain a $\delta (1+\sum_{i=1}^{n-1} {\widehat x}_i+{e^{-t}\over 
\prod_{i=1}^{n-1} {\widehat x}_i^{\,q_i}}-uv)$.  Thus we see that as far
as the BPS data is concerned
the mirror of the sigma model on
the non-compact CY,
which is originally the LG model, is equivalent to
the sigma model on another non-compact Calabi-Yau given by
\be
f({\widehat x}_i)=uv \qquad {\rm where} \qquad f({\widehat x}_i)=1+
\sum_{i=1}^{n-1} {\widehat x}_i+{e^{-t}\over 
\prod_{i=1}^{n-1} {\widehat x}_i^{\,q_i}}\,,
\ee
where ${\hat x}_i$ take values in ${\bf C}^*$ but $u,v$ are variables
in ${\bf C}$. 
Note that this non-compact mirror CY has dimension $n$ which
is the same as the dimension of the original non-compact CY.
The holomorphic $n$ form can be viewed as
\be
\Omega ={\prod_{i=1}^n{d{\widehat x}_i\over {\widehat x}_i}du dv\over df}=
{\prod_{i=1}^n{d{\widehat x}_i\over {\widehat x}_i}du \over u}\,.
\ee
It is this version of local mirror symmetry
that was first discovered in the literature \cite{kklmv,klmvw}.

We see a
striking resemblance between the non-conformal sigma model on the
Fano variety and the conformal sigma model on the total space
of the canonical line bundle over the Fano variety.  In particular
the $f({\hat x}_i)$ appearing in the above formula is precisely the superpotential
for the LG theory mirror to the non-conformal sigma model on the Fano
variety.

Even though we have presented the above discussion in the
context of a linear sigma model with a single $U(1)$, it can
be easily generalized to the case with more $U(1)$'s as well
as with extra superpotentials corresponding to complete
intersections.

\subsection*{\it BPS states and local mirror symmetry}

In computing the BPS states in such local contexts the idea
developed in \cite{klmvw} was to consider supersymmetric mid-dimensional
cycles on the mirror. Moreover one could simplify the counting
of such cycles by considering fibration structure of the non-compact
CY and studying the supersymmetric cycles on the fibers and consider
the effective tension of the branes as one varies over the base.  In
this way it was shown in \cite{klmvw}, 
in the context of local mirror of $SU(N)$ gauge
theories, how the problem is translated to finding minimal energy
string configurations (with varying tensions) on a Riemann
surface given by $f(x_1,x_2)=0$.  This was implemented in detail
for the $SU(2)$ case where various expected properties of BPS states
in the corresponding ${\cal N}=2$ gauge theory in 4 dimensions \cite{witten}
was recovered including
the decay of certain BPS states.  
Further applications along these lines have been
considered \cite{rabin,warners,gvw,shapv}.

We can also connect the above description to the BPS states
for the probes in the context of F-theory, which is the subject
of the next section.

\subsection{Local mirror symmetry and F-theory}

In this section we will see that the W-plane geometry of LG theory
mirror to sigma model with certain non-compact CY threefolds as target
space is closely related to some F-theory backgrounds \cite{vafa-Ftheory}. 
The link we find is as follows:
The BPS states
on the non-compact CY threefold side are D-branes wrapped on compact
even dimensional cycles.  As discussed above these get transformed
on the mirror side to certain 3-cycles in a non-compact 
CY 3-fold.  For the particular backgrounds of interest
the non-compact CY 3-fold itself has a simple ${\bf C}^*$ fibration
structure over CY 2-fold (a local description of elliptic
$K3$).  The image of the closed 3-cycles get mapped to minimal
2-cycles in this geometry, which can possibly have boundaries
where the ${\bf C}^*$ fibration degenerates.  This in turn
can be viewed as computation of BPS state in a certain F-theory
background with a 3-brane probe (placed
where the ${\bf C}^*$ fibration degenerates).

We will first review the probe theory description
for F-theory and its BPS states
and then give some examples of non-compact CY manifolds and the
corresponding probe theory.

\subsubsection{Probe theory and BPS states}

~~~~
Consider a manifold ${\cal X}$ which is an elliptic fibration over the
complex plane $B$
\beq
y^{2}\,=\,x^{3}+f(z)\,x+g(z)\,,\,\,\,\,z\in B\,,
\label{weis}
\eeq
provided with a non-vanishing holomorphic 2-form $\Omega$
\beq
\Omega\,=\,\lambda \,\,dz,
\eeq
where $\lambda=\frac{dx}{y}$ is the holomorphic 1-form on the elliptic fibers.
F-theory compactification on ${\cal X}$ is equivalent to type IIB
compactification on the base $B$ with a varying coupling constant
$\tau$ (defined up to $\sl2z$ transformations)
given by the complex structure of the elliptic fiber. The
position of the degenerate elliptic fibers on the base is given by the
zeroes of the discriminant, $\Delta(z)$, of the elliptic fibration
(\ref{weis}),
\be
\Delta(z)=4\,f(z)^{3}+27\,g(z)^{2}\,.
\ee
{}From Picard Lefshetz theory we know that as we go around the position
of a degenerate fiber in the base, the complex structure parameter
$\tau$ undergoes an $\sl2z$ transformation. As mentioned before this
complex structure parameter is identified with the coupling constant of
type IIB. Since in type IIB monodromies associated with 7-branes
transform $\tau$ by $\sl2z$ transformations, the position of
a degenerate fiber on the base, in type IIB, is associated with a
7-brane. $\sl2z$ symmetry of type IIB then implies the existence of a
family of 7-branes labelled by two relatively prime integers, $(p,q)$.
As for the case of 7-brane, a $(p,q)$ 7-brane at a point $z_{*}$ can
be associated, in F-theory, with an elliptic fiber over $z_{*}$,
$T^{\,2}_{z_{*}}$, whose degenerating 1-cycle is $p\alpha + q\beta \in
H_{1}(T^{\,2}_{z_{*}},\z)$. 
 
In ref.\cite{bds}, $\cal N$=2 $SU(2)$ Seiberg-Witten theory was
interpreted as the worldvolume theory of a D3-brane in the presence of
mutually non-local 7-branes. A BPS state of charge $(p,q)$ in the
D3-brane theory is a BPS string or a BPS string junction of total
asymptotic charge $(p,q)$ with support on 7-branes and ending on the
D3-brane. In the F-theory picture D3-brane lifts to a regular elliptic
curve of ${\cal X}$. 
Strings or string junctions stretched between the
7-branes and the D3-brane are, in F-theory, two real dimensional
curves in the manifold ${\cal X}$ with or without boundary depending
on whether the string junction ends on the D3-brane or not\footnote{
This description follows from the connection 
between F-theory compactified on a circle and the M-theory
in one lower dimension.  In the M-theory description
the D3-brane probe gets mapped to an M5 brane wrapped
over the corresponding elliptic fiber and the BPS states
are M2 branes wrapped over 2-cycles of the $K3$ geometry,
possibly ending on the M5 brane.  The image of the M2 brane
projected on the $z$-plane gives the string junction
description in the type IIB setup.}. If the
string junction ends on a D3-brane the corresponding curve in F-theory
has a boundary on the elliptic curve above the position of the
D3-brane. The homology cycle of the boundary is determined by the
$(p,q)$ charge of the string junction ending on the D3-brane.  BPS
string junctions correspond to curves holomorphic in the complex
structure whose k\"ahler form is $\Omega$. The mass of a BPS state of
charge $(p,q)$ is given by the area of the corresponding curve ${\cal
C}_{p,q}$,
\be
M_{p,q}= |\int_{{\cal C}_{p,q}}\Omega|\,.
\ee

\subsubsection{Superpotentials and F-theory backgrounds}

~~~~
We will see in this section that
the non-compact CY 3-folds with an equation of the form
$f(x_1,x_2)=uv$ where $x_1,x_2$ are ${\bf C}^*$ variables
and $u,v$ are ${\bf C}$ variables get related to the
F-theory probe description.  We first discuss
the structure of the BPS D3-branes in the non-compact
local Calabi-Yau description and then
relate it to the F-theory description.

 Instead of being general, we consider a concrete example.  The
 general case is similar. Consider the case of ${\cal O}(-3)$ over
 $\PP^{2}$. This non-compact CY threefold, which is the total space of
 ${\cal O}(-3)$ bundle over $\PP^{2}$, will be denoted by ${\cal M}$.
 This has linear sigma model description in terms of a single $U(1)$
 gauge theory with charges of the matter fields $(-3,1,1,1)$. The LG
 superpotential of the mirror theory is,
\be
W({x})=x_{0}+x_{1}+x_{2}+
e^{-t}\frac{x_{0}^{3}}{x_{1}x_{2}}\,.
\label{noncompact2}
\ee
As we
discussed before as far as the BPS data is concerned the non-compact
CY defined by eq.\,(\ref{noncompact2}) is equivalent to another
non-compact CY, $\widehat{\cal M}$, defined by,
\bea
1+x_{1}+x_{2}+\frac{e^{-t}}{x_{1}x_{2}}=-z\,,\,\, \qquad z=-uv\,.
\label{mirrorofM}
\eea
where $x_1,x_2$ are ${\bf C}^*$ variables and $u,v,z$ are
variables in ${\bf C}$.  In particular the relevant holomorphic
3-form is given by $\frac{dx_1dx_2du}{x_1x_2u}$ (by eliminating $z,v$
and noting that the denominator has $\partial uv/\partial v$).
 To better understand the
geometry of $\widehat{\cal M}$ we rewrite the defining equation
eq.\,(\ref{mirrorofM}) in the following form,
\bea
h(x_{1},x_{2},z)&:=&x_{1}^{2}x_{2}+x_{1}x_{2}^2+1+ \,z\, x_{1}x_{2}=0
\,\label{first}\\ 
z-e^{t/3}&=&-uv\,. \label{2nd}
\eea
where we have rescaled variables and shifted $z$.  In this form
the holomorphic 3-form becomes
\be
\Omega={dx_1dx_2du\over x_1x_2u}={ dx_1dx_2\over
\partial h/\partial z}\cdot {du\over u}=\Omega_2 {du\over u}
\ee
The first equation in eq.\,(\ref{first}) 
defines
an elliptic fibration (in terms
of the $x_1,x_2$ variables over the complex plane with coordinate $z$).
Moreover the corresponding two form $\Omega_2$ is the same
as would be for a K3 geometry where $x_1,x_2$ are now viewed
as variables in ${\bf C}$. 
 It is more convenient to homogenize the above elliptic curve
by introducing an extra variable $x_0$:
\be
x_1^2x_2+x_1x_2^2+x_0^3+zx_0x_1x_2=0
\ee

 We can convert this into the
Weierstrass form by the following coordinate transformation,
\bea
x_{1}=Y+\fracs{{U}}{{2}}+z\fracs{{X}}{{2}}\,
,\,x_{2}=-Y+\fracs{{U}}{{2}}+z\fracs{{X}}{{2}}\,,\,x_{0}=X\,,\\
UY^{2}=X^{3}+(\fracs{{z}}{{2}})^{2}UX^{2}+\fracs{{z}}{{2}}U^{2}X+\fracs{{1}}{{4}}U^{3}\,,
\eea
where now $U$ is a scaling variable and we can set it to $1$.
Thus the BPS data of $\widehat{\cal M}$ is the same as that
of
\be
Y^2=X^3+(\fracs{{z}}{{2}})^{2}X^2+\fracs{{z}}{{2}}\, X+\fracs{{1}}{{4}}\qquad,\,\,\, z-e^{t/3}=-uv\,,
\ee
with the holomorphic 3-form
\be
\Omega ={dX\over Y}dz {du\over u}.
\ee
Thus we see that $\widehat{\cal M}$ can be viewed as the
product of an elliptic fibration times a ${\bf C}^*$ fibration
over $z$. The elliptic fibration has a discriminant
given by $\Delta(z)=\frac{1}{16}(27-z^3)$, the
three degenerate fibers are located symmetrically at
$(27)^{\frac{1}{3}}\{1, e^{2\pi i/3},e^{4\pi
i/3}\}=:\{z_{1},z_{2},z_{3}\}$.  The ${\bf C}^*$ fibration
has one degenerate fiber given at $z=e^{t/3}$.

To determine supersymmetric 3-cycles in $\widehat{\cal M}$ we use
the circle of ${\bf C}^*$ fibration as one cycle times an additional 
2-cycle.
We note that there
are only two closed 2-cycles in this fibration. One is the elliptic
fiber of the fibration itself and the other closed 2-cycle is formed
by taking a path in the base that encloses $z_{i}$ and the $(1,0)$ cycle of
the elliptic fiber above this path.\footnote{ The existence of the 2nd type of
closed 2-cycle is precisely the reason this fibration can be used to
construct 5-brane web description of 5D $E_{0}$ field theory which can
also be obtained via geometric engineering from the CY threefold
${\cal M}$ \cite{DHIK}.} The fact that there are no other closed
2-cycles is not obvious even though we do not have degenerating cycles
of the same charge, since a cycle starting from a degenerate fiber can
undergo monodromy transformations when it goes around other
degenerating fibers. One can, however, construct 2-cycles with
boundaries such that the boundary is a 1-cycle of an elliptic fiber
above the point $z_{*}$.  We can construct closed 3-cycles using the
1-cycle of the $\c^{\,\,*}$ fibration and the 2-cycles (with
boundaries) if we choose the position of the boundary carefully. If
$z_{*}=e^{\frac{t}{3}}$ then the boundary of the 2-cycle is exactly
at the point on the base where the $\c^{\,\,*}$ fibration
degenerates. In this case the 2-cycle and the 1-cycle of the
$\c^{\,\,*}$ fibration together define a closed 3-cycle
\cite{estring,HI}. Since the 1-cycle of the $\c^{\,\,*}$ fibration is
always present the essential geometry of the 3-cycle is captured by
the 2-cycle with boundary at $z_{*}=e^{\frac{t}{3}}$. This
gets translated to finding minimal 2 surfaces in the corresponding
$K3$ geometry with boundary being a circle on a particular elliptic fiber.

The connection with F-theory is now rather clear.  In fact
this elliptic fibration defines an F-theory background studied before
in the context of non-BPS stable states in F-theory \cite{SZ} and
compactification of 5D $E_{n}$ field theories on a circle \cite{YY}.
As usual in the F-theory description, we can assign
 $(p,q)$ charges to the 7-branes
 which correspond to $(p,q)$ degenerating cycle of $T^2$ (
 which can be defined by choosing
 paths to
 a base point, $z_{0}$). In this case the charges are as shown in
\figref{b0-a} \cite{SZ,YY}.Note that the $(p,q)$ charge of the 
three degenerating cycles can be cyclically transformed by the $\sl2z$
matrix $ST$, $(ST)^{3}=1$ \cite{YY}. 
\onefigure{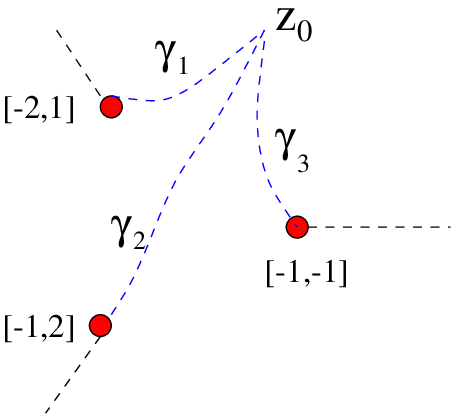}{}

The relation with the probe theory and its BPS states is now
clear. The degenerating fibers at $z_{i}$ define a 7-brane background
and the degenerating 1-cycle at $z_{*}=e^{\frac{t}{3}}$ of the
$\c^{\,\,*}$ fibration defines the position of the D3-brane. The
2-cycles with boundary are strings and string junctions stretched
between the 7-branes and the D3-brane. BPS states in the D3-brane
theory correspond to BPS string junctions which are the projections of
holomorphic 2-cycles. Thus we see that D-branes wrapped on even
dimensional cycles of ${\cal M}$ are mirror to states in the D3-brane
worldvolume theory. This connection between sheaves on a non-compact
CY and states in the field theory realized on a D3-brane in the
presence of 7-branes was also studied in \cite{HI}. 

\subsubsection{Soliton Numbers for $\PP^2$ and its Blowups}

~~~~
As noted before,
the elliptic fibration defined by eq.\,(\ref{first})
is exactly the W-plane geometry of the massive LG theory mirror to
sigma model with $\PP^{2}$ target space. The vanishing cycles
are just the cycles of the $T^2$ fiber, and so we can use
the knowledge of the degeneration types to find the intersection
number of vanishing cycles, and thus the soliton
numbers of this theory. From \figref{b0-a} it
follows that,
\be
\gamma_{i}\circ \gamma_{j}=\pmatrix{0 &-3 & ~~3 \cr 0 & ~~0 &  -3 \cr 0 & ~~0 & ~~0}\,.
\ee
Where we have written the intersection matrix as an upper triangular
matrix. Similarly, as mentioned before, we can consider other
non-compact CY threefolds ${\cal M}_{n}$ which are the total space of
the canonical line bundle over the toric del Pezzo, ${\cal B}_{n}$. As
for the case of $\PP^{2}$, in this case as well the mirror is an
elliptic fibration and a $\c^{\,\,*}$ fibration over the z-plane. The
elliptic fibration is defined by the superpotential of the
corresponding massive LG theory. The corresponding F-theory
backgrounds and the D3-brane theory were studied in \cite{7-brane,HI}. Since
the charges of the vanishing cycles are known for these cases we can
compute the soliton counting matrix and compare with the matrices
obtained from the collection of exceptional bundles. In the following we will
denote the superpotential of the LG theory mirror to sigma model on $X$ as $W_{X}$.

\noindent
{\bf ${\cal B}_{1}$:} There are four degenerate fibers in this case as
shown in \figref{b1-a}. 
\onefigure{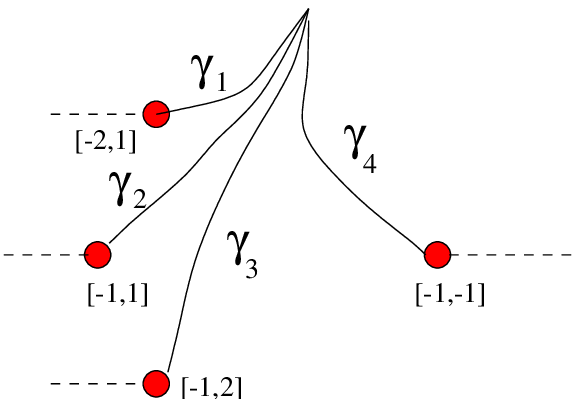}{Positions of the degenerate fibers of $W_{{\cal
B}_{1}}$}
Three out of four cycles are the same as before.  The new degenerate
fiber has charge $(-1,1)$. The intersection matrix is,
\be
\gamma_{i}\circ \gamma_{j}=\pmatrix{0& -1 & -3 & ~~3 \cr 0 & ~~0 & -1 &  ~~2 \cr  0 & ~~0& ~~0 & -3 
\cr 0&~~0&~~0&~~0}\,.
\ee
One can check that this matrix produces correct Ramond charges for the
chiral fields.

\noindent
{\bf ${\cal B}_{2}$:} In this case there are five degenerate fibers as
shown in \figref{b2-a}. There are two mutually local (with the same charge) fibers.
\onefigure{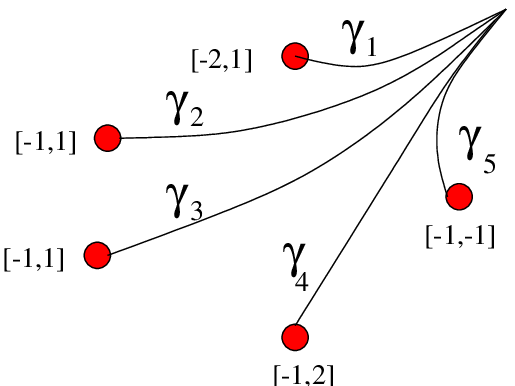}{Positions of the degenerate fibers of $W_{{\cal
B}_{2}}$.}
The intersection matrix is given by
\be
\gamma_{i}\circ \gamma_{j}=\pmatrix{0& -1 & -1 & -3 & ~~3 
\cr 0 & ~~0 & ~~0  & -1 & ~~2 \cr  0 & ~~0& ~~0 & -1 & ~~2 
\cr 0 & ~~0 &~~0 & ~~0 & -3\cr 0&~~0&~~0&~~0& ~~0}\,.\ee

\noindent
{\bf ${\cal B}_{3}$:}
The six degenerate fibers in this case are shown in \figref{b3-aa}. In 
this case there are three mutually local fibers.
\onefigure{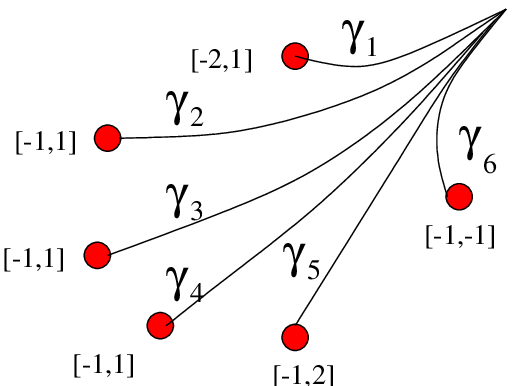}{Positions of the degenerate fibers of $W_{{\cal
B}_{3}}$.}
The intersection matrix is given by
\be
\gamma_{i}\circ \gamma_{j}=\pmatrix{0& -1 & -1 &-1 & -3 & ~~3 
\cr 0 & ~~0 & ~~0  &0 & -1 & ~~2 \cr  0 & ~~0& ~~0&0 & -1 & ~~2 
\cr  0 & ~~0& ~~0&0 & -1 & ~~2 
\cr 0 & 0& ~~0 &~~0 & ~~0 & -3\cr 0& 0&~~0&~~0&~~0& ~~0}\,.
\ee

\noindent
{\bf $F_{0}$:}
The four degenerate fibers in this case are shown in \figref{p1p1}. 
\onefigure{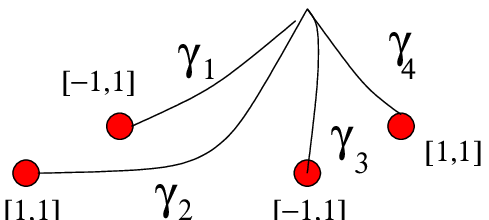}{Positions of the degenerate fibers of $W_{F_{0}}$.}
The intersection matrix is given by
\be
\gamma_{i}\circ \gamma_{j}=\pmatrix{0& -2 & 0 &2  
\cr 0 & ~~0 & 2  &0  \cr  0 & ~~0& ~~0&2
\cr  0 & ~~0& ~~0&0 }\,.
\ee
One can check that these matrices give the correct Ramond charge for the
chiral fields.

\section*{Acknowledgement}

\noindent
We would like to thank M. Aganagic, C. Bachas,
A. Chari, S. Coleman,
M. Gutperle, R. Myers,
A. Polishchuk, R. Thomas, S.T. Yau and E. Zaslow
for valuable discussions. AI would also like to thank Asad Naqvi for useful 
discussions.
K.H. would like to thank McGill University for hospitality.
The research of K.H. is supported in
part by NSF-DMS 9709694.
The research of A.I. is supported in part by the
US Department of Energy under contract
\#DE-FC02-94ER40818.
The research of C.V.
is supported in part by NSF grant PHY-98-02709.


\end{document}